\documentclass[citeautoscript,floatfix,aps,prl,twocolumn,superscriptaddress,longbibliography]{revtex4-1}  %
\usepackage{graphicx}  %
\usepackage{dcolumn}   %
\usepackage{bm}        %
\usepackage{amssymb}   %
\usepackage{amsmath}
\usepackage{amsthm}
\newtheorem{theorem}{Theorem}
\newtheorem{lemma}[theorem]{Lemma}
\newtheorem{fact}{Fact}[section]
\newtheorem{remark}{Remark}[section]
\theoremstyle{definition}
\newtheorem{definition}{Definition}[section]
\usepackage{multirow}
\usepackage{natbib}
\usepackage{braket}
\usepackage{amsfonts}
\usepackage{dsfont}
\usepackage{hyperref}
\usepackage{xcolor}

\usepackage{mathtools}

\usepackage{tikz}
\usepackage{tensor}
\usepackage{tikz-network}
\usepackage{paratensor}
\usepackage{paraKDHlattice}
\usepackage{color}
\definecolor{mypurple}{RGB}{255,0,255}

\makeatletter
\newcommand{\customlabel}[2]{%
	\protected@write \@auxout {}{\string \newlabel {#1}{{#2}{\thepage}{#2}{#1}{}} }%
	\hypertarget{#1}{#2}
}
\makeatother
\newcommand{\cc}{\xi}
\begin{document}

	\title{Particle exchange statistics beyond fermions and bosons}
	\author{Zhiyuan Wang}
	\affiliation{Department of Physics and Astronomy, Rice University, Houston, Texas 77005,
		USA}
	\affiliation{Rice Center for Quantum Materials, Rice University, Houston, Texas 77005, USA}
	\affiliation{Max-Planck-Institut f{\"{u}}r Quantenoptik, Hans-Kopfermann-Str. 1, 85748 Garching, Germany}
	\author{Kaden R.~A. Hazzard}
	\affiliation{Department of Physics and Astronomy, Rice University, Houston, Texas 77005,
		USA}
	\affiliation{Rice Center for Quantum Materials, Rice University, Houston, Texas 77005, USA}
	\date{\today}

	\begin{abstract}
		It is commonly believed that there are only two types of particle exchange statistics in quantum mechanics, fermions and bosons, with the exception of anyons in two dimension~\cite{Leinaas1977,Wilczek1982Magnetic,*Wilczek1982Quantum,Wilczek1990book,Nayak2008NAAnyons}. In principle, a second exception known as parastatistics, which extends outside of two dimensions, has been considered~\cite{Green1952} but was believed to be physically equivalent to fermions and bosons~\cite{doplicher1971local, *doplicher1974local,DR1972}. 
		Here we show that nontrivial parastatistics inequivalent to either fermions or bosons can exist in physical systems. 
		These new types of identical particles obey generalized exclusion  principles, leading to exotic free-particle thermodynamics distinct from any system of free fermions and bosons. 
		We formulate our theory by developing a second quantization of paraparticles, which naturally includes exactly solvable non-interacting theories, and incorporates physical constraints such as locality. 
		We then construct a family of exactly solvable quantum spin models in one and two dimensions where free paraparticles emerge as quasiparticle excitations, and their exchange statistics can be physically observed  and is notably distinct from fermions and bosons. 
		This demonstrates the possibility of a new type of quasiparticle in condensed matter systems, and, more speculatively, the potential for  previously unconsidered types of elementary particles.
	\end{abstract}
	
	\maketitle
	\paragraph{Introduction}
	It is commonly believed that there are only two types of particle exchange statistics --- fermions and bosons. The standard textbook argument for this dichotomy goes as follows. Each multiparticle quantum state is described by a wavefunction $\Psi(x_1,x_2,\ldots, x_n)$, a complex-valued function of particle coordinates  in a $d$ dimensional space $x_1,x_2,\ldots, x_n\in \mathbb{R}^d$. The particles are identical, meaning that when we exchange any two of them~(say $x_1,x_2$), the resulting wavefunction $\Psi(x_2,x_1,\ldots, x_n)$ must represent the same physical state, and therefore can change by at most a constant factor%
	\begin{equation}\label{eq:wavefuntion_exchange}
		\Psi(x_2,x_1,\ldots, x_n)=c \Psi(x_1,x_2,\ldots, x_n).
	\end{equation} 
	If we do a second exchange, %
	we have 
	\begin{eqnarray}\label{eq:wavefuntion_exchange_2nd}
		\Psi(x_1,x_2,\ldots, x_n)&=&c \Psi(x_2,x_1,\ldots, x_n)\nonumber\\
		&=&c^2 \Psi(x_1,x_2,\ldots, x_n),
	\end{eqnarray}
	leading to $c^2=1$, since the wavefunction cannot be constantly zero. This provides exactly two possibilities, bosons~($c=1$) and fermions~($c=-1$). 
	
	Despite being simple and convincing, there are two important exceptions to the fermion/boson dichotomy. The first is anyons  in two spatial dimension~(2D)~\cite{Leinaas1977,Wilczek1982Magnetic,*Wilczek1982Quantum,Wilczek1990book,Nayak2008NAAnyons,STERN2008204,Suppl}. %
	The second  %
	is parastatistics~\cite{Green1952,Araki1961,Greenberg1965,LANDSHOFF196772,druhl1970parastatistics,Taylor1970b}, which can be consistently defined in any dimension. The way this evades the above  argument is that the wavefunction can carry extra indices that transform nontrivially during an exchange. Consider an $n$-particle wavefunction $\Psi^I(x_1,x_2,\ldots, x_n)$, where $I$ is a collection of extra indices corresponding to some internal degrees of freedom inaccessible to local measurements. Under an exchange between particles $j$ and $j+1$~\footnote{Note that we only need to specify the behavior of the wavefunction under exchange of particles with adjacent labels, since exchange of particles with nonadjacent labels can always be decomposed into a series of adjacent exchanges. For example, under the exchange of particles $1$ and $3$, the wavefunction should multiply by the matrix $R_{1}R_{2}R_{1}$. }, the wavefunction may undergo a matrix transformation
		\begin{equation}\label{eq:wavefuntion_exchange_para}
		\Psi^I(\{x_i\}^n_{i=1})|_{x_j\leftrightarrow x_{j+1}}=\sum_J (R_{j})^I_{J} \Psi^J(\{x_i\}^n_{i=1}),
	\end{equation} 
	for $j=1,\ldots,n-1$, where %
	the summation is over all possible values of $J$. 
	Similar to  the $c^2=1$ constraint for Eq.~\eqref{eq:wavefuntion_exchange}, the matrices $(R_{j})^I_{J}$ have to satisfy some algebraic constraints to guarantee consistency of Eq.~\eqref{eq:wavefuntion_exchange_para}:
	\begin{eqnarray}\label{eq:SNbraidrelations}
		\begin{tikzpicture}[baseline={([yshift=.5ex]current bounding box.center)}, scale=0.6]
			\draw[red, very thick] (-\AL,-2*\AL) .. controls (\AL, 0) .. (-\AL,2*\AL);
			\draw[very thick] (\AL,-2*\AL) .. controls (-\AL, 0) .. (\AL,2*\AL);
			\node  at (-\AL,-2.5*\AL) {\footnotesize $j$};
			\node  at (\AL,-2.5*\AL) {\footnotesize $j+1$};
		\end{tikzpicture}\!\!\!\!\!&=&\!
		\begin{tikzpicture}[baseline={([yshift=.5ex]current bounding box.center)}, scale=0.6]
			\draw[red, very thick] (-\AL,-2*\AL) -- (-\AL,2*\AL);
			\draw[very thick] (.7*\AL,-2*\AL) -- (.7*\AL,2*\AL);
			\node  at (-\AL,-2.5*\AL) {\footnotesize $j$};
			\node  at (.7*\AL,-2.5*\AL) {\footnotesize $j+1$};
		\end{tikzpicture}\!\!,~~~
		\begin{tikzpicture}[baseline={([yshift=.5ex]current bounding box.center)}, scale=0.6]
			\draw[red, very thick] (-2*\AL,-2*\AL) .. controls (-\AL, 0) and (\AL, 0) .. (2*\AL,2*\AL);
			\draw[very thick] (0,-2*\AL) .. controls (-\AL, 0) .. (0*\AL,2*\AL);
			\draw[blue, very thick] (2*\AL,-2*\AL) .. controls (\AL, 0) and (-\AL, 0) .. (-2*\AL,2*\AL);
			\node  at (-2*\AL,-2.5*\AL) {\footnotesize $j-1$};
			\node  at (0*\AL,-2.5*\AL) {\footnotesize $j$};
			\node  at (2*\AL,-2.5*\AL) {\footnotesize $j+1$};
		\end{tikzpicture}=
		\begin{tikzpicture}[baseline={([yshift=.5ex]current bounding box.center)}, scale=0.6]
			\draw[red, very thick] (-2*\AL,-2*\AL) .. controls (-\AL, 0) and (\AL, 0) .. (2*\AL,2*\AL);
			\draw[very thick] (0,-2*\AL) .. controls (\AL, 0) .. (0*\AL,2*\AL);
			\draw[blue, very thick] (2*\AL,-2*\AL) .. controls (\AL, 0) and (-\AL, 0) .. (-2*\AL,2*\AL);
			\node  at (-2*\AL,-2.5*\AL) {\footnotesize $j-1$};
			\node  at (0*\AL,-2.5*\AL) {\footnotesize $j$};
			\node  at (2*\AL,-2.5*\AL) {\footnotesize $j+1$};
		\end{tikzpicture}\\
		R_{j}^2&=&\mathds{1},~~~~~~~~~~~R_{j-1}R_{j}R_{j-1}=R_{j}R_{j-1}R_{j},\nonumber
	\end{eqnarray}
	and $R_{i}R_{j}=R_{j}R_{i}$ for $|i-j|\geq 2$. 
	The derivation of the first equation is similar to Eq.~\eqref{eq:wavefuntion_exchange_2nd}, %
	the second equation is due to the equivalence of two different ways of swapping $x_{j-1},x_j,x_{j+1}$ to $x_{j+1},x_j,x_{j-1}$, and the last one is due to the commutativity of the swaps $x_{i}\leftrightarrow x_{i+1}$ and $x_{j}\leftrightarrow x_{j+1}$ for $|i-j|\geq 2$. %
	These constraints are equivalent to the requirement that $\{R_{j}\}^{n-1}_{j=1}$ generate a representation of the symmetric group $S_n$~\cite{kassel2008braid}. %
	If this representation is not one-dimensional, we say Eq.~\eqref{eq:wavefuntion_exchange_para} defines a type of parastatistical particles, or paraparticles for short. 
Notice that the first relation in Eq.~\eqref{eq:SNbraidrelations} is crucial for parastatistics to be consistently defined in any dimension; anyons generally do not satisfy this relation, and consequently they only form a representation of the braid group $B_n$~\cite{kassel2008braid} instead of the symmetric group $S_n$, and are therefore limited to 2D. 

Parastatistics, and their apparent absence in nature, has been discussed since the dawn of quantum mechanics~\cite{nobelphysics1946}. %
The first concrete %
theory of parastatistics was proposed and investigated by Green in 1953~\cite{Green1952}. %
This theory was subsequently studied in detail~\cite{Araki1961,Greenberg1965,LANDSHOFF196772,druhl1970parastatistics,Taylor1970b}, and also more generally and  rigorously~\cite{doplicher1971local, *doplicher1974local,DR1972, doplicher1990}  %
within the framework of algebraic quantum field theory~\cite{WightmanPCTbook, haag2012book}.
These works did not rule out the existence of paraparticles in nature, but led to the conclusion that under certain assumptions any theory of paraparticles~(in particular, Green's theory) is physically indistinguishable from theories of ordinary fermions and bosons. %
This seemingly obviated the need to consider paraparticle theories, as they give exactly the same physical predictions as  theories of ordinary particles.

In this paper we show that nontrivial paraparticles inequivalent to either fermions or bosons exist in physical systems, in a way compatible with spatial locality and Hermiticity. This poses no contradiction with earlier results, as the construction evades their restrictive assumptions. %
We demonstrate this by first introducing a second quantization formulation of parastatistics that is distinct from previous constructions~\cite{Suppl}, %
which includes exactly solvable theories of free paraparticles,
and in this formulation paraparticles can display non-Abelian permutation statistics~[Eq.~\eqref{eq:wavefuntion_exchange_para}] and generalized exclusion principles inequivalent to free fermions and bosons. 
Then we show that these  paraparticles emerge as quasiparticle excitations in a family of exactly solvable quantum spin models, %
explicitly demonstrating how to avoid  the aforementioned no-go theorems~\cite{doplicher1971local,doplicher1974local}, allowing nontrivial %
consequences of parastatistics to be physically observed.
Our second quantization 
formulation of paraparticles is valid in any spatial dimension, %
and can be extended to incorporate special relativity, %
hinting the potential existence of elementary paraparticles in nature. %

\paragraph{Basic formalism}
We first present our second quantization formulation of parastatistics. 
This formulation only realizes a subfamily of the parastatistics defined by the first quantization approach presented above, %
but the payoff is that it automatically guarantees the fundamental requirement of spatial locality, which is not ensured by the first quantization formulation~\footnote{It is hard to guarantee locality in the first quantization formulation, and without locality such theories are hard to be realized as elementary particles or as emergent quasiparticle excitations in locally interacting systems. Note that it is exactly for this reason that the Doplicher-Haag-Roberts~(DHR) no-go theorem~\cite{doplicher1971local,*doplicher1974local} does not apply to the first quantization formulation, since the former takes locality as a fundamental assumption, while the latter does not have locality built-in. See Sec.~\ref{SI:relation_first_quantization} in the SI for the relation between the first and second quantization formulation of parastatistics in this paper. 
}. 
In this formulation, each type of parastatistics is labeled by a four-index tensor $R^{ab}_{cd}$~(where $1\leq a,b,c,d\leq m$, $m\in \mathbb{Z}$) satisfying %
\begin{equation}\label{eq:YBE}
	\begin{tikzpicture}[baseline={([yshift=-.8ex]current bounding box.center)}, scale=0.5]
		\Rmatrix{0}{\AL}{R}
		\Rmatrix{0}{-\AL}{R}
		\node  at (-\AL,2.5*\AL) {\footnotesize $a$};
		\node  at (\AL,2.5*\AL) {\footnotesize $b$};
		\node  at (-\AL,-2.5*\AL) {\footnotesize $c$};
		\node  at (\AL,-2.5*\AL) {\footnotesize $d$};
	\end{tikzpicture}=
	\begin{tikzpicture}[baseline={([yshift=-.8ex]current bounding box.center)}, scale=0.5]
		\draw[thick] (-\AL,-2*\AL) -- (-\AL,2*\AL);
		\draw[thick] (\AL,-2*\AL) -- (\AL,2*\AL);
		\node  at (-\AL,2.5*\AL) {\footnotesize $a$};
		\node  at (\AL,2.5*\AL) {\footnotesize $b$};
		\node  at (-\AL,-2.5*\AL) {\footnotesize $c$};
		\node  at (\AL,-2.5*\AL) {\footnotesize $d$};
		\node  at (-1.5*\AL,0*\AL) {\footnotesize $\delta$};
		\node  at (1.5*\AL,0*\AL) {\footnotesize $\delta$};
	\end{tikzpicture}~,~~~
	\begin{tikzpicture}[baseline={([yshift=-.8ex]current bounding box.center)}, scale=0.5]
		\Rmatrix{-\AL}{2*\AL}{R}
		\Rmatrix{\AL}{0}{R}
		\Rmatrix{-\AL}{-2*\AL}{R}
		\draw[thick] (-2*\AL,-\AL) -- (-2*\AL,\AL);
		\draw[thick] (2*\AL,\AL) -- (2*\AL,3*\AL);
		\draw[thick] (2*\AL,-\AL) -- (2*\AL,-3*\AL);
		\node  at (-2*\AL,3.5*\AL) {\footnotesize $a$};
		\node  at (0*\AL,3.5*\AL) {\footnotesize $b$};
		\node  at (2*\AL,3.5*\AL) {\footnotesize $c$};
		\node  at (-2*\AL,-3.7*\AL) {\footnotesize $d$};
		\node  at (0*\AL,-3.7*\AL) {\footnotesize $e$};
		\node  at (2*\AL,-3.7*\AL) {\footnotesize $f$};
	\end{tikzpicture}
	=\begin{tikzpicture}[baseline={([yshift=-.8ex]current bounding box.center)}, scale=0.5]
		\Rmatrix{\AL}{2*\AL}{R}
		\Rmatrix{-\AL}{0}{R}
		\Rmatrix{\AL}{-2*\AL}{R}
		\draw[thick] (2*\AL,-\AL) -- (2*\AL,\AL);
		\draw[thick] (-2*\AL,\AL) -- (-2*\AL,3*\AL);
		\draw[thick] (-2*\AL,-\AL) -- (-2*\AL,-3*\AL);
		\node  at (-2*\AL,3.5*\AL) {\footnotesize $a$};
		\node  at (0*\AL,3.5*\AL) {\footnotesize $b$};
		\node  at (2*\AL,3.5*\AL) {\footnotesize $c$};
		\node  at (-2*\AL,-3.7*\AL) {\footnotesize $d$};
		\node  at (0*\AL,-3.7*\AL) {\footnotesize $e$};
		\node  at (2*\AL,-3.7*\AL) {\footnotesize $f$};
	\end{tikzpicture},
\end{equation}
where $R^{ab}_{cd}=\!\!\begin{tikzpicture}[baseline={([yshift=-.6ex]current bounding box.center)}, scale=0.45]
	\Rmatrix{0}{0}{R}
	\node  at (-1.5*\AL,\AL) {\footnotesize $a$};
	\node  at (1.5*\AL,\AL) {\footnotesize $b$};
	\node  at (-1.5*\AL,-\AL) {\footnotesize $c$};
	\node  at (1.5*\AL,-\AL) {\footnotesize $d$};
\end{tikzpicture}$,
and throughout this paper we use tensor graphical notation where open indices are identified on both sides of the equation and contracted indices are summed over, and a line segment represents a Kronecker $\delta$ function.
These two equations are reminiscent of Eq.~\eqref{eq:SNbraidrelations}, and  we describe their precise relation in the Supplementary Information~(SI)~\cite{Suppl}. %
	The second equation in Eq.~\eqref{eq:YBE} is known in the literature as the constant Yang-Baxter equation~(YBE)~\cite{Turaev1988,Majid1990,etingof1999set}, whose solutions are called $R$-matrices. %
	In Tab.~\ref{tab:Hilbert_series} we present some basic examples of $R$-matrices, and one can check by straightforward computation that they satisfy Eq.~\eqref{eq:YBE}. %
 \begin{table}%
	\centering
	{\renewcommand{\arraystretch}{1.5}
		\begin{tabular}{|c|c|c|c|c|}
			\hline
			Ex. & \customlabel{ex:decoupled}{1}	& 	\customlabel{ex:Green}{2} & \customlabel{ex:1m}{3} & \customlabel{ex:1m1}{4}  \\
			\hline
			$R^{ab}_{cd}$ & $-\delta_{ad}\delta_{bc}$ & $\delta_{ad}\delta_{bc}(-1)^{\delta_{ab}}$&$-\delta_{ac}\delta_{bd}$&$\lambda_{ab}\xi_{cd}-\delta_{ac}\delta_{bd}$\\
			\hline
			$z_R(x)$  & $(1+x)^m$	 & $(1+x)^m$ &$1+m x$ &$1+mx+x^2$\\
			\hline
		\end{tabular}
	}
	\caption{\label{tab:Hilbert_series} Examples of $R$-matrices and their single mode partition functions $z_R(x)$, as defined in Eq.~\eqref{eq:single_mode_Z}, where $x=e^{-\beta\epsilon}$. The $\lambda, \xi$ in Ex.~\ref{ex:1m1} are $m\times m$ constant matrices satisfying $	\lambda \xi \lambda^T \xi^T=\mathds{1}_{m}$ and $\mathrm{Tr}(\lambda \xi^T)=2$~\cite{footnote_lambdac}.}
\end{table}

For a given $R$-matrix, we define the paraparticle creation and annihilation operators $\hat{\psi}^\pm_{i,a}$ through the commutation relations~(CRs) 
	\begin{eqnarray}\label{eq:fundamental_Rcommu}
		\hat{\psi}^-_{i,a} \hat{\psi}^+_{j,b}&=&\sum_{cd}R^{ac}_{bd} \hat{\psi}_{j, c}^+ \hat{\psi}_{i,d}^-+\delta_{ab}\delta_{ij},\nonumber\\%\label{eq:RMQAmp}
		\hat{\psi}^+_{i,a} \hat{\psi}^+_{j,b}&=&\sum_{cd}R^{cd}_{ab} \hat{\psi}_{j,c}^+ \hat{\psi}_{i,d}^+,\nonumber\\%\label{eq:RMQApp}\\
		\hat{\psi}^-_{i,a} \hat{\psi}^-_{j,b}&=&\sum_{cd}R^{ba}_{dc} \hat{\psi}_{j,c}^- \hat{\psi}_{i,d}^-,%
	\end{eqnarray}
	where $i,j$ are mode indices~(e.g., position, momentum), and $a,b,c,d$ are internal indices.
Notice that $R^{ab}_{cd}=\pm\delta_{ad}\delta_{bc}$ %
gives back fermions~($-$) and bosons~($+$) with an internal degree of freedom. %
While our construction works for any $R$-matrix satisfying Eq.~\eqref{eq:YBE},
in this paper we mainly focus on unitary $R$-matrices for simplicity, i.e.,  $\sum_{a,b} R^{ab}_{cd} (R^{ab}_{ef})^*=\delta_{ce}\delta_{df}$, which is true for Exs.~(\ref{ex:decoupled}-\ref{ex:1m}) in Tab.~\ref{tab:Hilbert_series} and the $R$-matrix of the 2D solvable spin model~[Eqs.~(\ref{eq:seth-R},\ref{eq:set-thR}) in Methods]. With a unitary $R$, we have $\hat{\psi}^+_{i,a}=(\hat{\psi}^-_{i,a})^\dagger$~\cite{Suppl}, which guarantees the Hermiticity of physical observables as we show later~\footnote{As we show in the SI~\cite{Suppl}, even with a non-unitary $R$-matrix, such as Ex.~\ref{ex:1m1} in Tab.~\ref{tab:Hilbert_series}, Eq.~\eqref{eq:fundamental_Rcommu} is still consistently defined, and we can still define a Hermitian inner product on the state space, with respect to which all physical observables %
are Hermitian, and most of the main results of this paper still apply, including the generalized exclusion statistics, the exact solution of free particles, and the construction of solvable spin models with emergent paraparticles.}.

A crucial structure in our construction is the Lie algebra of contracted bilinear operators defined as
\begin{equation}\label{eq:def_e_ab}
	\hat{e}_{ij}\equiv \sum^m_{a=1} \hat{\psi}^+_{i,a}\hat{\psi}^-_{j,a}.
\end{equation}
We show that the space $\{\hat{e}_{ij}\}_{1\leq i,j\leq N}$ is closed under the commutator $[\hat{A},\hat{B}]=\hat{A}\hat{B}-\hat{B}\hat{A}$, and the corresponding Lie algebra is $\mathfrak{gl}_N$. First, using Eq.~\eqref{eq:fundamental_Rcommu}, we have
\begin{eqnarray}\label{eq:commu_Eab_psi_p}
	[\hat{e}_{ij}, \hat{\psi}^+_{k,b}]&=&\delta_{jk}\hat{\psi}^+_{i,b},\nonumber\\
	{}[\hat{e}_{ij}, \hat{\psi}^-_{k,b}]&=&-\delta_{ik}\hat{\psi}^-_{j,b},
\end{eqnarray}
which leads to
\begin{eqnarray}\label{eq:commu_Eab_Ecd}
	[\hat{e}_{ij}, \hat{e}_{kl}]=\delta_{jk}\hat{e}_{il}-\delta_{il}\hat{e}_{kj}.
\end{eqnarray}
(See Methods for detailed derivation.) Eq.~\eqref{eq:commu_Eab_Ecd} is the CR between the basis elements $\{\hat{e}_{ij}\}_{1\leq i,j\leq N}$ of the $\mathfrak{gl}_N$ Lie algebra, where $\hat{e}_{ij}$ represents the matrix that has $1$ in the $i$-th row and $j$-th column and zero everywhere else. We will see that this Lie algebra structure enables straightforward construction of theories of paraparticles that obey locality, Hermiticity, and free particle solvability. %

In the usual case of fermions, physical observables are composed of even products of fermionic operators. This comes from the physical requirement of locality --- local observables supported on disjoint regions~(or space-like regions in relativistic quantum field theory) must commute. We define an analog for parastatistics and show they have analogous properties:
for each  local region of space $S$, %
we define a local observable on $S$ to be a Hermitian operator that is a sum of products of $\hat{e}_{ij}$ where $i,j\in S$. For example, $\hat{O}_{S}=\hat{e}_{ij}\hat{e}_{ji}$ with $i,j\in S$ is a local observable in $S$, since $\hat{e}^\dagger_{ij}=\hat{e}_{ji}$.
Then, Eq.~\eqref{eq:commu_Eab_Ecd} immediately implies  the aforementioned locality condition $[\hat{O}_{S_1},\hat{O}_{S_2}]=0$ for $S_1\cap S_2=\emptyset$.  %
A locally-interacting Hamiltonian $\hat{H}$ is defined to be a sum of local observables $\hat{H}=\sum_{S} h_{S} \hat{O}_S$, where $h_{S}\in \mathbb{R}$ and the summation is over local regions $S$ whose diameters are smaller than some constant cutoff. %
This definition of local observables and Hamiltonians guarantees unitarity~(time evolution $\hat{U}=e^{-i\hat{H}t}$ generated by a Hamiltonian operator $\hat{H}$ is unitary) and microcausality~(no signal can travel faster than a finite speed) in both relativistic quantum field theory and non-relativistic lattice quantum systems~\footnote{For the former, the commutativity of local observables  at space-like separations rules out faster-than-light travel and communication; for the latter, Ref.~\cite{wang2020tightening} proved that as long as all the local Hamiltonian terms have uniformly bounded norms and their algebra has a local structure, then the Lieb-Robinson bound~\cite{Lieb1972} holds, which gives an effective lightcone of causality.}. %

A particularly important family of physical observables are the particle number operators %
	$\hat{n}_i\equiv \hat{e}_{ii}$. %
It follows from Eq.~\eqref{eq:commu_Eab_Ecd} that they mutually commute $[\hat{n}_i,\hat{n}_j]=0$, so they have a complete set of common eigenstates. %
Meanwhile, Eq.~\eqref{eq:commu_Eab_psi_p}  gives $[\hat{n}_{i}, \hat{\psi}^\pm_{j,b}]=\pm\delta_{ij}\hat{\psi}^\pm_{j,b}$, meaning that $\hat{\psi}^+_{j,b}$~($\hat{\psi}^-_{j,b}$) increases~(decreases) the eigenvalue of $\hat{n}_j$ by $1$, and does not change the eigenvalue of $\hat{n}_i$ for $j\neq i$. This justifies the terminology creation and annihilation operators, since $\hat{\psi}^+_{j,b}$~($\hat{\psi}^-_{j,b}$) creates~(annihilates) a particle in the mode $j$. We also define the total particle number operator $\hat{n}=\sum_{i=1}^N \hat{n}_i$, so we have $[\hat{n}, \hat{\psi}^\pm_{j,b}]=\pm\hat{\psi}^\pm_{j,b}$. These CRs involving the number operators are the same as for fermions and bosons. However, we will see later that due to the generalized CRs between $\{\hat{\psi}^\pm_{i,b}\}$ in Eqs.~\eqref{eq:fundamental_Rcommu}, the spectrum of $\{\hat{n}_i \}$ is different for paraparticles.
\begin{figure}
	\center{\includegraphics[width=1\linewidth]{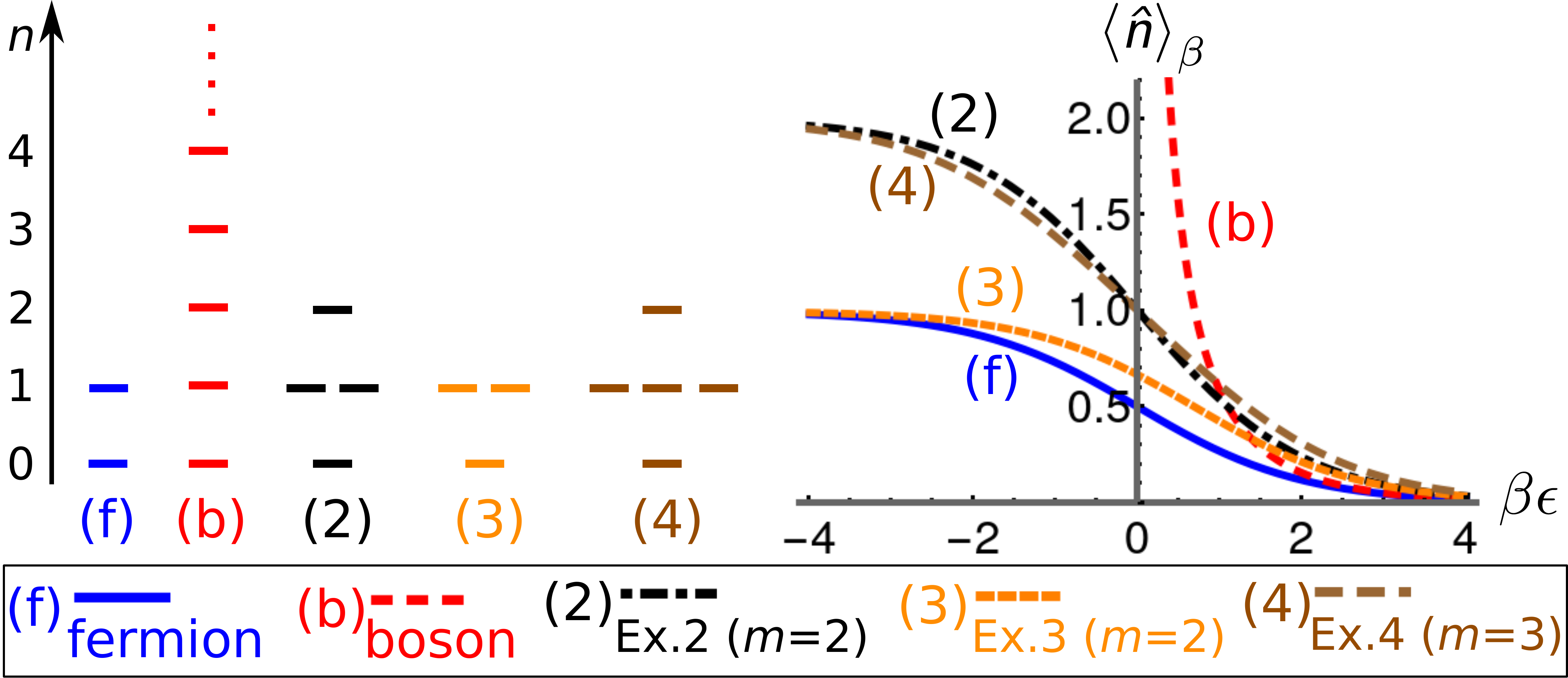}}
	\caption{\label{fig:level_statistics}\label{fig:Tempdep}  The generalized exclusion statistics and free particle thermodynamics of paraparticles defined by the $R$-matrices in Ex.~\ref{ex:decoupled}-\ref{ex:1m1} of Tab.~\ref{tab:Hilbert_series}, and a comparison to ordinary fermions and bosons. (left) the level degeneracy $\{d_n\}_{n\geq 0}$; (right)
	thermal expectation value of the single-mode occupation number $\langle \hat{n}\rangle_\beta$. %
	}
\end{figure}
\paragraph{Generalized exclusion statistics}
Paraparticles defined by the CRs in Eq.~\eqref{eq:fundamental_Rcommu} exhibit generalized exclusion statistics that is notably different from ordinary fermions and bosons. %
We demonstrate this phenomenon for the paraparticles defined by the $R$-matrix in Ex.~\ref{ex:1m} in Tab.~\ref{tab:Hilbert_series}, and present the general case in the SI~\cite{Suppl}.

Analogous to the Fock space of fermions and bosons, there is a vacuum state $|0\rangle$ satisfying $\hat{\psi}^-_{i,a}\ket{0}=0,~\forall i,a$, so the vacuum contains no particles, $\hat{n}|0\rangle=0$. 
The second line of Eq.~\eqref{eq:fundamental_Rcommu} with $R^{ab}_{cd}=-\delta_{ac}\delta_{bd}$~(Ex.~\ref{ex:1m} in Tab.~\ref{tab:Hilbert_series}) reads 
\begin{equation}\label{eq:CR_Ex3}
	\hat{\psi}^+_{i,a} \hat{\psi}^+_{j,b}=-\hat{\psi}_{j,a}^+ \hat{\psi}_{i,b}^+,~\forall i,j,a,b.
\end{equation}
Taking $i=j$ in Eq.~\eqref{eq:CR_Ex3}, we get $\hat{\psi}^+_{i,a} \hat{\psi}^+_{i,b}=0$, which means that any mode $i$ cannot be occupied by two paraparticles  even if they have different labels $a\neq b$, in contrast to fermions. Meanwhile, Eq.~\eqref{eq:CR_Ex3} does not imply any exclusion between paraparticles in different modes $i\neq j$, and the first line in Eq.~\eqref{eq:fundamental_Rcommu} implies that the one particle states $\hat{\psi}^+_{i,a}\ket{0}$ are orthonormal $\bra{0}\hat{\psi}^-_{j,b}\hat{\psi}^+_{i,a}\ket{0}=\delta_{ij}\delta_{ab}$. 
The whole %
state space is $(m+1)^N$-dimensional, spanned by orthonormal basis states of the form  
\begin{equation}\label{eq:state_space_Ex3}
|\Psi\rangle=\hat{\psi}^+_{i_1,a_1}\hat{\psi}^+_{i_2,a_2}\ldots \hat{\psi}^+_{i_n,a_n}|0\rangle,
\end{equation}
where $0\leq n\leq N$ and $1\leq i_1<i_2<\ldots<i_n\leq N$, and the action of $\hat{\psi}^\pm_{i,a}$ on these basis states is completely determined by the CRs in Eq.~\eqref{eq:fundamental_Rcommu}~\cite{Suppl}. %

For a general $R$-matrix, a single mode $i$ can be occupied by multiple particles, and the space of $n$-particle states $\hat{\psi}^+_{i,a_1}\hat{\psi}^+_{i,a_2}\ldots\hat{\psi}^+_{i,a_n}\ket{0}$ is $d_n$-dimensional, where $\{d_n\}_{n\geq 0}$ are non-negative integers that define the generalized exclusion statistics for the paraparticles associated with $R$, as illustrated in Fig.~\ref{fig:level_statistics}. In the above example, we have $d_0=1,d_1=m$, and $d_n=0 ,\forall n\geq 2$. This generalizes  %
Fermi-Dirac statistics
~(where $d_0=d_1=1$, and $d_n=0 ,\forall n\geq 2$), 
and Bose-Einstein statistics~(where $d_n=1, \forall n\geq0$)~\cite{Suppl}.  %
In the SI~\cite{Suppl} we show how to calculate $\{d_n\}_{n\geq 0}$ for a general $R$-matrix. %

The numbers $\{d_n\}_{n\geq 0}$ allow us to compute the grand canonical partition function for a single mode at temperature $T$. Suppose that each particle in this mode carries energy $\epsilon$~(i.e., the   Hamiltonian is $\hat{H}=\epsilon\hat{n}$). Then 
\begin{equation}\label{eq:single_mode_Z}
	z_R(e^{-\beta\epsilon})\equiv \mathrm{Tr} [e^{-\beta \epsilon \hat{n}}]=\sum_{n=0}^\infty d_n e^{-n\beta \epsilon },
\end{equation}
where $\beta=1/(k_B T)$, $k_B$ is Boltzmann's constant, and we have absorbed the chemical potential $\mu$ into $\epsilon$.
The  single mode partition functions $z_R(e^{-\beta\epsilon})$ for the $R$-matrices in Exs.~\ref{ex:decoupled}-\ref{ex:1m1} are given in Tab.~\ref{tab:Hilbert_series}.
Multi-mode partition functions factorize into products of single-mode partition functions exactly as for fermions and bosons.

The single mode partition function $z_R(x)$~(where $x=e^{-\beta\epsilon}$) provides a straightforward demonstration of the non-triviality~(i.e. distinct from fermions and bosons) of the parastatistics for some $R$-matrices. %
If the paraparticle system defined by $R$ can be transformed into a system of $p$ flavors of free fermions and $q$ flavors of free bosons, %
then $z_R(x)=(1+x)^p (1-x)^{-q}$. Therefore the $R$-matrix given in Ex.~\ref{ex:1m} must define a non-trivial type of parastatistics for $m\geq 2$, since $z_R(x)=1+mx$ is not equal to  %
$(1+x)^p (1-x)^{-q}$ for any integers $p,q$. 
Ex.~\ref{ex:1m1} is similarly non-trivial for $m\geq 3$~\footnote{A caveat, however, is that $z_R(x)$ only gives a sufficient condition for non-triviality, %
	and having a trivial $z_R(x)$ does not imply that the corresponding paraparticle theory is completely equivalent to fermions or bosons. %
For example, the emergent paraparticles in our 2D solvable spin model are defined by the $R$-matrix given in Eqs.~(\ref{eq:seth-R},\ref{eq:set-thR}), whose partition function $z_R(x)=(1+x)^4$ is the same as free fermions with an SU$(4)$ symmetry, but the exchange statistics of these emergent paraparticles are still physically~(observably) distinct from fermions and boson, as we show in Methods.  %
}. 

\paragraph{Particle exchange statistics}
In addition to generalized exclusion statistics, paraparticles defined by Eq.~\eqref{eq:fundamental_Rcommu} also display exotic exchange statistics defined by the $R$-matrix that results from physically exchanging paraparticles. 
Consider a state with two paraparticles at different positions $i\neq j$: $|0;ia,jb\rangle\equiv \hat{\psi}^+_{i,a}\hat{\psi}^+_{j,b}\ket{0}$. Let $\hat{E}_{ij}$ be a unitary operator that exchanges  the positions of the paraparticles at $i$ and $j$:
\begin{equation}\label{eq:Uij_action}
	\hat{E}_{ij}\hat{\psi}^+_{i,a}\hat{E}_{ij}^\dagger=\hat{\psi}^+_{j,a},\quad	\hat{E}_{ij}\hat{\psi}^+_{j,a}\hat{E}_{ij}^\dagger=\hat{\psi}^+_{i,a},\quad \forall a.
\end{equation}
Note that such an operator can always be constructed from a product of local unitaries of the form $e^{i\frac{\pi}{2} (\hat{e}_{kl}+\hat{e}_{lk})}$, which exchanges $\hat{\psi}^+_{k,a}\leftrightarrow \hat{\psi}^+_{l,a}$. 
The exchange operator $\hat{E}_{ij}$ acts on the two particle states $\ket{0;ia,jb}$ as 
\begin{eqnarray}\label{eq:braiding_derivation}
	\hat{E}_{ij}|0;ia,jb\rangle%
	&=&( \hat{E}_{ij}\hat{\psi}^+_{i,a}\hat{E}_{ij}^\dagger)(\hat{E}_{ij}\hat{\psi}^+_{j,b}\hat{E}_{ij}^\dagger)\hat{E}_{ij}\ket{0}\nonumber\\
	&=&\hat{\psi}^+_{j,a}\hat{\psi}^+_{i,b}\ket{0}\nonumber\\
	&=&\sum_{a',b'}R^{b'a'}_{ab}|0;ib',ja'\rangle,
\end{eqnarray}
where in the second line we applied Eq.~\eqref{eq:Uij_action} and the invariance of $\ket{0}$ under $\hat{E}_{ij}$, and in the third line we used the fundamental CR~\eqref{eq:fundamental_Rcommu}  between $\hat{\psi}^+_{j,a}$ and $\hat{\psi}^+_{i,b}$. %
Eq.~\eqref{eq:braiding_derivation} defines the physical meaning of the $R$-matrix as the unitary rotation of the two-particle state space %
that results from physically exchanging paraparticles, and in the solvable spin models with emergent paraparticles we present later, the effect of such a unitary rotation can be directly probed using local operations and measurements, which shows a striking difference from ordinary fermions and bosons, as we illustrate in Methods. The above derivation is valid in any spatial dimension and can be directly generalized to states with many paraparticles. %

\paragraph{Exact solution of free paraparticles}
In our second quantization framework, the general bilinear Hamiltonian describing free paraparticles,
\begin{equation}\label{eq:u1_sym_H}
	\hat{H}=\sum_{1\leq i,j\leq N} h_{ij} \hat{e}_{ij} =\sum_{\substack{1\leq i,j\leq N\\1\leq a\leq m}} h_{ij}  \hat{\psi}^+_{i,a}\hat{\psi}^-_{j,a},
\end{equation} 
can be solved analogously to bosons and fermions. We sketch this here; details can be found in Methods. We require $h_{ij}^*=h_{ji}$ so that $\hat{H}^\dagger=\hat{H}$. Using a canonical transformation of $\{\hat{\psi}^\pm_{i,a}\}$,  %
the Hamiltonian becomes %
	$\hat{H}=\sum^N_{k=1} \epsilon_{k} \tilde{n}_{k}$, 
where $\{\epsilon_{k}\}^N_{k=1}$ are the eigenvalues of the %
coefficient matrix $h_{ij}$,
and $\{\tilde{n}_{k}\}^N_{k=1}$ are mutually commuting occupation number operators for each mode $k$. 
The partition function of the whole system, $\mathrm{Tr} [e^{-\beta \hat{H}}]$, factorizes as a product of single mode partition functions in Eq.~\eqref{eq:single_mode_Z}, from which we obtain
the average occupation number of mode $k$ %
\begin{equation}\label{eq:n_k_expectation}
	\langle \tilde{n}_{k}\rangle_\beta\equiv \frac{\mathrm{Tr} [\tilde{n}_{k} e^{-\beta \hat{H}}]}{ \mathrm{Tr} [e^{-\beta \hat{H}}]}=\frac{z'_R(e^{-\beta\epsilon_k})e^{-\beta\epsilon_k}}{z_R(e^{-\beta\epsilon_k})}.
\end{equation}
Fig.~\ref{fig:Tempdep} plots $\langle \tilde{n}_{k}\rangle_\beta$ as a function of $\beta\epsilon_k$ for the $R$-matrices in Exs.~\ref{ex:1m} and \ref{ex:1m1}~(Tab.~\ref{tab:Hilbert_series}) with $m=5$, showing the distinct finite-temperature thermodynamics of paraparticles compared to ordinary fermions and bosons, characterizing a new type of ideal gas. %
\paragraph{Emergent paraparticles in condensed matter systems}%
We finally discuss the potential impacts of paraparticles, including routes to observe them in nature, starting with the promising setting for paraparticles  as quasiparticle excitations in condensed matter systems. Significant insight to this direction and a proof-of-principle that such excitations can occur in physical systems  are provided by a family of exactly solvable quantum spin systems where free paraparticles emerge as quasiparticle excitations. Here we present the one-dimensional~(1D) case for simplicity, and we also discuss a 2D model whose details are presented in Methods and SI~\cite{Suppl}. %
For each $R$-matrix, we define a Hamiltonian 
	\begin{equation}\label{eq:Hamil1Dspin}
		\hat{H}=\sum_{i,a}%
		 J_i(\hat{x}^+_{i,a}\hat{y}^-_{i+1,a}+\hat{x}^-_{i,a}\hat{y}^+_{i+1,a}) -\sum_{i,a} \mu_i \hat{y}^+_{i,a}\hat{y}^-_{i,a},
	\end{equation}
where $\{\hat{x}^\pm_{i,a},\hat{y}^\pm_{i,a}\}^m_{a=1}$ are local spin operators~(i.e. operators on different sites commute) acting on the $i$-th site, whose definition depends on the $R$-matrix. The index $i$ runs from $1$ to $N$, with $N$ being the system size, and we use open boundary condition $J_N=0$. The model has a total conserved charge $\hat{n}=\sum_{i,a}  \hat{y}^+_{i,a}\hat{y}^-_{i,a}$ which will be mapped to the paraparticle number operator, and $\hat{x}^+_{i,a},\hat{y}^+_{i,a}$~($\hat{x}^-_{i,a},\hat{y}^-_{i,a}$) increase~(decrease) $\hat{n}$ by $1$. For example, with the $R$-matrix in Ex.~\ref{ex:1m}, the local Hilbert space $\mathfrak{V}$ is $m+1$-dimensional, with basis states $|0\rangle, \{|1,b\rangle\}^m_{b=1}$, the $\hat{y}^\pm_a$ are defined as~(omitting the site label)  $\hat{y}^+_a|0\rangle=|1,a\rangle$, $\hat{y}^-_a|1,b\rangle=\delta_{ab}|0\rangle$,  $\hat{y}^-_a|0\rangle=\hat{y}^+_a|1,b\rangle=0$, and $\hat{x}^\pm_a=\hat{y}^\pm_a$. This is a simple, nearest-neighbor spin model that is realized in 3-level Rydberg atom or molecule systems~\cite{Sundar2018Synthetic,Sundar2019Strings}. %
For the definition of $\hat{x}^\pm_a$ and $\hat{y}^\pm_a$ in general, see SI~\cite{Suppl}. 
This model can be solved using a significant generalization of the Jordan-Wigner transformation~(JWT)~\cite{JW1928} that we introduce here, in which the  products of operators~(``strings'') are replaced with matrix product operators~(MPOs)~\cite{Cirac2021Matrix}. Specifically, we introduce operators 
		\begin{eqnarray}\label{eq:JWT_string}
	\hat{\psi}_{ia}^-&=&\begin{tikzpicture}[baseline={([yshift=.4ex]current bounding box.center)}, scale=0.64]
		\node  at (-1.5*\AL,0.*\AL) {\footnotesize $a$};
		\Tmmatrix{0}{0}{}
		\node  at (0,-1.6*\AL) {\footnotesize $1$};
		\Tmmatrix{2*\AL}{0}{}
		\node  at (2*\AL,-1.6*\AL) {\footnotesize $2$};
		\Tmmatrix{4*\AL}{0}{}
		\node  at (4*\AL,-1.6*\AL) {\footnotesize $3$};
		\draw[dotted, thick] (5*\AL, 0) -- (7*\AL,0);
		\Tmmatrix{8*\AL}{0}{}
		\node  at (8*\AL,-1.6*\AL) {\footnotesize $i-1$};
		\ytriangle{10*\AL}{0}{-}{}
		\node  at (10*\AL,-1.6*\AL) {\footnotesize $i$};
	\end{tikzpicture},\nonumber\\
	\hat{\psi}_{ia}^+&=&\begin{tikzpicture}[baseline={([yshift=.4ex]current bounding box.center)}, scale=0.64]
		\node  at (-1.5*\AL,0.*\AL) {\footnotesize $a$};
		\Tpmatrix{0}{0}{}
		\node  at (0,-1.6*\AL) {\footnotesize $1$};
		\Tpmatrix{2*\AL}{0}{}
		\node  at (2*\AL,-1.6*\AL) {\footnotesize $2$};
		\Tpmatrix{4*\AL}{0}{}
		\node  at (4*\AL,-1.6*\AL) {\footnotesize $3$};
		\draw[dotted, thick] (5*\AL, 0) -- (7*\AL,0);
		\Tpmatrix{8*\AL}{0}{}
		\node  at (8*\AL,-1.6*\AL) {\footnotesize $i-1$};
		\ytriangle{10*\AL}{0}{+}{}
		\node  at (10*\AL,-1.6*\AL) {\footnotesize $i$};
	\end{tikzpicture},
\end{eqnarray}
where $\hat{y}^\pm_{ja}\equiv\!\!\begin{tikzpicture}[baseline={([yshift=-.2ex]current bounding box.center)}, scale=0.64]
	\ytriangle{0}{0}{\pm}{}
	\node  at (0,-1.6*\AL) {\footnotesize $ j$};
	\node  at (-1.4*\AL,0) {\footnotesize $a$};
\end{tikzpicture}\!\!$, and $\hat{T}^\pm_{j,ab}\equiv\!\!\begin{tikzpicture}[baseline={([yshift=-.2ex]current bounding box.center)}, scale=0.64]
	\Tpmmatrix{0}{0}{}
	\node  at (0,-1.6*\AL) {\footnotesize $j$};
	\node  at (-1.4*\AL,0) {\footnotesize $a$};
	\node  at (1.4*\AL,0) {\footnotesize $b$};
\end{tikzpicture}\!\!=\mp [\hat{y}^\pm_{j,a},\hat{x}^\mp_{j,b}]$ 
are local spin operators acting on site $j$. Both $\hat{\psi}_{ia}^\pm$ act non-trivially on sites $1,2,\ldots,i$ and act as identity on the rest of the chain.  %
For example,  with the $R$-matrix in Ex.~\ref{ex:1m}, $\hat{T}^\pm_{ab}$ %
act as $\hat{T}^\pm_{ab}\ket{0}=\delta_{ab}\ket{0}$, $\hat{T}^-_{ab}\ket{1,c}=-\delta_{ac}\ket{1,b}$,  and  $\hat{T}^+_{ab}\ket{1,c}=-\delta_{bc}\ket{1,a}$. 
In the special case $m=1$, $R=-1$, %
$\hat{H}$ in Eq.~\eqref{eq:Hamil1Dspin} is the  Hamiltonian for the spin-1/2 XY model, the operators $\hat{\psi}^\pm_{i,a}$ are fermion creation and annihilation operators, and the MPO JWT simplifies to the ordinary JWT.

The $\hat{\psi}^\pm_{i,a}$ constructed in Eq.~\eqref{eq:JWT_string} satisfy the parastatistical CRs in Eq.~\eqref{eq:fundamental_Rcommu}, as we prove in the SI~\cite{Suppl} using tensor network manipulations. %
Moreover, the Hamiltonian in Eq.~\eqref{eq:Hamil1Dspin} can be rewritten in terms of $\{\hat{\psi}^\pm_{i,a}\}$ %
as 
\begin{equation}\label{eq:Hamil1Dspin_para}
	\hat{H}=\sum_{i,a} J_i(\hat{\psi}^+_{i,a}\hat{\psi}^-_{i+1,a}+\hat{\psi}^+_{i+1,a}\hat{\psi}^-_{i,a}) -\sum_{i} \mu_i \hat{n}_i,
\end{equation}
therefore $\hat{\psi}^\pm_{i,a_1}$ create/annihilate free emergent paraparticles. %
Using a canonical transformation of $\{\hat{\psi}^\pm_{i,a}\}$, the free paraparticle Hamiltonian in Eq.~\eqref{eq:Hamil1Dspin_para} can be diagonalized into the form $\hat{H}=\sum^N_{k=1} \epsilon_{k} \tilde{n}_{k}$, %
and the full spectrum can be exactly obtained for arbitrary coupling constants $\{J_i\}$~(even with disorder). %
Exactly solvable quantum spin models with free emergent paraparticles can also be found in 2D.
In Methods we present the key features of these models through %
a specific example with $m=4$. 
These 2D models %
realize a special family of paraparticles which, despite having trivial exclusion statistics~[i.e. %
the same partition function as $m$ flavors of fermions], have non-trivial exchange statistics that is physically~(observably) distinct from fermions and bosons. 
Similar to the 1D case, these %
models are mapped to free paraparticle Hamiltonians of the form in Eq.~\eqref{eq:u1_sym_H}, using an MPO JWT defined in Eq.~\eqref{eq:JWT_string_2D} which generalizes Eq.~\eqref{eq:JWT_string}. In 2D, $\hat{\psi}^\pm_{i,a_1}$ are still MPO string operators, with the additional remarkable property that their actions on the low energy sector~(e.g. the ground states) are independent of the paths on which they are defined, which is reminiscent of the path independence property of the string~(ribbon) operators that create anyons in Kitaev's quantum double model~\cite{kitaev2003fault}. %

In summary, these results imply a new type of quasiparticle statistics, which can be searched for in condensed matter systems, %
and a starting point is the exactly solvable quantum spin model defined in Eq.~\eqref{eq:Hamil1Dspin} and its 2D generalizations, defined in Eq.~\eqref{def:2DsolvableH} in Methods. %
Systems with such excitations may display a wealth of new phenomena, and the exactly solvable models constructed above provide an efficient way to study them. 
Depending on the spectrum of the resulting free paraparticle systems the spin models are mapped to, novel phases of matter and phase transitions  can be discovered. %
For example, in 2D, if the free paraparticle system has a non-trivial topological band structure~(having a nonzero Chern number), then the spin model can be in a new chiral topological phase that is hard to study with previous techniques~\footnote{%
The chiral topological phases of our 2D models are expected to lie beyond those found in previous solvable models~\cite{kitaev2006anyons,YaoKivelson,Chapman2020characterizationof}, as explained in Methods. The study of chiral topological order using tensor network techniques is known to be hard~\cite{DubailRead2015}. It has been conjectured that  chiral topological states cannot have a gapped frustration-free parent Hamiltonian with strictly short-range interactions, although this is known to be possible in continuum with exponentially-decaying interactions~\cite{Wang2017Number,Wang2018}. %
}. %
If the free paraparticle system has a gapless spectrum, the spin model can realize a phase transition point or a gapless topological phase~\cite{Pollmann2021gapless,Scaffidi2017gaplessSPT,Vishwanath2021Intrinsic}, which are interesting and difficult areas of research even in 1D systems. Furthermore, allowing the tunneling constants $\{J_i\}$ to be spatially disordered may lead to new localized phases. 

\paragraph{Speculations about elementary paraparticles}
In addition to the possibility of emergent parastatistical excitations in interacting quantum matter, a natural, albeit highly speculative, question is to 
ask if paraparticles may exist as elementary particles in nature. 
We have seen that our second quantized theory of paraparticles satisfies the fundamental requirements of locality and Hermiticity, and is consistently defined in all dimensions.  It is also straightforward to incorporate relativity to get a fully consistent relativistic quantum field theory of elementary paraparticles, in which the canonical quantization of field operators are defined by the $R$-CRs in Eq.~\eqref{eq:fundamental_Rcommu}. Most fundamental field-theoretical concepts and tools~\cite{weinberg_1995} generalize straightforwardly to parastatistics. %
In order to consider paraparticles as elementary particles, it is important to consider their superselection rules. We discuss this  issue in Sec.~\ref{SI:superselection} of the SI~\cite{Suppl}, where we explain how superselection rules fundamentally constrain the observability of parastatistics, which is reminiscent of the previous no-go theorems~\cite{doplicher1971local,*doplicher1974local}. %
We then discuss how our proposed realization of emergent paraparticles in condensed matter systems  breaks these superselection rules, which motivates routes to construct theories of elementary paraparticles   observably distinct from fermions and bosons, %
evading the no-go theorems~\cite{doplicher1971local,*doplicher1974local}.

\textit{Note added:} more than a year after the initial submission of this paper, we became aware of a related work~\cite{Dakic2024reconstructionof} that appeared on arXiv two months before our work. Ref.~\cite{Dakic2024reconstructionof} also predicted a new family of identical particles with exotic exclusion statistics~(termed ``transtatistics'' there), including ones with the same partition function as our Ex.~\ref{ex:1m} and \ref{ex:1m1}. Ref.~\cite{Dakic2024reconstructionof} achieved this using a very different approach, based on a set of axioms motivated by quantum information theory. In particular, the second quantization formulation of the parastatistics based on the $R$-matrix CRs and all its quantum spin model realizations in our work are new and original.	
	
\acknowledgments
	We thank Alexei Kitaev, J.~Ignacio Cirac, Kevin Slagle, Andrew Long, Mustafa Amin, Alexander Hahn, and Paul Fendley for discussions. We acknowledge support from the Robert A. Welch Foundation~(C-1872), the National Science Foundation~(PHY-1848304), the Office of Naval Research (N00014-20-1-2695), and the W. M. Keck
Foundation~(Grant No. 995764). K.H.'s contribution benefited from discussions at the Aspen Center for Physics, supported by the National Science Foundation grant PHY1066293, and the KITP, which was supported in part by the
National Science Foundation under Grant No. NSF PHY1748958. ZW is supported by the Munich Quantum Valley~(MQV), which is supported by the Bavarian state government with funds from the Hightech Agenda Bayern Plus.

\bibliography{LA,/home/lagrenge/Documents/Mendeley_bib/library}

%merlin.mbs apsrev4-1.bst 2010-07-25 4.21a (PWD, AO, DPC) hacked
%Control: key (0)
%Control: author (72) initials jnrlst
%Control: editor formatted (1) identically to author
%Control: production of article title (-1) disabled
%Control: page (0) single
%Control: year (1) truncated
%Control: production of eprint (0) enabled
\begin{thebibliography}{103}%
\makeatletter
\providecommand \@ifxundefined [1]{%
 \@ifx{#1\undefined}
}%
\providecommand \@ifnum [1]{%
 \ifnum #1\expandafter \@firstoftwo
 \else \expandafter \@secondoftwo
 \fi
}%
\providecommand \@ifx [1]{%
 \ifx #1\expandafter \@firstoftwo
 \else \expandafter \@secondoftwo
 \fi
}%
\providecommand \natexlab [1]{#1}%
\providecommand \enquote  [1]{``#1''}%
\providecommand \bibnamefont  [1]{#1}%
\providecommand \bibfnamefont [1]{#1}%
\providecommand \citenamefont [1]{#1}%
\providecommand \href@noop [0]{\@secondoftwo}%
\providecommand \href [0]{\begingroup \@sanitize@url \@href}%
\providecommand \@href[1]{\@@startlink{#1}\@@href}%
\providecommand \@@href[1]{\endgroup#1\@@endlink}%
\providecommand \@sanitize@url [0]{\catcode `\\12\catcode `\$12\catcode
  `\&12\catcode `\#12\catcode `\^12\catcode `\_12\catcode `\%12\relax}%
\providecommand \@@startlink[1]{}%
\providecommand \@@endlink[0]{}%
\providecommand \url  [0]{\begingroup\@sanitize@url \@url }%
\providecommand \@url [1]{\endgroup\@href {#1}{\urlprefix }}%
\providecommand \urlprefix  [0]{URL }%
\providecommand \Eprint [0]{\href }%
\providecommand \doibase [0]{http://dx.doi.org/}%
\providecommand \selectlanguage [0]{\@gobble}%
\providecommand \bibinfo  [0]{\@secondoftwo}%
\providecommand \bibfield  [0]{\@secondoftwo}%
\providecommand \translation [1]{[#1]}%
\providecommand \BibitemOpen [0]{}%
\providecommand \bibitemStop [0]{}%
\providecommand \bibitemNoStop [0]{.\EOS\space}%
\providecommand \EOS [0]{\spacefactor3000\relax}%
\providecommand \BibitemShut  [1]{\csname bibitem#1\endcsname}%
\let\auto@bib@innerbib\@empty
%</preamble>
\bibitem [{\citenamefont {Leinaas}\ and\ \citenamefont
  {Myrheim}(1977)}]{Leinaas1977}%
  \BibitemOpen
  \bibfield  {author} {\bibinfo {author} {\bibfnamefont {J.~M.}\ \bibnamefont
  {Leinaas}}\ and\ \bibinfo {author} {\bibfnamefont {J.}~\bibnamefont
  {Myrheim}},\ }\href {\doibase 10.1007/BF02727953} {\bibfield  {journal}
  {\bibinfo  {journal} {Nuovo Cim. B}\ }\textbf {\bibinfo {volume} {37}},\
  \bibinfo {pages} {1} (\bibinfo {year} {1977})}\BibitemShut {NoStop}%
\bibitem [{\citenamefont {Wilczek}(1982{\natexlab{a}})}]{Wilczek1982Magnetic}%
  \BibitemOpen
  \bibfield  {author} {\bibinfo {author} {\bibfnamefont {F.}~\bibnamefont
  {Wilczek}},\ }\href {\doibase 10.1103/PhysRevLett.48.1144} {\bibfield
  {journal} {\bibinfo  {journal} {Phys. Rev. Lett.}\ }\textbf {\bibinfo
  {volume} {48}},\ \bibinfo {pages} {1144} (\bibinfo {year}
  {1982}{\natexlab{a}})}\BibitemShut {NoStop}%
\bibitem [{\citenamefont {Wilczek}(1982{\natexlab{b}})}]{Wilczek1982Quantum}%
  \BibitemOpen
  \bibfield  {author} {\bibinfo {author} {\bibfnamefont {F.}~\bibnamefont
  {Wilczek}},\ }\href {\doibase 10.1103/PhysRevLett.49.957} {\bibfield
  {journal} {\bibinfo  {journal} {Phys. Rev. Lett.}\ }\textbf {\bibinfo
  {volume} {49}},\ \bibinfo {pages} {957} (\bibinfo {year}
  {1982}{\natexlab{b}})}\BibitemShut {NoStop}%
\bibitem [{\citenamefont {Wilczek}(1990)}]{Wilczek1990book}%
  \BibitemOpen
  \bibfield  {author} {\bibinfo {author} {\bibfnamefont {F.}~\bibnamefont
  {Wilczek}},\ }\href {\doibase 10.1142/0961} {\emph {\bibinfo {title}
  {Fractional Statistics and Anyon Superconductivity}}}\ (\bibinfo  {publisher}
  {World Scientific},\ \bibinfo {address} {Singapore},\ \bibinfo {year}
  {1990})\BibitemShut {NoStop}%
\bibitem [{\citenamefont {Nayak}\ \emph {et~al.}(2008)\citenamefont {Nayak},
  \citenamefont {Simon}, \citenamefont {Stern}, \citenamefont {Freedman},\ and\
  \citenamefont {Das~Sarma}}]{Nayak2008NAAnyons}%
  \BibitemOpen
  \bibfield  {author} {\bibinfo {author} {\bibfnamefont {C.}~\bibnamefont
  {Nayak}}, \bibinfo {author} {\bibfnamefont {S.~H.}\ \bibnamefont {Simon}},
  \bibinfo {author} {\bibfnamefont {A.}~\bibnamefont {Stern}}, \bibinfo
  {author} {\bibfnamefont {M.}~\bibnamefont {Freedman}}, \ and\ \bibinfo
  {author} {\bibfnamefont {S.}~\bibnamefont {Das~Sarma}},\ }\href {\doibase
  10.1103/RevModPhys.80.1083} {\bibfield  {journal} {\bibinfo  {journal} {Rev.
  Mod. Phys.}\ }\textbf {\bibinfo {volume} {80}},\ \bibinfo {pages} {1083}
  (\bibinfo {year} {2008})}\BibitemShut {NoStop}%
\bibitem [{\citenamefont {Green}(1953)}]{Green1952}%
  \BibitemOpen
  \bibfield  {author} {\bibinfo {author} {\bibfnamefont {H.~S.}\ \bibnamefont
  {Green}},\ }\href {\doibase 10.1103/PhysRev.90.270} {\bibfield  {journal}
  {\bibinfo  {journal} {Phys. Rev.}\ }\textbf {\bibinfo {volume} {90}},\
  \bibinfo {pages} {270} (\bibinfo {year} {1953})}\BibitemShut {NoStop}%
\bibitem [{\citenamefont {Doplicher}\ \emph {et~al.}(1971)\citenamefont
  {Doplicher}, \citenamefont {Haag},\ and\ \citenamefont
  {Roberts}}]{doplicher1971local}%
  \BibitemOpen
  \bibfield  {author} {\bibinfo {author} {\bibfnamefont {S.}~\bibnamefont
  {Doplicher}}, \bibinfo {author} {\bibfnamefont {R.}~\bibnamefont {Haag}}, \
  and\ \bibinfo {author} {\bibfnamefont {J.~E.}\ \bibnamefont {Roberts}},\
  }\href {\doibase 10.1007/BF01877742} {\bibfield  {journal} {\bibinfo
  {journal} {Commun. Math. Phys.}\ }\textbf {\bibinfo {volume} {23}},\ \bibinfo
  {pages} {199} (\bibinfo {year} {1971})}\BibitemShut {NoStop}%
\bibitem [{\citenamefont {Doplicher}\ \emph {et~al.}(1974)\citenamefont
  {Doplicher}, \citenamefont {Haag},\ and\ \citenamefont
  {Roberts}}]{doplicher1974local}%
  \BibitemOpen
  \bibfield  {author} {\bibinfo {author} {\bibfnamefont {S.}~\bibnamefont
  {Doplicher}}, \bibinfo {author} {\bibfnamefont {R.}~\bibnamefont {Haag}}, \
  and\ \bibinfo {author} {\bibfnamefont {J.~E.}\ \bibnamefont {Roberts}},\
  }\href {\doibase 10.1007/BF01646454} {\bibfield  {journal} {\bibinfo
  {journal} {Commun. Math. Phys.}\ }\textbf {\bibinfo {volume} {35}},\ \bibinfo
  {pages} {49} (\bibinfo {year} {1974})}\BibitemShut {NoStop}%
\bibitem [{\citenamefont {Doplicher}\ and\ \citenamefont
  {Roberts}(1972)}]{DR1972}%
  \BibitemOpen
  \bibfield  {author} {\bibinfo {author} {\bibfnamefont {S.}~\bibnamefont
  {Doplicher}}\ and\ \bibinfo {author} {\bibfnamefont {J.~E.}\ \bibnamefont
  {Roberts}},\ }\href {\doibase cmp/1103858447} {\bibfield  {journal} {\bibinfo
   {journal} {Commun. Math. Phys.}\ }\textbf {\bibinfo {volume} {28}},\
  \bibinfo {pages} {331 } (\bibinfo {year} {1972})}\BibitemShut {NoStop}%
\bibitem [{\citenamefont {Stern}(2008)}]{STERN2008204}%
  \BibitemOpen
  \bibfield  {author} {\bibinfo {author} {\bibfnamefont {A.}~\bibnamefont
  {Stern}},\ }\href {\doibase https://doi.org/10.1016/j.aop.2007.10.008}
  {\bibfield  {journal} {\bibinfo  {journal} {Ann. Phys.}\ }\textbf {\bibinfo
  {volume} {323}},\ \bibinfo {pages} {204} (\bibinfo {year} {2008})},\ \bibinfo
  {note} {{J}anuary Special Issue 2008}\BibitemShut {NoStop}%
\bibitem [{Sup()}]{Suppl}%
  \BibitemOpen
  \href@noop {} {}\bibinfo {note} {See Supplementary Information~(SI) for the
  difference between parastatistics and other known types of particle
  statistics and the difference between our theory of parastatistics and
  previous theories~(Sec.~\ref{SI:relation_to_others}), the construction of the
  state space and the action of $\hat{\psi}^\pm_{i,a}$ for a general
  $R$-matrix~(including non-unitaries ones)~(Sec.~\ref{SI:state_space}), the
  relation between the first and the second quantization formulation of
  parastatistics~(Sec.~\ref{SI:relation_first_quantization}), the detailed
  construction of the solvable spin models with emergent paraparticles and the
  proof that the MPO JWT maps the spin model to free
  paraparticles~(Sec.~\ref{SI:1Dspin_model} for the 1D model and
  Sec.~\ref{SI:2DparaKDH} for the 2D model), and the discussion of
  superselection rules and the observability of elementary
  paraparticles~(Sec.~\ref{SI:superselection}).}\BibitemShut {Stop}%
\bibitem [{\citenamefont {Araki}(1961)}]{Araki1961}%
  \BibitemOpen
  \bibfield  {author} {\bibinfo {author} {\bibfnamefont {H.}~\bibnamefont
  {Araki}},\ }\href {\doibase 10.1063/1.1703710} {\bibfield  {journal}
  {\bibinfo  {journal} {J. Math. Phys.}\ }\textbf {\bibinfo {volume} {2}},\
  \bibinfo {pages} {267} (\bibinfo {year} {1961})}\BibitemShut {NoStop}%
\bibitem [{\citenamefont {Greenberg}\ and\ \citenamefont
  {Messiah}(1965)}]{Greenberg1965}%
  \BibitemOpen
  \bibfield  {author} {\bibinfo {author} {\bibfnamefont {O.~W.}\ \bibnamefont
  {Greenberg}}\ and\ \bibinfo {author} {\bibfnamefont {A.~M.~L.}\ \bibnamefont
  {Messiah}},\ }\href {\doibase 10.1103/PhysRev.138.B1155} {\bibfield
  {journal} {\bibinfo  {journal} {Phys. Rev.}\ }\textbf {\bibinfo {volume}
  {138}},\ \bibinfo {pages} {B1155} (\bibinfo {year} {1965})}\BibitemShut
  {NoStop}%
\bibitem [{\citenamefont {Landshoff}\ and\ \citenamefont
  {Stapp}(1967)}]{LANDSHOFF196772}%
  \BibitemOpen
  \bibfield  {author} {\bibinfo {author} {\bibfnamefont {P.}~\bibnamefont
  {Landshoff}}\ and\ \bibinfo {author} {\bibfnamefont {H.~P.}\ \bibnamefont
  {Stapp}},\ }\href {\doibase https://doi.org/10.1016/0003-4916(67)90317-X}
  {\bibfield  {journal} {\bibinfo  {journal} {Ann. Phys.}\ }\textbf {\bibinfo
  {volume} {45}},\ \bibinfo {pages} {72} (\bibinfo {year} {1967})}\BibitemShut
  {NoStop}%
\bibitem [{\citenamefont {Dr{\"u}hl}\ \emph {et~al.}(1970)\citenamefont
  {Dr{\"u}hl}, \citenamefont {Haag},\ and\ \citenamefont
  {Roberts}}]{druhl1970parastatistics}%
  \BibitemOpen
  \bibfield  {author} {\bibinfo {author} {\bibfnamefont {K.}~\bibnamefont
  {Dr{\"u}hl}}, \bibinfo {author} {\bibfnamefont {R.}~\bibnamefont {Haag}}, \
  and\ \bibinfo {author} {\bibfnamefont {J.~E.}\ \bibnamefont {Roberts}},\
  }\href {\doibase https://doi.org/10.1007/BF01649433} {\bibfield  {journal}
  {\bibinfo  {journal} {Commun. Math. Phys.}\ }\textbf {\bibinfo {volume}
  {18}},\ \bibinfo {pages} {204} (\bibinfo {year} {1970})}\BibitemShut
  {NoStop}%
\bibitem [{\citenamefont {Stolt}\ and\ \citenamefont
  {Taylor}(1970{\natexlab{a}})}]{Taylor1970b}%
  \BibitemOpen
  \bibfield  {author} {\bibinfo {author} {\bibfnamefont {R.~H.}\ \bibnamefont
  {Stolt}}\ and\ \bibinfo {author} {\bibfnamefont {J.~R.}\ \bibnamefont
  {Taylor}},\ }\href {\doibase https://doi.org/10.1016/0550-3213(70)90024-6}
  {\bibfield  {journal} {\bibinfo  {journal} {Nucl. Phys. B}\ }\textbf
  {\bibinfo {volume} {19}},\ \bibinfo {pages} {1} (\bibinfo {year}
  {1970}{\natexlab{a}})}\BibitemShut {NoStop}%
\bibitem [{Note1()}]{Note1}%
  \BibitemOpen
  \bibinfo {note} {Note that we only need to specify the behavior of the
  wavefunction under exchange of particles with adjacent labels, since exchange
  of particles with nonadjacent labels can always be decomposed into a series
  of adjacent exchanges. For example, under the exchange of particles $1$ and
  $3$, the wavefunction should multiply by the matrix
  $R_{1}R_{2}R_{1}$.}\BibitemShut {Stop}%
\bibitem [{\citenamefont {Kassel}\ and\ \citenamefont
  {Turaev}(2008)}]{kassel2008braid}%
  \BibitemOpen
  \bibfield  {author} {\bibinfo {author} {\bibfnamefont {C.}~\bibnamefont
  {Kassel}}\ and\ \bibinfo {author} {\bibfnamefont {V.}~\bibnamefont
  {Turaev}},\ }\href {\doibase https://doi.org/10.1007/978-0-387-68548-9}
  {\emph {\bibinfo {title} {Braid groups}}},\ \bibinfo {series} {Graduate Texts
  in Mathematics}, Vol.\ \bibinfo {volume} {247}\ (\bibinfo  {publisher}
  {Springer New York},\ \bibinfo {year} {2008})\BibitemShut {NoStop}%
\bibitem [{\citenamefont {Pauli}(1946)}]{nobelphysics1946}%
  \BibitemOpen
  \bibfield  {author} {\bibinfo {author} {\bibfnamefont {W.}~\bibnamefont
  {Pauli}},\ }\enquote {\bibinfo {title} {Exclusion principle and quantum
  mechanics},}\ in\ \href
  {https://www.nobelprize.org/prizes/physics/1945/pauli/lecture/} {\emph
  {\bibinfo {booktitle} {Nobel Lectures. Physics 1942--1962}}}\ (\bibinfo
  {publisher} {Elsevier Publishing Company, Amsterdam},\ \bibinfo {year}
  {1946})\ pp.\ \bibinfo {pages} {27--43}\BibitemShut {NoStop}%
\bibitem [{\citenamefont {Doplicher}\ and\ \citenamefont
  {Roberts}(1990)}]{doplicher1990}%
  \BibitemOpen
  \bibfield  {author} {\bibinfo {author} {\bibfnamefont {S.}~\bibnamefont
  {Doplicher}}\ and\ \bibinfo {author} {\bibfnamefont {J.~E.}\ \bibnamefont
  {Roberts}},\ }\href {\doibase cmp/1104200703} {\bibfield  {journal} {\bibinfo
   {journal} {Commun. Math. Phys.}\ }\textbf {\bibinfo {volume} {131}},\
  \bibinfo {pages} {51 } (\bibinfo {year} {1990})}\BibitemShut {NoStop}%
\bibitem [{\citenamefont {Streater}\ and\ \citenamefont
  {Wightman}(2001)}]{WightmanPCTbook}%
  \BibitemOpen
  \bibfield  {author} {\bibinfo {author} {\bibfnamefont {R.~F.}\ \bibnamefont
  {Streater}}\ and\ \bibinfo {author} {\bibfnamefont {A.~S.}\ \bibnamefont
  {Wightman}},\ }\href
  {https://press.princeton.edu/books/paperback/9780691070629/pct-spin-and-statistics-and-all-that}
  {\emph {\bibinfo {title} {PCT, spin and statistics, and all that}}}\
  (\bibinfo  {publisher} {Princeton University Press},\ \bibinfo {year}
  {2001})\BibitemShut {NoStop}%
\bibitem [{\citenamefont {Haag}(1996)}]{haag2012book}%
  \BibitemOpen
  \bibfield  {author} {\bibinfo {author} {\bibfnamefont {R.}~\bibnamefont
  {Haag}},\ }\href {\doibase https://doi.org/10.1007/978-3-642-61458-3} {\emph
  {\bibinfo {title} {Local quantum physics: Fields, particles, algebras}}}\
  (\bibinfo  {publisher} {Springer-Verlag, Berlin, Heidelberg},\ \bibinfo
  {year} {1996})\BibitemShut {NoStop}%
\bibitem [{Note2()}]{Note2}%
  \BibitemOpen
  \bibinfo {note} {It is hard to guarantee locality in the first quantization
  formulation, and without locality such theories are hard to be realized as
  elementary particles or as emergent quasiparticle excitations in locally
  interacting systems. Note that it is exactly for this reason that the
  Doplicher-Haag-Roberts~(DHR) no-go theorem~\cite
  {doplicher1971local,*doplicher1974local} does not apply to the first
  quantization formulation, since the former takes locality as a fundamental
  assumption, while the latter does not have locality built-in. See Sec.~\ref
  {SI:relation_first_quantization} in the SI for the relation between the first
  and second quantization formulation of parastatistics in this
  paper.}\BibitemShut {Stop}%
\bibitem [{\citenamefont {Turaev}(1988)}]{Turaev1988}%
  \BibitemOpen
  \bibfield  {author} {\bibinfo {author} {\bibfnamefont {V.~G.}\ \bibnamefont
  {Turaev}},\ }\href {\doibase 10.1007/BF01393746} {\bibfield  {journal}
  {\bibinfo  {journal} {Invent. Math.}\ }\textbf {\bibinfo {volume} {92}},\
  \bibinfo {pages} {527} (\bibinfo {year} {1988})}\BibitemShut {NoStop}%
\bibitem [{\citenamefont {Majid}(1990)}]{Majid1990}%
  \BibitemOpen
  \bibfield  {author} {\bibinfo {author} {\bibfnamefont {S.}~\bibnamefont
  {Majid}},\ }\href {\doibase 10.1142/S0217751X90000027} {\bibfield  {journal}
  {\bibinfo  {journal} {Int. J. Mod. Phys. A}\ }\textbf {\bibinfo {volume}
  {05}},\ \bibinfo {pages} {1} (\bibinfo {year} {1990})}\BibitemShut {NoStop}%
\bibitem [{\citenamefont {Etingof}\ \emph {et~al.}(1999)\citenamefont
  {Etingof}, \citenamefont {Schedler},\ and\ \citenamefont
  {Soloviev}}]{etingof1999set}%
  \BibitemOpen
  \bibfield  {author} {\bibinfo {author} {\bibfnamefont {P.}~\bibnamefont
  {Etingof}}, \bibinfo {author} {\bibfnamefont {T.}~\bibnamefont {Schedler}}, \
  and\ \bibinfo {author} {\bibfnamefont {A.}~\bibnamefont {Soloviev}},\ }\href
  {\doibase 10.1215/S0012-7094-99-10007-X} {\bibfield  {journal} {\bibinfo
  {journal} {Duke Math. J.}\ }\textbf {\bibinfo {volume} {100}},\ \bibinfo
  {pages} {169 } (\bibinfo {year} {1999})}\BibitemShut {NoStop}%
\bibitem [{foo()}]{footnote_lambdac}%
  \BibitemOpen
  \href@noop {} {}\bibinfo {note} {For example, we can take
  $\lambda=e^{-M},\xi=-e^M$, where $M$ is an $m\times m$ antisymmetric matrix,
  $M^T=-M$, with complex entries satisfying $\mathrm{Tr}[e^{-2M}]=-2$~(one can
  get an explicit solution using a block-diagonal ansatz for $M$ with maximum
  block size $2$). A small caveat here is that this example of $R$-matrix is
  not unitary for $m\geq 3$, and the corresponding 1D spin model with emergent
  paraparticles defined later in the paper is not Hermitian, but PT-symmetric,
  see the discussion in Sec.~\ref{SI:spin_model_def} of the SI.}\BibitemShut
  {Stop}%
\bibitem [{Note3()}]{Note3}%
  \BibitemOpen
  \bibinfo {note} {As we show in the SI~\cite {Suppl}, even with a non-unitary
  $R$-matrix, such as Ex.~\ref {ex:1m1} in Tab.~\ref {tab:Hilbert_series},
  Eq.~\protect \textup {\hbox {\mathsurround \z@ \protect \normalfont
  (\ignorespaces \ref {eq:fundamental_Rcommu}\unskip \@@italiccorr )}} is still
  consistently defined, and we can still define a Hermitian inner product on
  the state space, with respect to which all physical observables are
  Hermitian, and most of the main results of this paper still apply, including
  the generalized exclusion statistics, the exact solution of free particles,
  and the construction of solvable spin models with emergent
  paraparticles.}\BibitemShut {Stop}%
\bibitem [{Note4()}]{Note4}%
  \BibitemOpen
  \bibinfo {note} {For the former, the commutativity of local observables at
  space-like separations rules out faster-than-light travel and communication;
  for the latter, Ref.~\cite {wang2020tightening} proved that as long as all
  the local Hamiltonian terms have uniformly bounded norms and their algebra
  has a local structure, then the Lieb-Robinson bound~\cite {Lieb1972} holds,
  which gives an effective lightcone of causality.}\BibitemShut {Stop}%
\bibitem [{Note5()}]{Note5}%
  \BibitemOpen
  \bibinfo {note} {A caveat, however, is that $z_R(x)$ only gives a sufficient
  condition for non-triviality, and having a trivial $z_R(x)$ does not imply
  that the corresponding paraparticle theory is completely equivalent to
  fermions or bosons. For example, the emergent paraparticles in our 2D
  solvable spin model are defined by the $R$-matrix given in Eqs.~(\ref
  {eq:seth-R},\ref {eq:set-thR}), whose partition function $z_R(x)=(1+x)^4$ is
  the same as free fermions with an SU$(4)$ symmetry, but the exchange
  statistics of these emergent paraparticles are still physically~(observably)
  distinct from fermions and boson, as we show in Methods.}\BibitemShut {Stop}%
\bibitem [{\citenamefont {Sundar}\ \emph {et~al.}(2018)\citenamefont {Sundar},
  \citenamefont {Gadway},\ and\ \citenamefont {Hazzard}}]{Sundar2018Synthetic}%
  \BibitemOpen
  \bibfield  {author} {\bibinfo {author} {\bibfnamefont {B.}~\bibnamefont
  {Sundar}}, \bibinfo {author} {\bibfnamefont {B.}~\bibnamefont {Gadway}}, \
  and\ \bibinfo {author} {\bibfnamefont {K.~R.~A.}\ \bibnamefont {Hazzard}},\
  }\href {\doibase 10.1038/s41598-018-21699-x} {\bibfield  {journal} {\bibinfo
  {journal} {Sci. Rep.}\ }\textbf {\bibinfo {volume} {8}},\ \bibinfo {pages}
  {3422} (\bibinfo {year} {2018})}\BibitemShut {NoStop}%
\bibitem [{\citenamefont {Sundar}\ \emph {et~al.}(2019)\citenamefont {Sundar},
  \citenamefont {Thibodeau}, \citenamefont {Wang}, \citenamefont {Gadway},\
  and\ \citenamefont {Hazzard}}]{Sundar2019Strings}%
  \BibitemOpen
  \bibfield  {author} {\bibinfo {author} {\bibfnamefont {B.}~\bibnamefont
  {Sundar}}, \bibinfo {author} {\bibfnamefont {M.}~\bibnamefont {Thibodeau}},
  \bibinfo {author} {\bibfnamefont {Z.}~\bibnamefont {Wang}}, \bibinfo {author}
  {\bibfnamefont {B.}~\bibnamefont {Gadway}}, \ and\ \bibinfo {author}
  {\bibfnamefont {K.~R.~A.}\ \bibnamefont {Hazzard}},\ }\href {\doibase
  10.1103/PhysRevA.99.013624} {\bibfield  {journal} {\bibinfo  {journal} {Phys.
  Rev. A}\ }\textbf {\bibinfo {volume} {99}},\ \bibinfo {pages} {013624}
  (\bibinfo {year} {2019})}\BibitemShut {NoStop}%
\bibitem [{\citenamefont {Jordan}\ and\ \citenamefont {Wigner}(1928)}]{JW1928}%
  \BibitemOpen
  \bibfield  {author} {\bibinfo {author} {\bibfnamefont {P.}~\bibnamefont
  {Jordan}}\ and\ \bibinfo {author} {\bibfnamefont {E.}~\bibnamefont
  {Wigner}},\ }\href {\doibase 10.1007/BF01331938} {\bibfield  {journal}
  {\bibinfo  {journal} {Z. Phys.}\ }\textbf {\bibinfo {volume} {47}},\ \bibinfo
  {pages} {631} (\bibinfo {year} {1928})}\BibitemShut {NoStop}%
\bibitem [{\citenamefont {Cirac}\ \emph {et~al.}(2021)\citenamefont {Cirac},
  \citenamefont {Perez-Garcia}, \citenamefont {Schuch},\ and\ \citenamefont
  {Verstraete}}]{Cirac2021Matrix}%
  \BibitemOpen
  \bibfield  {author} {\bibinfo {author} {\bibfnamefont {J.~I.}\ \bibnamefont
  {Cirac}}, \bibinfo {author} {\bibfnamefont {D.}~\bibnamefont {Perez-Garcia}},
  \bibinfo {author} {\bibfnamefont {N.}~\bibnamefont {Schuch}}, \ and\ \bibinfo
  {author} {\bibfnamefont {F.}~\bibnamefont {Verstraete}},\ }\href {\doibase
  10.1103/RevModPhys.93.045003} {\bibfield  {journal} {\bibinfo  {journal}
  {Rev. Mod. Phys.}\ }\textbf {\bibinfo {volume} {93}},\ \bibinfo {pages}
  {45003} (\bibinfo {year} {2021})}\BibitemShut {NoStop}%
\bibitem [{\citenamefont {Kitaev}(2003)}]{kitaev2003fault}%
  \BibitemOpen
  \bibfield  {author} {\bibinfo {author} {\bibfnamefont {A.}~\bibnamefont
  {Kitaev}},\ }\href {\doibase https://doi.org/10.1016/S0003-4916(02)00018-0}
  {\bibfield  {journal} {\bibinfo  {journal} {Ann. Phys.}\ }\textbf {\bibinfo
  {volume} {303}},\ \bibinfo {pages} {2} (\bibinfo {year} {2003})}\BibitemShut
  {NoStop}%
\bibitem [{Note6()}]{Note6}%
  \BibitemOpen
  \bibinfo {note} {The chiral topological phases of our 2D models are expected
  to lie beyond those found in previous solvable models~\cite
  {kitaev2006anyons,YaoKivelson,Chapman2020characterizationof}, as explained in
  Methods. The study of chiral topological order using tensor network
  techniques is known to be hard~\cite {DubailRead2015}. It has been
  conjectured that chiral topological states cannot have a gapped
  frustration-free parent Hamiltonian with strictly short-range interactions,
  although this is known to be possible in continuum with
  exponentially-decaying interactions~\cite
  {Wang2017Number,Wang2018}.}\BibitemShut {Stop}%
\bibitem [{\citenamefont {Verresen}\ \emph {et~al.}(2021)\citenamefont
  {Verresen}, \citenamefont {Thorngren}, \citenamefont {Jones},\ and\
  \citenamefont {Pollmann}}]{Pollmann2021gapless}%
  \BibitemOpen
  \bibfield  {author} {\bibinfo {author} {\bibfnamefont {R.}~\bibnamefont
  {Verresen}}, \bibinfo {author} {\bibfnamefont {R.}~\bibnamefont {Thorngren}},
  \bibinfo {author} {\bibfnamefont {N.~G.}\ \bibnamefont {Jones}}, \ and\
  \bibinfo {author} {\bibfnamefont {F.}~\bibnamefont {Pollmann}},\ }\href
  {\doibase 10.1103/PhysRevX.11.041059} {\bibfield  {journal} {\bibinfo
  {journal} {Phys. Rev. X}\ }\textbf {\bibinfo {volume} {11}},\ \bibinfo
  {pages} {041059} (\bibinfo {year} {2021})}\BibitemShut {NoStop}%
\bibitem [{\citenamefont {Scaffidi}\ \emph {et~al.}(2017)\citenamefont
  {Scaffidi}, \citenamefont {Parker},\ and\ \citenamefont
  {Vasseur}}]{Scaffidi2017gaplessSPT}%
  \BibitemOpen
  \bibfield  {author} {\bibinfo {author} {\bibfnamefont {T.}~\bibnamefont
  {Scaffidi}}, \bibinfo {author} {\bibfnamefont {D.~E.}\ \bibnamefont
  {Parker}}, \ and\ \bibinfo {author} {\bibfnamefont {R.}~\bibnamefont
  {Vasseur}},\ }\href {\doibase 10.1103/PhysRevX.7.041048} {\bibfield
  {journal} {\bibinfo  {journal} {Phys. Rev. X}\ }\textbf {\bibinfo {volume}
  {7}},\ \bibinfo {pages} {041048} (\bibinfo {year} {2017})}\BibitemShut
  {NoStop}%
\bibitem [{\citenamefont {Thorngren}\ \emph {et~al.}(2021)\citenamefont
  {Thorngren}, \citenamefont {Vishwanath},\ and\ \citenamefont
  {Verresen}}]{Vishwanath2021Intrinsic}%
  \BibitemOpen
  \bibfield  {author} {\bibinfo {author} {\bibfnamefont {R.}~\bibnamefont
  {Thorngren}}, \bibinfo {author} {\bibfnamefont {A.}~\bibnamefont
  {Vishwanath}}, \ and\ \bibinfo {author} {\bibfnamefont {R.}~\bibnamefont
  {Verresen}},\ }\href {\doibase 10.1103/PhysRevB.104.075132} {\bibfield
  {journal} {\bibinfo  {journal} {Phys. Rev. B}\ }\textbf {\bibinfo {volume}
  {104}},\ \bibinfo {pages} {075132} (\bibinfo {year} {2021})}\BibitemShut
  {NoStop}%
\bibitem [{\citenamefont {Weinberg}(1995)}]{weinberg_1995}%
  \BibitemOpen
  \bibfield  {author} {\bibinfo {author} {\bibfnamefont {S.}~\bibnamefont
  {Weinberg}},\ }\href {\doibase 10.1017/CBO9781139644167} {\emph {\bibinfo
  {title} {The Quantum Theory of Fields}}},\ Vol.~\bibinfo {volume} {1}\
  (\bibinfo  {publisher} {Cambridge University Press},\ \bibinfo {year}
  {1995})\BibitemShut {NoStop}%
\bibitem [{\citenamefont {S{\'{a}}nchez}\ and\ \citenamefont
  {Daki{\'{c}}}(2024)}]{Dakic2024reconstructionof}%
  \BibitemOpen
  \bibfield  {author} {\bibinfo {author} {\bibfnamefont {N.~M.}\ \bibnamefont
  {S{\'{a}}nchez}}\ and\ \bibinfo {author} {\bibfnamefont {B.}~\bibnamefont
  {Daki{\'{c}}}},\ }\href {\doibase 10.22331/q-2024-09-12-1473} {\bibfield
  {journal} {\bibinfo  {journal} {Quantum}\ }\textbf {\bibinfo {volume} {8}},\
  \bibinfo {pages} {1473} (\bibinfo {year} {2024})}\BibitemShut {NoStop}%
\bibitem [{\citenamefont {Wang}\ and\ \citenamefont
  {Hazzard}(2020)}]{wang2020tightening}%
  \BibitemOpen
  \bibfield  {author} {\bibinfo {author} {\bibfnamefont {Z.}~\bibnamefont
  {Wang}}\ and\ \bibinfo {author} {\bibfnamefont {K.~R.~A.}\ \bibnamefont
  {Hazzard}},\ }\href {\doibase 10.1103/PRXQuantum.1.010303} {\bibfield
  {journal} {\bibinfo  {journal} {PRX Quantum}\ }\textbf {\bibinfo {volume}
  {1}},\ \bibinfo {pages} {010303} (\bibinfo {year} {2020})}\BibitemShut
  {NoStop}%
\bibitem [{\citenamefont {Lieb}\ and\ \citenamefont
  {Robinson}(1972)}]{Lieb1972}%
  \BibitemOpen
  \bibfield  {author} {\bibinfo {author} {\bibfnamefont {E.~H.}\ \bibnamefont
  {Lieb}}\ and\ \bibinfo {author} {\bibfnamefont {D.~W.}\ \bibnamefont
  {Robinson}},\ }\href {\doibase 10.1007/BF01645779} {\bibfield  {journal}
  {\bibinfo  {journal} {Commun. Math. Phys.}\ }\textbf {\bibinfo {volume}
  {28}},\ \bibinfo {pages} {251} (\bibinfo {year} {1972})}\BibitemShut
  {NoStop}%
\bibitem [{\citenamefont {Kitaev}(2006)}]{kitaev2006anyons}%
  \BibitemOpen
  \bibfield  {author} {\bibinfo {author} {\bibfnamefont {A.}~\bibnamefont
  {Kitaev}},\ }\href {\doibase https://doi.org/10.1016/j.aop.2005.10.005}
  {\bibfield  {journal} {\bibinfo  {journal} {Ann. Phys.}\ }\textbf {\bibinfo
  {volume} {321}},\ \bibinfo {pages} {2} (\bibinfo {year} {2006})}\BibitemShut
  {NoStop}%
\bibitem [{\citenamefont {Yao}\ and\ \citenamefont
  {Kivelson}(2007)}]{YaoKivelson}%
  \BibitemOpen
  \bibfield  {author} {\bibinfo {author} {\bibfnamefont {H.}~\bibnamefont
  {Yao}}\ and\ \bibinfo {author} {\bibfnamefont {S.~A.}\ \bibnamefont
  {Kivelson}},\ }\href {\doibase 10.1103/PhysRevLett.99.247203} {\bibfield
  {journal} {\bibinfo  {journal} {Phys. Rev. Lett.}\ }\textbf {\bibinfo
  {volume} {99}},\ \bibinfo {pages} {247203} (\bibinfo {year}
  {2007})}\BibitemShut {NoStop}%
\bibitem [{\citenamefont {Chapman}\ and\ \citenamefont
  {Flammia}(2020)}]{Chapman2020characterizationof}%
  \BibitemOpen
  \bibfield  {author} {\bibinfo {author} {\bibfnamefont {A.}~\bibnamefont
  {Chapman}}\ and\ \bibinfo {author} {\bibfnamefont {S.~T.}\ \bibnamefont
  {Flammia}},\ }\href {\doibase 10.22331/q-2020-06-04-278} {\bibfield
  {journal} {\bibinfo  {journal} {{Quantum}}\ }\textbf {\bibinfo {volume}
  {4}},\ \bibinfo {pages} {278} (\bibinfo {year} {2020})}\BibitemShut {NoStop}%
\bibitem [{\citenamefont {Dubail}\ and\ \citenamefont
  {Read}(2015)}]{DubailRead2015}%
  \BibitemOpen
  \bibfield  {author} {\bibinfo {author} {\bibfnamefont {J.}~\bibnamefont
  {Dubail}}\ and\ \bibinfo {author} {\bibfnamefont {N.}~\bibnamefont {Read}},\
  }\href {\doibase 10.1103/PhysRevB.92.205307} {\bibfield  {journal} {\bibinfo
  {journal} {Phys. Rev. B}\ }\textbf {\bibinfo {volume} {92}},\ \bibinfo
  {pages} {205307} (\bibinfo {year} {2015})}\BibitemShut {NoStop}%
\bibitem [{\citenamefont {Wang}\ \emph {et~al.}(2017)\citenamefont {Wang},
  \citenamefont {Xu}, \citenamefont {Pu},\ and\ \citenamefont
  {Hazzard}}]{Wang2017Number}%
  \BibitemOpen
  \bibfield  {author} {\bibinfo {author} {\bibfnamefont {Z.}~\bibnamefont
  {Wang}}, \bibinfo {author} {\bibfnamefont {Y.}~\bibnamefont {Xu}}, \bibinfo
  {author} {\bibfnamefont {H.}~\bibnamefont {Pu}}, \ and\ \bibinfo {author}
  {\bibfnamefont {K.~R.~A.}\ \bibnamefont {Hazzard}},\ }\href {\doibase
  10.1103/PhysRevB.96.115110} {\bibfield  {journal} {\bibinfo  {journal} {Phys.
  Rev. B}\ }\textbf {\bibinfo {volume} {96}},\ \bibinfo {pages} {115110}
  (\bibinfo {year} {2017})},\ \Eprint {http://arxiv.org/abs/1703.01249}
  {arXiv:1703.01249} \BibitemShut {NoStop}%
\bibitem [{\citenamefont {Wang}\ and\ \citenamefont
  {Hazzard}(2018)}]{Wang2018}%
  \BibitemOpen
  \bibfield  {author} {\bibinfo {author} {\bibfnamefont {Z.}~\bibnamefont
  {Wang}}\ and\ \bibinfo {author} {\bibfnamefont {K.~R.}\ \bibnamefont
  {Hazzard}},\ }\href {\doibase 10.1103/PhysRevB.97.104501} {\bibfield
  {journal} {\bibinfo  {journal} {Phys. Rev. B}\ }\textbf {\bibinfo {volume}
  {97}} (\bibinfo {year} {2018}),\ 10.1103/PhysRevB.97.104501},\ \Eprint
  {http://arxiv.org/abs/1712.09904} {arXiv:1712.09904} \BibitemShut {NoStop}%
\bibitem [{\citenamefont {Wang}\ \emph {et~al.}(2021)\citenamefont {Wang},
  \citenamefont {Foss-Feig},\ and\ \citenamefont {Hazzard}}]{Wang2021Bounding}%
  \BibitemOpen
  \bibfield  {author} {\bibinfo {author} {\bibfnamefont {Z.}~\bibnamefont
  {Wang}}, \bibinfo {author} {\bibfnamefont {M.}~\bibnamefont {Foss-Feig}}, \
  and\ \bibinfo {author} {\bibfnamefont {K.~R.~A.}\ \bibnamefont {Hazzard}},\
  }\href {\doibase 10.1103/PhysRevResearch.3.L032047} {\bibfield  {journal}
  {\bibinfo  {journal} {Phys. Rev. Res.}\ }\textbf {\bibinfo {volume} {3}},\
  \bibinfo {pages} {L032047} (\bibinfo {year} {2021})}\BibitemShut {NoStop}%
\bibitem [{\citenamefont {Wang}(2025)}]{wang2024hopf}%
  \BibitemOpen
  \bibfield  {author} {\bibinfo {author} {\bibfnamefont {Z.}~\bibnamefont
  {Wang}},\ }\href {\doibase 10.1103/PhysRevB.111.104315} {\bibfield  {journal}
  {\bibinfo  {journal} {Phys. Rev. B}\ }\textbf {\bibinfo {volume} {111}},\
  \bibinfo {pages} {104315} (\bibinfo {year} {2025})}\BibitemShut {NoStop}%
\bibitem [{\citenamefont {Etingof}\ and\ \citenamefont
  {Gelaki}(1998)}]{etingof1998THAconstruction}%
  \BibitemOpen
  \bibfield  {author} {\bibinfo {author} {\bibfnamefont {P.}~\bibnamefont
  {Etingof}}\ and\ \bibinfo {author} {\bibfnamefont {S.}~\bibnamefont
  {Gelaki}},\ }\href@noop {} {\bibfield  {journal} {\bibinfo  {journal} {Math.
  Res. Lett.}\ }\textbf {\bibinfo {volume} {5}},\ \bibinfo {pages} {551}
  (\bibinfo {year} {1998})}\BibitemShut {NoStop}%
\bibitem [{\citenamefont {Molnar}\ \emph {et~al.}(2022)\citenamefont {Molnar},
  \citenamefont {Ruiz-de Alarc{\'{o}}n}, \citenamefont {Garre-Rubio},
  \citenamefont {Schuch}, \citenamefont {Cirac},\ and\ \citenamefont
  {P{\'e}rez-Garc{\'i}a}}]{molnar2022matrix}%
  \BibitemOpen
  \bibfield  {author} {\bibinfo {author} {\bibfnamefont {A.}~\bibnamefont
  {Molnar}}, \bibinfo {author} {\bibfnamefont {A.}~\bibnamefont {Ruiz-de
  Alarc{\'{o}}n}}, \bibinfo {author} {\bibfnamefont {J.}~\bibnamefont
  {Garre-Rubio}}, \bibinfo {author} {\bibfnamefont {N.}~\bibnamefont {Schuch}},
  \bibinfo {author} {\bibfnamefont {J.~I.}\ \bibnamefont {Cirac}}, \ and\
  \bibinfo {author} {\bibfnamefont {D.}~\bibnamefont {P{\'e}rez-Garc{\'i}a}},\
  }\href@noop {} {\bibfield  {journal} {\bibinfo  {journal} {arXiv Prepr.
  arXiv2204.05940}\ } (\bibinfo {year} {2022})}\BibitemShut {NoStop}%
\bibitem [{\citenamefont {Andruskiewitsch}\ \emph {et~al.}(2001)\citenamefont
  {Andruskiewitsch}, \citenamefont {Etingof},\ and\ \citenamefont
  {Gelaki}}]{andruskiewitsch2001THAChevalley}%
  \BibitemOpen
  \bibfield  {author} {\bibinfo {author} {\bibfnamefont {N.}~\bibnamefont
  {Andruskiewitsch}}, \bibinfo {author} {\bibfnamefont {P.}~\bibnamefont
  {Etingof}}, \ and\ \bibinfo {author} {\bibfnamefont {S.}~\bibnamefont
  {Gelaki}},\ }\href@noop {} {\bibfield  {journal} {\bibinfo  {journal}
  {Michigan Math. J.}\ }\textbf {\bibinfo {volume} {49}},\ \bibinfo {pages}
  {277} (\bibinfo {year} {2001})}\BibitemShut {NoStop}%
\bibitem [{\citenamefont {Etingof}\ and\ \citenamefont
  {Gelaki}(2001)}]{etingof2001classificationTHAchevalley}%
  \BibitemOpen
  \bibfield  {author} {\bibinfo {author} {\bibfnamefont {P.}~\bibnamefont
  {Etingof}}\ and\ \bibinfo {author} {\bibfnamefont {S.}~\bibnamefont
  {Gelaki}},\ }\href@noop {} {\bibfield  {journal} {\bibinfo  {journal} {Math.
  Res. Lett.}\ }\textbf {\bibinfo {volume} {8}},\ \bibinfo {pages} {249}
  (\bibinfo {year} {2001})}\BibitemShut {NoStop}%
\bibitem [{\citenamefont {Kac}(1977)}]{Kac1977LieSuper}%
  \BibitemOpen
  \bibfield  {author} {\bibinfo {author} {\bibfnamefont {V.~G.}\ \bibnamefont
  {Kac}},\ }\href {\doibase https://doi.org/10.1016/0001-8708(77)90017-2}
  {\bibfield  {journal} {\bibinfo  {journal} {Adv. Math. (N. Y).}\ }\textbf
  {\bibinfo {volume} {26}},\ \bibinfo {pages} {8} (\bibinfo {year}
  {1977})}\BibitemShut {NoStop}%
\bibitem [{\citenamefont {Rittenberg}\ and\ \citenamefont
  {Wyler}(1978{\natexlab{a}})}]{Rittenberg1978}%
  \BibitemOpen
  \bibfield  {author} {\bibinfo {author} {\bibfnamefont {V.}~\bibnamefont
  {Rittenberg}}\ and\ \bibinfo {author} {\bibfnamefont {D.}~\bibnamefont
  {Wyler}},\ }\href {\doibase https://doi.org/10.1016/0550-3213(78)90186-4}
  {\bibfield  {journal} {\bibinfo  {journal} {Nucl. Phys. B}\ }\textbf
  {\bibinfo {volume} {139}},\ \bibinfo {pages} {189} (\bibinfo {year}
  {1978}{\natexlab{a}})}\BibitemShut {NoStop}%
\bibitem [{\citenamefont {Rittenberg}\ and\ \citenamefont
  {Wyler}(1978{\natexlab{b}})}]{Rittenberg1978a}%
  \BibitemOpen
  \bibfield  {author} {\bibinfo {author} {\bibfnamefont {V.}~\bibnamefont
  {Rittenberg}}\ and\ \bibinfo {author} {\bibfnamefont {D.}~\bibnamefont
  {Wyler}},\ }\href {\doibase 10.1063/1.523552} {\bibfield  {journal} {\bibinfo
   {journal} {J. Math. Phys.}\ }\textbf {\bibinfo {volume} {19}},\ \bibinfo
  {pages} {2193} (\bibinfo {year} {1978}{\natexlab{b}})}\BibitemShut {NoStop}%
\bibitem [{\citenamefont {Toppan}(2021{\natexlab{a}})}]{Toppan2021Z2Z2}%
  \BibitemOpen
  \bibfield  {author} {\bibinfo {author} {\bibfnamefont {F.}~\bibnamefont
  {Toppan}},\ }\href {\doibase 10.1088/1751-8121/abe2f2} {\bibfield  {journal}
  {\bibinfo  {journal} {J. Phys. A Math. Theor.}\ }\textbf {\bibinfo {volume}
  {54}},\ \bibinfo {pages} {115203} (\bibinfo {year}
  {2021}{\natexlab{a}})}\BibitemShut {NoStop}%
\bibitem [{\citenamefont
  {Toppan}(2021{\natexlab{b}})}]{Toppan2021Inequivalent}%
  \BibitemOpen
  \bibfield  {author} {\bibinfo {author} {\bibfnamefont {F.}~\bibnamefont
  {Toppan}},\ }\href {\doibase 10.1088/1751-8121/ac17a5} {\bibfield  {journal}
  {\bibinfo  {journal} {J. Phys. A Math. Theor.}\ }\textbf {\bibinfo {volume}
  {54}},\ \bibinfo {pages} {355202} (\bibinfo {year}
  {2021}{\natexlab{b}})}\BibitemShut {NoStop}%
\bibitem [{Mma()}]{MmaCode}%
  \BibitemOpen
  \href@noop {} {}\bibinfo {note} {We provide accompanying Mathematica codes
  for computational verification of several important mathemtical facts in the
  main text and SI at
  \url{https://github.com/lagrenge94/Mathematica-codes-for-parastatistics}.
  There are two notebook files in total, their goals are: (1) In
  RMatricesAnd1DSpinModel.nb we construct all the $R$-matrices that appear in
  this paper, verify that they satisfy the YBE~\eqref{eq:YBE}, and compute the
  exclusion statistics and single mode partition function $z_R(x)$ for each
  $R$-matrix; then we construct the local operators of the 1D solvable spin
  model for each $R$-matrix, and verify that they satisfy the defining
  relations in Eq.~\eqref{eq:XYQA}; (2) In Verify2DSpinModel.nb, we first
  construct the tensors $u_L, u_R,v_L,v_R$ in the 2D spin model, and verify
  they satisfy all the CRs in Fig.~\ref{fig:CRuLvLuRvR}; then we verify
  Fact.~\ref{fact:paravacuum} that the $\hat{H}_1$ reduces to Kitaev's quantum
  double model when all black sites are in the state $\ket{0}$, and has a
  unique ground state on a $3\times 3$ lattice; then we verify that the action
  of the JW string on the ground state is path independent, as claimed in
  Fact.~\ref{lemma:path-indep}; finally we verify that the JW string connecting
  upper left and lower right corner acts on the ground state as a delta
  function $\hat{W}_{bc}\ket{G}=\delta_{bc}\ket{G}$, a property we used with
  only a partial proof in Sec.~\ref{SI:paracreation_measure}. Note that since
  the quantum double ground state has zero correlation length, finite size
  error vanishes~\cite{Wang2021Bounding} and all the properties above hold
  exactly even on a small lattice with $L=3$, as we see in the
  code.}\BibitemShut {Stop}%
\bibitem [{\citenamefont {Wang}\ and\ \citenamefont
  {Hazzard}(2023)}]{Wang2022IK}%
  \BibitemOpen
  \bibfield  {author} {\bibinfo {author} {\bibfnamefont {Z.}~\bibnamefont
  {Wang}}\ and\ \bibinfo {author} {\bibfnamefont {K.~R.~A.}\ \bibnamefont
  {Hazzard}},\ }\href {\doibase 10.1103/PhysRevResearch.5.013086} {\bibfield
  {journal} {\bibinfo  {journal} {Phys. Rev. Res.}\ }\textbf {\bibinfo {volume}
  {5}},\ \bibinfo {pages} {13086} (\bibinfo {year} {2023})}\BibitemShut
  {NoStop}%
\bibitem [{\citenamefont {Fendley}(2014)}]{Fendley2014Freepara}%
  \BibitemOpen
  \bibfield  {author} {\bibinfo {author} {\bibfnamefont {P.}~\bibnamefont
  {Fendley}},\ }\href {\doibase 10.1088/1751-8113/47/7/075001} {\bibfield
  {journal} {\bibinfo  {journal} {J. Phys. A Math. Theor.}\ }\textbf {\bibinfo
  {volume} {47}},\ \bibinfo {pages} {75001} (\bibinfo {year}
  {2014})}\BibitemShut {NoStop}%
\bibitem [{\citenamefont {Alicea}\ and\ \citenamefont
  {Fendley}(2016)}]{Fendley2016review}%
  \BibitemOpen
  \bibfield  {author} {\bibinfo {author} {\bibfnamefont {J.}~\bibnamefont
  {Alicea}}\ and\ \bibinfo {author} {\bibfnamefont {P.}~\bibnamefont
  {Fendley}},\ }\href {\doibase 10.1146/annurev-conmatphys-031115-011336}
  {\bibfield  {journal} {\bibinfo  {journal} {Annu. Rev. Condens. Matter
  Phys.}\ }\textbf {\bibinfo {volume} {7}},\ \bibinfo {pages} {119} (\bibinfo
  {year} {2016})}\BibitemShut {NoStop}%
\bibitem [{\citenamefont {Stoilova}\ and\ \citenamefont {{Van der
  Jeugt}}(2020)}]{Stoilova2020}%
  \BibitemOpen
  \bibfield  {author} {\bibinfo {author} {\bibfnamefont {N.~I.}\ \bibnamefont
  {Stoilova}}\ and\ \bibinfo {author} {\bibfnamefont {J.}~\bibnamefont {{Van
  der Jeugt}}},\ }\href {\doibase
  https://doi.org/10.1016/j.physleta.2020.126421} {\bibfield  {journal}
  {\bibinfo  {journal} {Phys. Lett. A}\ }\textbf {\bibinfo {volume} {384}},\
  \bibinfo {pages} {126421} (\bibinfo {year} {2020})}\BibitemShut {NoStop}%
\bibitem [{\citenamefont {Aneva}\ and\ \citenamefont
  {Popov}(2005)}]{Aneva2005deformedGreen}%
  \BibitemOpen
  \bibfield  {author} {\bibinfo {author} {\bibfnamefont {B.}~\bibnamefont
  {Aneva}}\ and\ \bibinfo {author} {\bibfnamefont {T.}~\bibnamefont {Popov}},\
  }\href {\doibase 10.1088/0305-4470/38/29/004} {\bibfield  {journal} {\bibinfo
   {journal} {J. Phys. A. Math. Gen.}\ }\textbf {\bibinfo {volume} {38}},\
  \bibinfo {pages} {6473} (\bibinfo {year} {2005})}\BibitemShut {NoStop}%
\bibitem [{\citenamefont {Kanakoglou}\ and\ \citenamefont
  {Daskaloyannis}(2007)}]{Kanakoglou2007braidedGreen}%
  \BibitemOpen
  \bibfield  {author} {\bibinfo {author} {\bibfnamefont {K.}~\bibnamefont
  {Kanakoglou}}\ and\ \bibinfo {author} {\bibfnamefont {C.}~\bibnamefont
  {Daskaloyannis}},\ }\href {\doibase 10.1063/1.2816258} {\bibfield  {journal}
  {\bibinfo  {journal} {J. Math. Phys.}\ }\textbf {\bibinfo {volume} {48}},\
  \bibinfo {pages} {113516} (\bibinfo {year} {2007})}\BibitemShut {NoStop}%
\bibitem [{\citenamefont {Alderete}\ and\ \citenamefont
  {Rodriguez-Lara}(2017)}]{alderete2017Greenexperiment}%
  \BibitemOpen
  \bibfield  {author} {\bibinfo {author} {\bibfnamefont {C.~H.}\ \bibnamefont
  {Alderete}}\ and\ \bibinfo {author} {\bibfnamefont {B.~M.}\ \bibnamefont
  {Rodriguez-Lara}},\ }\href@noop {} {\bibfield  {journal} {\bibinfo  {journal}
  {Phys. Rev. A}\ }\textbf {\bibinfo {volume} {95}},\ \bibinfo {pages} {013820}
  (\bibinfo {year} {2017})}\BibitemShut {NoStop}%
\bibitem [{\citenamefont {Alderete}\ \emph {et~al.}(2021)\citenamefont
  {Alderete}, \citenamefont {Green}, \citenamefont {Nguyen}, \citenamefont
  {Zhu}, \citenamefont {Rodriguez-Lara},\ and\ \citenamefont
  {Linke}}]{alderete2021experimental}%
  \BibitemOpen
  \bibfield  {author} {\bibinfo {author} {\bibfnamefont {C.~H.}\ \bibnamefont
  {Alderete}}, \bibinfo {author} {\bibfnamefont {A.~M.}\ \bibnamefont {Green}},
  \bibinfo {author} {\bibfnamefont {N.~H.}\ \bibnamefont {Nguyen}}, \bibinfo
  {author} {\bibfnamefont {Y.}~\bibnamefont {Zhu}}, \bibinfo {author}
  {\bibfnamefont {B.~M.}\ \bibnamefont {Rodriguez-Lara}}, \ and\ \bibinfo
  {author} {\bibfnamefont {N.~M.}\ \bibnamefont {Linke}},\ }\href@noop {}
  {\bibfield  {journal} {\bibinfo  {journal} {arXiv Prepr. arXiv2108.05471}\ }
  (\bibinfo {year} {2021})}\BibitemShut {NoStop}%
\bibitem [{\citenamefont {Biedenharn}(1989)}]{Biedenharn1989qboson}%
  \BibitemOpen
  \bibfield  {author} {\bibinfo {author} {\bibfnamefont {L.~C.}\ \bibnamefont
  {Biedenharn}},\ }\href {\doibase 10.1088/0305-4470/22/18/004} {\bibfield
  {journal} {\bibinfo  {journal} {J. Phys. A. Math. Gen.}\ }\textbf {\bibinfo
  {volume} {22}},\ \bibinfo {pages} {L873} (\bibinfo {year}
  {1989})}\BibitemShut {NoStop}%
\bibitem [{\citenamefont {Gentile~j.}(1940)}]{Gentile1940}%
  \BibitemOpen
  \bibfield  {author} {\bibinfo {author} {\bibfnamefont {G.}~\bibnamefont
  {Gentile~j.}},\ }\href {\doibase 10.1007/BF02960187} {\bibfield  {journal}
  {\bibinfo  {journal} {Nuovo Cim.}\ }\textbf {\bibinfo {volume} {17}},\
  \bibinfo {pages} {493} (\bibinfo {year} {1940})}\BibitemShut {NoStop}%
\bibitem [{\citenamefont {Haldane}(1991)}]{Haldane1991}%
  \BibitemOpen
  \bibfield  {author} {\bibinfo {author} {\bibfnamefont {F.~D.~M.}\
  \bibnamefont {Haldane}},\ }\href {\doibase 10.1103/PhysRevLett.67.937}
  {\bibfield  {journal} {\bibinfo  {journal} {Phys. Rev. Lett.}\ }\textbf
  {\bibinfo {volume} {67}},\ \bibinfo {pages} {937} (\bibinfo {year}
  {1991})}\BibitemShut {NoStop}%
\bibitem [{\citenamefont {Laughlin}(1983)}]{Laughlin1983}%
  \BibitemOpen
  \bibfield  {author} {\bibinfo {author} {\bibfnamefont {R.~B.}\ \bibnamefont
  {Laughlin}},\ }\href@noop {} {\bibfield  {journal} {\bibinfo  {journal}
  {Phys. Rev. Lett.}\ }\textbf {\bibinfo {volume} {50}},\ \bibinfo {pages}
  {1395} (\bibinfo {year} {1983})}\BibitemShut {NoStop}%
\bibitem [{\citenamefont {Schoutens}(1997)}]{Schoutens1997CFT}%
  \BibitemOpen
  \bibfield  {author} {\bibinfo {author} {\bibfnamefont {K.}~\bibnamefont
  {Schoutens}},\ }\href {\doibase 10.1103/PhysRevLett.79.2608} {\bibfield
  {journal} {\bibinfo  {journal} {Phys. Rev. Lett.}\ }\textbf {\bibinfo
  {volume} {79}},\ \bibinfo {pages} {2608} (\bibinfo {year}
  {1997})}\BibitemShut {NoStop}%
\bibitem [{\citenamefont {Bouwknegt}\ and\ \citenamefont
  {Schoutens}(1999)}]{Bouwknegt1999CFT}%
  \BibitemOpen
  \bibfield  {author} {\bibinfo {author} {\bibfnamefont {P.}~\bibnamefont
  {Bouwknegt}}\ and\ \bibinfo {author} {\bibfnamefont {K.}~\bibnamefont
  {Schoutens}},\ }\href {\doibase
  https://doi.org/10.1016/S0550-3213(99)00095-4} {\bibfield  {journal}
  {\bibinfo  {journal} {Nucl. Phys. B}\ }\textbf {\bibinfo {volume} {547}},\
  \bibinfo {pages} {501} (\bibinfo {year} {1999})}\BibitemShut {NoStop}%
\bibitem [{Note7()}]{Note7}%
  \BibitemOpen
  \bibinfo {note} {First, notice that $\protect \mathaccentV {hat}05E{\psi
  }^+_{i,a}=(\protect \mathaccentV {hat}05E{\psi }^-_{i,a})^\dagger $ is fully
  consistent with the fundamental CRs in Eq.~\protect \textup {\hbox
  {\mathsurround \z@ \protect \normalfont (\ignorespaces \ref
  {eq:fundamental_Rcommu}\unskip \@@italiccorr )}}, as taking the Hermitian
  conjugate on both sides leaves Eq.~\protect \textup {\hbox {\mathsurround \z@
  \protect \normalfont (\ignorespaces \ref {eq:fundamental_Rcommu}\unskip
  \@@italiccorr )}} invariant~(maps the first line to itself and swaps the
  second and the third lines). Then it can be checked straightforwardly that
  the explicit matrix representation of $\protect \mathaccentV {hat}05E{\psi
  }^\pm _{i,a}$ defined in Sec.~\ref {SI:action_parafield_basis} indeed
  satisfies $\protect \mathaccentV {hat}05E{\psi }^+_{i,a}=(\protect
  \mathaccentV {hat}05E{\psi }^-_{i,a})^\dagger $. By contrast, if $R$ is not
  unitary, the relation $\protect \mathaccentV {hat}05E{\psi
  }^+_{i,a}=(\protect \mathaccentV {hat}05E{\psi }^-_{i,a})^\dagger $ is not
  consistent with Eq.~\protect \textup {\hbox {\mathsurround \z@ \protect
  \normalfont (\ignorespaces \ref {eq:fundamental_Rcommu}\unskip \@@italiccorr
  )}} as taking the Hermitian conjugate on both sides leads to extra relations
  that result in an algebraic inconsistency, and the representation of
  $\protect \mathaccentV {hat}05E{\psi }^\pm _{i,a}$ constructed in Sec.~\ref
  {SI:action_parafield_basis} does not satisfy $\protect \mathaccentV
  {hat}05E{\psi }^+_{i,a}=(\protect \mathaccentV {hat}05E{\psi
  }^-_{i,a})^\dagger $. However, as we show in Sec.~\ref
  {SI:proof_hermiticity}, even for a non-unitary $R$, we can still define a
  Hermitian inner product on the Fock space such that $\protect \mathaccentV
  {hat}05E{e}^\dagger _{ij}=\protect \mathaccentV {hat}05E{e}_{ji}$ is
  satisfied, and consequently all physical observables are Hermitian with
  respect to this inner product.}\BibitemShut {Stop}%
\bibitem [{\citenamefont {Polishchuk}\ and\ \citenamefont
  {Positselski}(2005)}]{polishchuk2005quadratic}%
  \BibitemOpen
  \bibfield  {author} {\bibinfo {author} {\bibfnamefont {A.}~\bibnamefont
  {Polishchuk}}\ and\ \bibinfo {author} {\bibfnamefont {L.}~\bibnamefont
  {Positselski}},\ }\href {\doibase 10.1090/ulect/037} {\emph {\bibinfo {title}
  {Quadratic algebras}}},\ Vol.~\bibinfo {volume} {37}\ (\bibinfo  {publisher}
  {American Mathematical Society},\ \bibinfo {year} {2005})\BibitemShut
  {NoStop}%
\bibitem [{\citenamefont {Giaquinto}\ and\ \citenamefont
  {Zhang}(1995)}]{GIAQUINTO1995QWA}%
  \BibitemOpen
  \bibfield  {author} {\bibinfo {author} {\bibfnamefont {A.}~\bibnamefont
  {Giaquinto}}\ and\ \bibinfo {author} {\bibfnamefont {J.}~\bibnamefont
  {Zhang}},\ }\href {\doibase https://doi.org/10.1006/jabr.1995.1276}
  {\bibfield  {journal} {\bibinfo  {journal} {J. Algebra}\ }\textbf {\bibinfo
  {volume} {176}},\ \bibinfo {pages} {861} (\bibinfo {year}
  {1995})}\BibitemShut {NoStop}%
\bibitem [{\citenamefont {Etingof}\ \emph {et~al.}(2011)\citenamefont
  {Etingof}, \citenamefont {Golberg}, \citenamefont {Hensel}, \citenamefont
  {Liu}, \citenamefont {Schwendner}, \citenamefont {Vaintrob},\ and\
  \citenamefont {Yudovina}}]{etingof2011introduction}%
  \BibitemOpen
  \bibfield  {author} {\bibinfo {author} {\bibfnamefont {P.~I.}\ \bibnamefont
  {Etingof}}, \bibinfo {author} {\bibfnamefont {O.}~\bibnamefont {Golberg}},
  \bibinfo {author} {\bibfnamefont {S.}~\bibnamefont {Hensel}}, \bibinfo
  {author} {\bibfnamefont {T.}~\bibnamefont {Liu}}, \bibinfo {author}
  {\bibfnamefont {A.}~\bibnamefont {Schwendner}}, \bibinfo {author}
  {\bibfnamefont {D.}~\bibnamefont {Vaintrob}}, \ and\ \bibinfo {author}
  {\bibfnamefont {E.}~\bibnamefont {Yudovina}},\ }\href {\doibase
  10.1090/stml/059} {\emph {\bibinfo {title} {Introduction to representation
  theory}}},\ Vol.~\bibinfo {volume} {59}\ (\bibinfo  {publisher} {American
  Mathematical Society},\ \bibinfo {year} {2011})\BibitemShut {NoStop}%
\bibitem [{\citenamefont {Bourbaki}(2005)}]{bourbaki2005lie_chap9}%
  \BibitemOpen
  \bibfield  {author} {\bibinfo {author} {\bibfnamefont {N.}~\bibnamefont
  {Bourbaki}},\ }in\ \href@noop {} {\emph {\bibinfo {booktitle} {Lie groups and
  {L}ie algebras: chapters 7-9}}}\ (\bibinfo  {publisher} {Springer},\ \bibinfo
  {address} {Berlin Heidelberg},\ \bibinfo {year} {2005})\ Chap.~\bibinfo
  {chapter} {9}, pp.\ \bibinfo {pages} {281--377}\BibitemShut {NoStop}%
\bibitem [{\citenamefont {Humphreys}(1972)}]{humphreys_LA}%
  \BibitemOpen
  \bibfield  {author} {\bibinfo {author} {\bibfnamefont {J.~E.}\ \bibnamefont
  {Humphreys}},\ }\href {\doibase https://doi.org/10.1007/978-1-4612-6398-2}
  {\emph {\bibinfo {title} {Introduction to Lie algebras and representation
  theory}}}\ (\bibinfo  {publisher} {Springer-Verlag (New York)},\ \bibinfo
  {year} {1972})\BibitemShut {NoStop}%
\bibitem [{Note8()}]{Note8}%
  \BibitemOpen
  \bibinfo {note} {Note that it is exactly for this reason that the
  Doplicher-Haag-Roberts~(DHR) no-go theorem~\cite
  {doplicher1971local,*doplicher1974local} does not apply to the first
  quantization formulation, since the former takes locality as a fundamental
  assumption, while the latter does not have locality built-in.}\BibitemShut
  {Stop}%
\bibitem [{\citenamefont {Hartle}\ and\ \citenamefont
  {Taylor}(1969)}]{hartle1969}%
  \BibitemOpen
  \bibfield  {author} {\bibinfo {author} {\bibfnamefont {J.~B.}\ \bibnamefont
  {Hartle}}\ and\ \bibinfo {author} {\bibfnamefont {J.~R.}\ \bibnamefont
  {Taylor}},\ }\href {\doibase 10.1103/PhysRev.178.2043} {\bibfield  {journal}
  {\bibinfo  {journal} {Phys. Rev.}\ }\textbf {\bibinfo {volume} {178}},\
  \bibinfo {pages} {2043} (\bibinfo {year} {1969})}\BibitemShut {NoStop}%
\bibitem [{\citenamefont {Stolt}\ and\ \citenamefont
  {Taylor}(1970{\natexlab{b}})}]{Taylor1970a}%
  \BibitemOpen
  \bibfield  {author} {\bibinfo {author} {\bibfnamefont {R.~H.}\ \bibnamefont
  {Stolt}}\ and\ \bibinfo {author} {\bibfnamefont {J.~R.}\ \bibnamefont
  {Taylor}},\ }\href {\doibase 10.1103/PhysRevD.1.2226} {\bibfield  {journal}
  {\bibinfo  {journal} {Phys. Rev. D}\ }\textbf {\bibinfo {volume} {1}},\
  \bibinfo {pages} {2226} (\bibinfo {year} {1970}{\natexlab{b}})}\BibitemShut
  {NoStop}%
\bibitem [{\citenamefont {Lechner}\ \emph {et~al.}(2019)\citenamefont
  {Lechner}, \citenamefont {Pennig},\ and\ \citenamefont
  {Wood}}]{LECHNER2019106769}%
  \BibitemOpen
  \bibfield  {author} {\bibinfo {author} {\bibfnamefont {G.}~\bibnamefont
  {Lechner}}, \bibinfo {author} {\bibfnamefont {U.}~\bibnamefont {Pennig}}, \
  and\ \bibinfo {author} {\bibfnamefont {S.}~\bibnamefont {Wood}},\ }\href
  {\doibase https://doi.org/10.1016/j.aim.2019.106769} {\bibfield  {journal}
  {\bibinfo  {journal} {Adv. Math.}\ }\textbf {\bibinfo {volume} {355}},\
  \bibinfo {pages} {106769} (\bibinfo {year} {2019})}\BibitemShut {NoStop}%
\bibitem [{Note9()}]{Note9}%
  \BibitemOpen
  \bibinfo {note} {The model is still well-defined for $m=2$, where $\protect
  \mathaccentV {hat}05E{H}$ is Hermitian. But that case is trivial: when $m=2$,
  $\protect \mathaccentV {hat}05E{H}$ is equal to the sum of two decoupled
  chains of XY models.}\BibitemShut {Stop}%
\bibitem [{\citenamefont {Bender}(2007)}]{Bender_2007}%
  \BibitemOpen
  \bibfield  {author} {\bibinfo {author} {\bibfnamefont {C.~M.}\ \bibnamefont
  {Bender}},\ }\href {\doibase 10.1088/0034-4885/70/6/r03} {\bibfield
  {journal} {\bibinfo  {journal} {Rep. Prog. Phys.}\ }\textbf {\bibinfo
  {volume} {70}},\ \bibinfo {pages} {947} (\bibinfo {year} {2007})}\BibitemShut
  {NoStop}%
\bibitem [{\citenamefont {El-Ganainy}\ \emph {et~al.}(2018)\citenamefont
  {El-Ganainy}, \citenamefont {Makris}, \citenamefont {Khajavikhan},
  \citenamefont {Musslimani}, \citenamefont {Rotter},\ and\ \citenamefont
  {Christodoulides}}]{el2018non}%
  \BibitemOpen
  \bibfield  {author} {\bibinfo {author} {\bibfnamefont {R.}~\bibnamefont
  {El-Ganainy}}, \bibinfo {author} {\bibfnamefont {K.~G.}\ \bibnamefont
  {Makris}}, \bibinfo {author} {\bibfnamefont {M.}~\bibnamefont {Khajavikhan}},
  \bibinfo {author} {\bibfnamefont {Z.~H.}\ \bibnamefont {Musslimani}},
  \bibinfo {author} {\bibfnamefont {S.}~\bibnamefont {Rotter}}, \ and\ \bibinfo
  {author} {\bibfnamefont {D.~N.}\ \bibnamefont {Christodoulides}},\ }\href
  {\doibase https://doi.org/10.1038/nphys4323} {\bibfield  {journal} {\bibinfo
  {journal} {Nat. Phys.}\ }\textbf {\bibinfo {volume} {14}},\ \bibinfo {pages}
  {11} (\bibinfo {year} {2018})}\BibitemShut {NoStop}%
\bibitem [{Note10()}]{Note10}%
  \BibitemOpen
  \bibinfo {note} {Note that when $R'=X$, Eq.~\protect \textup {\hbox
  {\mathsurround \z@ \protect \normalfont (\ignorespaces \ref
  {eq:uuRSuugraphical}\unskip \@@italiccorr )}} is very similar to Eq.~(1) of
  Ref.~\cite {wang2024hopf}, where $R$ corresponds to the solvable gate $U$,
  and the tensors $u,v$ correspond to the tensors $\rho ,v$. Indeed, one can
  use Eq.~(2) of Ref.~\cite {wang2024hopf} to construct the tensors $u,v$ here
  if we know the underlying Hopf algebra for the $R$-matrix.}\BibitemShut
  {Stop}%
\bibitem [{\citenamefont {Radford}(1993)}]{Radford1993MQHA}%
  \BibitemOpen
  \bibfield  {author} {\bibinfo {author} {\bibfnamefont {D.~E.}\ \bibnamefont
  {Radford}},\ }\href {\doibase https://doi.org/10.1006/jabr.1993.1102}
  {\bibfield  {journal} {\bibinfo  {journal} {J. Algebr.}\ }\textbf {\bibinfo
  {volume} {157}},\ \bibinfo {pages} {285} (\bibinfo {year}
  {1993})}\BibitemShut {NoStop}%
\bibitem [{\citenamefont {Majid}(1995)}]{Majid1995BookFoundationQG}%
  \BibitemOpen
  \bibfield  {author} {\bibinfo {author} {\bibfnamefont {S.}~\bibnamefont
  {Majid}},\ }\href {\doibase DOI: 10.1017/CBO9780511613104} {\emph {\bibinfo
  {title} {{Foundations of Quantum Group Theory}}}}\ (\bibinfo  {publisher}
  {Cambridge University Press},\ \bibinfo {address} {Cambridge},\ \bibinfo
  {year} {1995})\BibitemShut {NoStop}%
\bibitem [{\citenamefont {Etingof}\ and\ \citenamefont
  {Gelaki}(2000)}]{etingof2000semisimpleTHAclassification}%
  \BibitemOpen
  \bibfield  {author} {\bibinfo {author} {\bibfnamefont {P.}~\bibnamefont
  {Etingof}}\ and\ \bibinfo {author} {\bibfnamefont {S.}~\bibnamefont
  {Gelaki}},\ }\href@noop {} {\bibfield  {journal} {\bibinfo  {journal} {Int.
  Math. Res. Not.}\ }\textbf {\bibinfo {volume} {2000}},\ \bibinfo {pages}
  {223} (\bibinfo {year} {2000})}\BibitemShut {NoStop}%
\bibitem [{Note11()}]{Note11}%
  \BibitemOpen
  \bibinfo {note} {The minimal $\protect \mathbb {C}^*$ triangular Hopf algebra
  $\protect \mathcal {H}_{64}$ is constructed using the method introduced in
  Ref.~\cite {etingof1998THAconstruction}, and it plays a key role in the
  construction of the tensors $u_L,v_L,u_R,v_R$. Our construction of
  $u_L,v_L,u_R,v_R$ from a minimal triangular Hopf algebra is partially
  motivated by the relation between MPO and Hopf algebra introduced in
  Ref.~\cite {molnar2022matrix}. We here mention that it may also be
  interesting to realize emergent paraparticles in higher dimensional fermionic
  systems, where triangular Hopf superalgebras~\cite
  {andruskiewitsch2001THAChevalley,etingof2001classificationTHAchevalley} may
  be useful tools in constructing exactly solvable models~(this is partially
  motivated by the connection between Lie superalgebras~\cite
  {Kac1977LieSuper,Rittenberg1978,Rittenberg1978a} and Green's parastatistics
  described in some recent works~\cite
  {Toppan2021Z2Z2,Toppan2021Inequivalent}).}\BibitemShut {Stop}%
\bibitem [{\citenamefont {Buerschaper}\ and\ \citenamefont
  {Aguado}(2009)}]{Buerschaper2009QDMSN}%
  \BibitemOpen
  \bibfield  {author} {\bibinfo {author} {\bibfnamefont {O.}~\bibnamefont
  {Buerschaper}}\ and\ \bibinfo {author} {\bibfnamefont {M.}~\bibnamefont
  {Aguado}},\ }\href {\doibase 10.1103/PhysRevB.80.155136} {\bibfield
  {journal} {\bibinfo  {journal} {Phys. Rev. B}\ }\textbf {\bibinfo {volume}
  {80}},\ \bibinfo {pages} {155136} (\bibinfo {year} {2009})}\BibitemShut
  {NoStop}%
\bibitem [{\citenamefont {Beigi}\ \emph {et~al.}(2011)\citenamefont {Beigi},
  \citenamefont {Shor},\ and\ \citenamefont {Whalen}}]{Beigi2011QDMBoundary}%
  \BibitemOpen
  \bibfield  {author} {\bibinfo {author} {\bibfnamefont {S.}~\bibnamefont
  {Beigi}}, \bibinfo {author} {\bibfnamefont {P.~W.}\ \bibnamefont {Shor}}, \
  and\ \bibinfo {author} {\bibfnamefont {D.}~\bibnamefont {Whalen}},\ }\href
  {\doibase 10.1007/s00220-011-1294-x} {\bibfield  {journal} {\bibinfo
  {journal} {Commun. Math. Phys.}\ }\textbf {\bibinfo {volume} {306}},\
  \bibinfo {pages} {663} (\bibinfo {year} {2011})}\BibitemShut {NoStop}%
\bibitem [{\citenamefont {Buerschaper}\ \emph {et~al.}(2013)\citenamefont
  {Buerschaper}, \citenamefont {Mombelli}, \citenamefont {Christandl},\ and\
  \citenamefont {Aguado}}]{Buerschaper2013HATC}%
  \BibitemOpen
  \bibfield  {author} {\bibinfo {author} {\bibfnamefont {O.}~\bibnamefont
  {Buerschaper}}, \bibinfo {author} {\bibfnamefont {J.~M.}\ \bibnamefont
  {Mombelli}}, \bibinfo {author} {\bibfnamefont {M.}~\bibnamefont
  {Christandl}}, \ and\ \bibinfo {author} {\bibfnamefont {M.}~\bibnamefont
  {Aguado}},\ }\href {\doibase 10.1063/1.4773316} {\bibfield  {journal}
  {\bibinfo  {journal} {J. Math. Phys.}\ }\textbf {\bibinfo {volume} {54}},\
  \bibinfo {pages} {12201} (\bibinfo {year} {2013})}\BibitemShut {NoStop}%
\bibitem [{\citenamefont {Meusburger}(2017)}]{Meusburger2017QDMHopfGauge}%
  \BibitemOpen
  \bibfield  {author} {\bibinfo {author} {\bibfnamefont {C.}~\bibnamefont
  {Meusburger}},\ }\href {\doibase 10.1007/s00220-017-2860-7} {\bibfield
  {journal} {\bibinfo  {journal} {Commun. Math. Phys.}\ }\textbf {\bibinfo
  {volume} {353}},\ \bibinfo {pages} {413} (\bibinfo {year}
  {2017})}\BibitemShut {NoStop}%
\bibitem [{\citenamefont {Jia}\ \emph {et~al.}(2023)\citenamefont {Jia},
  \citenamefont {Tan}, \citenamefont {Kaszlikowski},\ and\ \citenamefont
  {Chang}}]{Jia2023WHAQDM}%
  \BibitemOpen
  \bibfield  {author} {\bibinfo {author} {\bibfnamefont {Z.}~\bibnamefont
  {Jia}}, \bibinfo {author} {\bibfnamefont {S.}~\bibnamefont {Tan}}, \bibinfo
  {author} {\bibfnamefont {D.}~\bibnamefont {Kaszlikowski}}, \ and\ \bibinfo
  {author} {\bibfnamefont {L.}~\bibnamefont {Chang}},\ }\href {\doibase
  10.1007/s00220-023-04792-9} {\bibfield  {journal} {\bibinfo  {journal}
  {Commun. Math. Phys.}\ }\textbf {\bibinfo {volume} {402}},\ \bibinfo {pages}
  {3045} (\bibinfo {year} {2023})}\BibitemShut {NoStop}%
\bibitem [{\citenamefont {Cowtan}\ and\ \citenamefont
  {Majid}(2023)}]{Cowtan2023QDMBoundary}%
  \BibitemOpen
  \bibfield  {author} {\bibinfo {author} {\bibfnamefont {A.}~\bibnamefont
  {Cowtan}}\ and\ \bibinfo {author} {\bibfnamefont {S.}~\bibnamefont {Majid}},\
  }\href {\doibase 10.1063/5.0127285} {\bibfield  {journal} {\bibinfo
  {journal} {J. Math. Phys.}\ }\textbf {\bibinfo {volume} {64}},\ \bibinfo
  {pages} {102203} (\bibinfo {year} {2023})}\BibitemShut {NoStop}%
\bibitem [{\citenamefont {Yan}\ \emph {et~al.}(2022)\citenamefont {Yan},
  \citenamefont {Chen},\ and\ \citenamefont {Cui}}]{Yan2022Ribbon}%
  \BibitemOpen
  \bibfield  {author} {\bibinfo {author} {\bibfnamefont {B.}~\bibnamefont
  {Yan}}, \bibinfo {author} {\bibfnamefont {P.}~\bibnamefont {Chen}}, \ and\
  \bibinfo {author} {\bibfnamefont {S.~X.}\ \bibnamefont {Cui}},\ }\href
  {\doibase 10.1088/1751-8121/ac552c} {\bibfield  {journal} {\bibinfo
  {journal} {J. Phys. A Math. Theor.}\ }\textbf {\bibinfo {volume} {55}},\
  \bibinfo {pages} {185201} (\bibinfo {year} {2022})}\BibitemShut {NoStop}%
\bibitem [{Note12()}]{Note12}%
  \BibitemOpen
  \bibinfo {note} {This is very natural from a physical viewpoint, because
  $W_{bc}=\delta _{bc}$ means that when there is only one paraparticle
  excitation, the index $b$ of the paraparticle does not change when we
  transport it from the upper left to the lower right corner using $\protect
  \mathaccentV {hat}05E{E}_{ij}$: $\protect \mathaccentV
  {hat}05E{y}^+_{i,b}\mathinner {|{G}\delimiter "526930B }=\protect
  \mathaccentV {hat}05E{\psi }^+_{i,b}\mathinner {|{G}\delimiter "526930B
  }\protect \xrightarrow {\protect \mathaccentV {hat}05E{E}_{ij}}\protect
  \mathaccentV {hat}05E{\psi }^+_{j,b}\mathinner {|{G}\delimiter "526930B
  }=\protect \mathaccentV {hat}05E{y}^+_{i,b}\mathinner {|{G}\delimiter
  "526930B }$. The proof involves technical results about the quantum double
  ground state, which we do not prove in this paper, but we verify it in a
  small system in the accompanying code~\cite {MmaCode}}\BibitemShut {NoStop}%
\bibitem [{\citenamefont {Naaijkens}(2017)}]{naaijkens2017quantum}%
  \BibitemOpen
  \bibfield  {author} {\bibinfo {author} {\bibfnamefont {P.}~\bibnamefont
  {Naaijkens}},\ }\href {\doibase https://doi.org/10.1007/978-3-319-51458-1}
  {\emph {\bibinfo {title} {Quantum spin systems on infinite lattices}}},\
  Lecture Notes in Physics\ (\bibinfo  {publisher} {Springer Cham},\ \bibinfo
  {year} {2017})\BibitemShut {NoStop}%
\bibitem [{\citenamefont {Witten}(1989)}]{Witten1989}%
  \BibitemOpen
  \bibfield  {author} {\bibinfo {author} {\bibfnamefont {E.}~\bibnamefont
  {Witten}},\ }\href {\doibase 10.1007/BF01217730} {\bibfield  {journal}
  {\bibinfo  {journal} {Commun. Math. Phys.}\ }\textbf {\bibinfo {volume}
  {121}},\ \bibinfo {pages} {351} (\bibinfo {year} {1989})}\BibitemShut
  {NoStop}%
\end{thebibliography}%
\bibliographystyle{apsrev4-1}

\clearpage
\newpage
\clearpage 
\twocolumngrid
\subsection{Methods}
\paragraph{Derivations for Eqs.~(\ref{eq:commu_Eab_psi_p},\ref{eq:commu_Eab_Ecd})}%
The commutator between $\hat{e}_{ij}$ and $\hat{\psi}^\pm_{k,b}$ is 
\begin{eqnarray}\label{eq:commu_Eab_psi_p-supp}
	[\hat{e}_{ij}, \hat{\psi}^+_{k,b}]&=&\sum_a\left(\hat{\psi}^+_{i,a}\hat{\psi}^-_{j,a}\hat{\psi}^+_{k,b}-\hat{\psi}^+_{k,b}\hat{\psi}^+_{i,a}\hat{\psi}^-_{j,a}\right)\nonumber\\
	&=&\sum_a\hat{\psi}^+_{i,a}\left(\sum_{c,d}R^{ac}_{bd}\hat{\psi}^+_{k,c}\hat{\psi}^-_{j,d}+\delta_{jk}\delta_{ab}\right)\nonumber\\
	&&-\sum_a\hat{\psi}^+_{k,b}\hat{\psi}^+_{i,a}\hat{\psi}^-_{j,a}\nonumber\\
	&=&\sum_{a,c,d}(R^{ac}_{bd}\hat{\psi}^+_{i,a}\hat{\psi}^+_{k,c})\hat{\psi}^-_{j,d}+\delta_{jk}\hat{\psi}^+_{i,b}\nonumber\\
	&&-\sum_a\hat{\psi}^+_{k,b}\hat{\psi}^+_{i,a}\hat{\psi}^-_{j,a}\nonumber\\
	&=&\delta_{jk}\hat{\psi}^+_{i,b},
\end{eqnarray}
where in the second~(third) line we used the first~(second) line of Eq.~\eqref{eq:fundamental_Rcommu}. %
Similarly, we have
\begin{eqnarray}\label{eq:commu_Eab_psi_m-supp}
	[\hat{e}_{ij}, \hat{\psi}^-_{k,b}]=-\delta_{ik}\hat{\psi}^-_{j,b}. 
\end{eqnarray}
Now we can compute the commutator %
\begin{eqnarray}\label{eq:commu_Eab_Ecd-supp}
	[\hat{e}_{ij}, \hat{e}_{kl}]%
	&=&\sum_b[\hat{e}_{ij}, \hat{\psi}^+_{k,b}]\hat{\psi}^-_{l,b}+\sum_b\hat{\psi}^+_{k,b}[\hat{e}_{ij}, \hat{\psi}^-_{l,b}]\nonumber\\
	&=&\sum_b\delta_{jk}\hat{\psi}^+_{i,b}\hat{\psi}^-_{l,b}-\sum_b\hat{\psi}^+_{k,b}\delta_{il}\hat{\psi}^-_{j,b}\nonumber\\
	&=&\delta_{jk}\hat{e}_{il}-\delta_{il}\hat{e}_{kj},
\end{eqnarray}
where in the second line we used Eqs.~\eqref{eq:commu_Eab_psi_p-supp} and \eqref{eq:commu_Eab_psi_m-supp}. %

\paragraph{Exact solution of free paraparticles}\label{sec:solution_me}
Here we present details for solving the general bilinear Hamiltonian in Eq.~\eqref{eq:u1_sym_H}.
Analogous to usual free bosons and fermions, we consider $U(N)$ transformations of $\{\hat{\psi}^\pm_{i,a}\}$:
\begin{eqnarray}\label{eq:fourier_transform}
	\hat{\psi}^-_{i,a}&=& \sum^N_{k=1} U_{ki}^{*} \tilde{\psi}^-_{k,a},\nonumber\\
	\hat{\psi}^+_{i,a}&=& \sum^N_{k=1} U_{ki} \tilde{\psi}^+_{k,a},
\end{eqnarray}
where $ U_{ki}$ is an $N\times N$ unitary matrix, and we use operators with a tilde $\tilde{\psi}^\pm_{k,a}$ to denote eigenmode operators.  %
Inserting Eq.~\eqref{eq:fourier_transform} into Eq.~\eqref{eq:fundamental_Rcommu}, we see that the operators  $\{\tilde{\psi}^\pm_{k,a}\}$ satisfy exactly the same CRs as $\{\hat{\psi}^\pm_{i,a}\}$. Notice that most of our discussions regarding the second quantization formulation and the state space %
only assume the CRs in Eq.~\eqref{eq:fundamental_Rcommu}, so the results obtained for $\{\hat{\psi}^\pm_{i,a}\}$~(in particular the Lie algebra of bilinear operators  and the structure of the state space) must also apply to $\{\tilde{\psi}^\pm_{k,a}\}$. %

Inserting Eq.~\eqref{eq:fourier_transform} into Eq.~\eqref{eq:u1_sym_H}, we obtain
\begin{equation}\label{eq:u1_sym_H_kspace}
	\hat{H}=\sum_{\substack{1\leq k,p\leq N\\1\leq a \leq m}} h'_{kp}  \tilde{\psi}^+_{k,a}\tilde{\psi}^-_{p,a}\equiv\sum_{1\leq k,p\leq N} h'_{kp} \tilde{e}_{kp},
\end{equation} 
where 
$	h'_{kp}  =\sum_{1\leq i,j\leq N}U_{ki} h_{ij}U_{pj}^*=[UhU^\dagger]_{kp}.$
We can therefore choose the unitary matrix $U$ such that $h'_{kp}=\epsilon_k\delta_{kp}$, %
where $\{\epsilon_{k}\}^N_{k=1}$ are eigenvalues of $h_{ij}$.  
With this choice the Hamiltonian becomes diagonal $\hat{H}=\sum^N_{k=1} \epsilon_{k} \tilde{n}_{k}$, 
and its eigenstates %
can be taken as the common eigenstates $|{}^{\alpha_1}_{\tilde{n}_1},{}_{\tilde{n}_2}^{\alpha_2},\ldots, {}_{\tilde{n}_N}^{\alpha_N}\rangle$~[defined in Eq.~\eqref{eq:full_basis} of the SI~\cite{Suppl}] of the mutually commuting operators $\{\tilde{n}_{k}\}^N_{k=1}$, with
energy eigenvalues $E=\sum^N_{k=1} \epsilon_{k} \tilde{n}_{k}$, where $\{\tilde{n}_{k}\}^N_{k=1}$ are independent non-negative integers and $1\leq \alpha_k\leq d_{\tilde{n}_k}$ encodes the single particle exclusion statistics. 

We now calculate physical observables at temperature $T$. %
The partition function is a product of single-mode partition functions in Eq.~\eqref{eq:single_mode_Z}
\begin{equation}\label{eq:part_func_Z}
	Z(\beta)\equiv \mathrm{Tr} [e^{-\beta \hat{H}}]=\prod_k z_R(e^{-\beta\epsilon_k}),
\end{equation}
so the free energy is
\begin{equation}\label{eq:free_energy}
	F(\beta)= -\frac{1}{\beta}\ln Z(\beta)=-\frac{1}{\beta}\sum_k \ln z_R(e^{-\beta\epsilon_k}).
\end{equation}
The partition function allows us to compute the thermal average of observables $\tilde{n}^l_k$ and $\tilde{e}_{kp}$
\begin{eqnarray}\label{eq:avg_nkl}
	\langle \tilde{n}^l_k\rangle_\beta&=&\frac{\mathrm{Tr} [\tilde{n}_{k}^l e^{-\beta \hat{H}}]}{ \mathrm{Tr} [e^{-\beta \hat{H}}]}=\left.\frac{(x\partial_x)^l z_R(x)}{z_R(x)}\right|_{x=e^{-\beta\epsilon_k}},\nonumber\\
	\braket{\tilde{e}_{kp}}_\beta &=& \delta_{kp} \braket{\tilde{n}_{k}}_\beta.
\end{eqnarray}

The thermal average for physical operators $ \hat{e}_{ij}$ are obtained by transforming creation and annihilation operators to the eigenmode basis using Eq.~\eqref{eq:fourier_transform}, and using the result for $\langle \tilde{e}_{kp}\rangle_\beta$ given in Eq.~\eqref{eq:avg_nkl}, which yields  
\begin{equation}
	\langle \hat{e}_{ij}\rangle_\beta=\sum_k U_{ki} U_{kj}^*\langle \tilde{n}_{k}\rangle_\beta. 
\end{equation} 
The thermal average for other physical observables, including correlation functions in and out of equilibrium, can all be calculated exactly in a similar way.
\paragraph{$R$-matrix for the 2D solvable spin model}
In the following we present a unitary $R$-matrix with trivial exclusion statistics but non-trivial exchange statistics, which the 2D solvable spin model is based on. %
We define the $R$ matrix with $m=4$ in the following way. Let $S=\{1,2,3,4\}$ and $r:S\times S\to S\times S$ be an injective map defined as 
\begin{equation}\label{eq:seth-R}
	r(a,b)=
	\left(
	\begin{array}{cccc}
		43 & 12 & 24 & 31 \\
		21 & 34 & 42 & 13 \\
		14 & 41 & 33 & 22 \\
		32 & 23 & 11 & 44 \\
	\end{array}
	\right)_{ab},
\end{equation}
where we use $ab$ as a short hand for $(a,b)$. For example, $r(1,1)=(4,3)$ and $r(3,2)=(4,1)$. The map $r$ in Eq.~\eqref{eq:seth-R} satisfies the set-theoretical YBE~\cite{etingof1999set}
\begin{equation}\label{eq:sethYBE}
r^2=\mathrm{id}_{S\times S},\quad	r_{12}r_{23}r_{12}=r_{23}r_{12}r_{23},
\end{equation}
where in the second equation both sides are injective maps from the set $S\times S\times S$ to itself,  $r_{12}=r\times \mathrm{id}_S$, and $r_{23}=\mathrm{id}_S\times r$. 
Now we define the $R$ matrix as
\begin{equation}\label{eq:set-thR}
	R\ket{a,b}=-\ket{b',a'},~~~\forall a,b\in S,
\end{equation}
where $(b',a')=r(a,b)$. 
It then follows from %
Eq.~\eqref{eq:sethYBE} that $R$ satisfies the YBE~\eqref{eq:YBE}. %
The  single mode partition function $z_R(x)$ of this $R$-matrix is~\cite{Suppl}
	$z_R(x)=(1+x)^4$,
meaning that the exclusion statistics of this type of paraparticles is the same as 4 decoupled flavors of ordinary fermions. %
Despite having trivial exclusion statistics, the permutation statistics defined by this $R$-matrix is notably distinct from fermions, as is manifest in the paraparticle exchange process in the 2D solvable spin model that we demonstrate later. %

\paragraph{Solvable 2D spin models with emergent free paraparticles}
	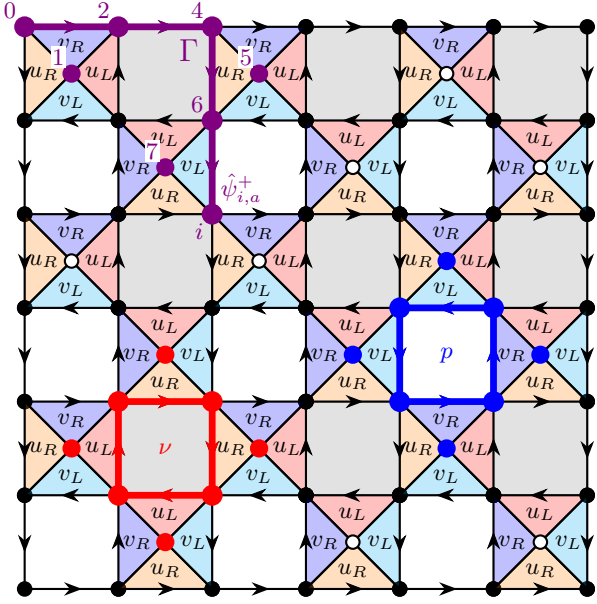
\begin{figure}
	\centering
	%\tikzsetnextfilename{MainLattice}
	\begin{tikzpicture}[baseline={([yshift=.4ex]current bounding box.center)}, scale=0.89]
		\ShadedTCLattice{0}{0}{2}{0.23}
		\OBClattice{0}{0}{2}
		\node  at (9*\hL,-7*\hL) {{\color{blue}$p$}};
		\node  at (3*\hL,-9*\hL) {{\color{red}$\nu$}};
		\ParticlePath{{2/-8,4/-8,4/-10,2/-10,2/-8/no}}{red}
		\IsoDots{{3/-7,5/-9,3/-11,1/-9}}{red}
		\ParticlePath{{8/-6/no,8/-8,10/-8,10/-6,8/-6}}{blue}
		\IsoDots{{9/-5,11/-7,9/-9,7/-7}}{blue}
		\ParticlePath{{0/0,2/0,4/0,4/-2,4/-4}}{parapurple}
		\IsoDots{{1/-1,5/-1,3/-3}}{parapurple}%
		\StringLabels{{0/0/0/above left,2/0/2/above left,4/0/4/above left,4/-2/6/above left,4/-4/i/below left}}{parapurple}%
		\node [fill=white,inner sep =0.5pt,minimum height=8pt] at (1*\hL,-1*\hL) [above left=0.1cm] {\small {\color{parapurple}1}};
		\node [fill=white,inner sep =0.5pt,minimum height=8pt] at (5*\hL,-1*\hL) [above left=0.1cm] {\small {\color{parapurple}5}};
		\node [fill=white,inner sep =0.5pt,minimum height=8pt] at (3*\hL,-3*\hL) [above left=0.1cm] {\small {\color{parapurple}7}};
		\node  at (4*\hL,-4*\hL)[above right] {${\color{parapurple}\hat{\psi}^+_{i,a}}$};
		\node  at (3.5*\hL,-0.5*\hL) {\large ${\color{parapurple}\Gamma}$};
	\end{tikzpicture}
	\caption{\label{fig:mainlattice} The 2D exactly solvable spin model on a $7\times 7$ lattice with open boundary conditions. Each black dot represents a 16 dimensional qudit on which the local operators $\hat{x}^\pm_{i,a},\hat{y}^\pm_{i,a}$ act, and each open circle represents a 64 dimensional auxiliary qudit on which the local operators $\hat{w}^\pm_{ab}$ act, for $w=u_L,u_R,v_L$, or $v_R$. Each colored triangle represents a 3-body interaction between qudits on its 3 vertices. In addition, we have 8-body interactions around every even plaquette~(i.e. the white and gray plaquettes). Eq.~\eqref{eq:JWT_string_2D} gives example of a paraparticle operator $\hat{\psi}_{i,a}^\pm$ defined on the string $\Gamma$~(shown in purple), which is an MPO acting consecutively on all the purple dots.  %
	}
\end{figure}
In the following we present a solvable 2D quantum spin model with emergent free paraparticles, based on the set theoretical $R$-matrix in Eq.~\eqref{eq:seth-R}. Here we only sketch the key definitions and the main results, and the technical details are found in the SI~\cite{Suppl}. 
The model is defined on a square lattice with two types of lattice sites and open boundary conditions, as illustrated in Fig.~\ref{fig:mainlattice}. 
The Hamiltonian consists of two parts $\hat{H}=\hat{H}_1+\hat{H}_2$, where
\begin{eqnarray}\label{def:2DsolvableH}
	\hat{H}_1&=&\sum_{\nu}\hat{A}_\nu+\sum_p \hat{B}_p,\nonumber\\
	\hat{H}_2&=&-\sum_{\langle ij\rangle}\hat{h}_{ij}-\sum_l \mu_l \hat{y}^+_{l,a}\hat{y}^-_{l,a},
\end{eqnarray}
where $\nu,p$ denote the shaded and the white plaquettes, respectively, $l$ runs over all black dots, and $\langle ij\rangle$ runs over all neighboring pairs of black dots~(each pair appears only once).
Here  $\hat{h}_{ij}$ is a 3-body interaction between the vertices of the triangle containing the directed edge $\langle ij\rangle$, defined as
\begin{eqnarray}\label{def:3bodyterm}
	\scalebox{0.8}{%\tikzsetnextfilename{TriangleIllu}
	\begin{tikzpicture}[baseline={([yshift=-.4ex]current bounding box.center)}, scale=.75]
		\TriangleTermIllustration{0*\hL}{-0*\hL}
	\end{tikzpicture}}=
	J_{ij}%\tikzsetnextfilename{TriangleTermMPOU}
	\scalebox{0.8}{\begin{tikzpicture}[baseline={([yshift=.4ex]current bounding box.center)}, scale=.8]
		\xtriangle{0}{0}{-}{i}\umatrix{1}{0}{w^+}{}\ytriangle{2}{0}{+}{j}
		\node at (1,-0.5) [below] {\small $k$};
	\end{tikzpicture}}+\mathrm{h.c.}\equiv \hat{h}_{ij},%
\end{eqnarray}
where $\hat{y}^\pm_{j,a}\equiv\!\!\begin{tikzpicture}[baseline={([yshift=.4ex]current bounding box.center)}, scale=0.7]
	\ytriangle{0}{0}{\pm}{}
	\node  at (0,-1.6*\AL) {\footnotesize $j$};
	\node  at (-1.4*\AL,0) {\footnotesize $a$};
\end{tikzpicture}$,
$\hat{x}^\pm_{j,a}\equiv\!\!\begin{tikzpicture}[baseline={([yshift=.4ex]current bounding box.center)}, scale=0.7]
	\xtriangle{0}{0}{\pm}{}
	\node  at (0,-1.6*\AL) {\footnotesize $j$};
	\node  at (1.4*\AL,0) {\footnotesize $a$};
\end{tikzpicture}$
are the same spin operators that appeared in the 1D model,
 $w$ is one of $u_L,u_R,v_L$, or $v_R$ depending on the type of triangle in the lattice, and $\hat{w}^\pm_{ab}=
	\scalebox{1}{\begin{tikzpicture}[baseline={([yshift=-.8ex]current bounding box.center)}, scale=.8]
	\umatr{0}{0}{w^\pm}{}
	\paraindices{0}{0}{a}{b}
\end{tikzpicture}}
$ is an operator acting on an auxiliary site~(open circles in Fig.~\ref{fig:mainlattice}), for $a,b=1,\ldots,4$. The definition of the tensors $u^\pm_L,u^\pm_R,v^\pm_L,v^\pm_R$ are given in SI~\cite{Suppl}. The operators $\hat{A}_\nu,\hat{B}_p$ in Eq.~\eqref{def:2DsolvableH} are 8-body interaction terms %
defined as
\begin{eqnarray}
	\!\!\!\!\!\!\!\scalebox{0.75}{
		%\tikzsetnextfilename{DiamIlluP}
		\begin{tikzpicture}[baseline={([yshift=.4ex]current bounding box.center)}, scale=.85]
		\DiamTermIllustrationP{0}{0}
		\node at (0,0) {$\hat{B}_p$};
	\end{tikzpicture}}\!\!&=&\scalebox{0.8}{%\tikzsetnextfilename{DiamIlluMPOP}
	\begin{tikzpicture}[baseline={([yshift=.4ex]current bounding box.center)}, scale=.8]
		\umatrix{0}{0}{v_L^+}{1}\Tpmatrix{1}{0}{2}\umatrix{2}{0}{v_L^+}{3}\Tpmatrix{3}{0}{4}
		\umatrix{4}{0}{v_R^+}{5}\Tpmatrix{5}{0}{6}\umatrix{6}{0}{v_R^+}{7}\Tpmatrix{7}{0}{8}
		\PBCcontract{-0.5}{0}{7.5}{1.5}
	\end{tikzpicture}}+\mathrm{h.c.}\nonumber\\
	\!\!\!\!\!\!\!\scalebox{0.75}{%\tikzsetnextfilename{DiamIlluV}
	\begin{tikzpicture}[baseline={([yshift=.4ex]current bounding box.center)}, scale=.85]
		\node at (0,0) {$\hat{A}_\nu$};
		\DiamTermIllustrationV{0*\hL}{-0*\hL}
	\end{tikzpicture}}\!\!&=&\scalebox{0.8}{%\tikzsetnextfilename{DiamIlluMPOV}
	\begin{tikzpicture}[baseline={([yshift=.4ex]current bounding box.center)}, scale=.8]
		\umatrix{0}{0}{u_R^+}{1}\Tpmatrix{1}{0}{2}\umatrix{2}{0}{u_R^+}{3}\Tpmatrix{3}{0}{4}
		\umatrix{4}{0}{u_L^+}{5}\Tpmatrix{5}{0}{6}\umatrix{6}{0}{u_L^+}{7}\Tpmatrix{7}{0}{8}
		\PBCcontract{-0.5}{0}{7.5}{1.5}
	\end{tikzpicture}}+\mathrm{h.c.}\nonumber
\end{eqnarray}
(If a loop term lies on the boundary, then one or more of its white circles will be absent. In this case the tensors $u^\pm_L,u^\pm_R,v^\pm_L,v^\pm_R$ on the absent site is replaced by a $\delta$ tensor, i.e., $\hat{w}^\pm_{ab}=\delta_{ab}$, for $w=u_L,u_R,v_L,v_R$.)

The loop terms $\hat{A}_\nu,\hat{B}_p$ are constructed such that they mutually commute, and commute with each individual 3 body term in $\hat{H}_2$, therefore, they are conserved quantities and eigenstates of $\hat{H}$ can be labeled by their common eigenvalues. In this paper we are mainly interested in the subspace of states in which all $\hat{A}_\nu,\hat{B}_p$ have maximal eigenvalues~(i.e. the space of ground states of $\hat{H}_1$), henceforth referred to as the zero-vortex sector $\Phi_0$. The Hilbert space dimension of this sector is $16^N$ where $N$ is the total number of black dots in the lattice.

To solve the spectrum in the zero-vortex sector, we define paraparticle creation and annihilation operators via a generalized MPO JWT, which generalizes the 1D case given in Eq.~\eqref{eq:JWT_string}. Each paraparticle operator $\hat{\psi}_{i,a}^\pm$ is defined on a string $\Gamma$ connecting the lattice origin to the site $i$, and $\hat{\psi}_{i,a}^\pm$ is an MPO acting consecutively on all the black dots on $\Gamma$~(including the start and end points) and all the open circles adjacent to $\Gamma$, see Fig.~\ref{fig:mainlattice} for an example. The MPO representation of $\hat{\psi}_{i,a}^\pm$ is similar to the 1D case given in Eq.~\eqref{eq:JWT_string}, but now with the tensors $w^\pm$ inserted between neighboring $T^\pm$, where $w$ is one of $u_L,v_L,u_R,v_R$ depending on the type of the triangle between the neighboring black dots. %
For example, for the string $\Gamma$ in Fig.~\ref{fig:mainlattice} that starts at point $0$~(lattice origin) and ends at point $i$,  $\hat{\psi}_{i,a}^\pm$ acts on all the purple dots and is defined as
\begin{eqnarray}\label{eq:JWT_string_2D}
	\hat{\psi}_{i,a}^+%
	&=&	%\tikzsetnextfilename{MPOJWTpsip}
	\begin{tikzpicture}[baseline={([yshift=.4ex]current bounding box.center)}, scale=.74]
		\Tpmatrix{0}{0}{0}\umatrix{1}{0}{v_R^+}{1}\Tpmatrix{2}{0}{2}
		\Tpmatrix{3}{0}{4}\umatrix{4}{0}{u_R^+}{5}\Tpmatrix{5}{0}{6}\umatrix{6}{0}{v_L^+}{7}%
		\ytriangle{7}{0}{+}{i}
		\node at (-0.5,0) [left=-.3mm] {\small $a$};
	\end{tikzpicture},\nonumber\\
	\hat{\psi}_{i,a}^-%
	&=&	%\tikzsetnextfilename{MPOJWTpsim}
	\begin{tikzpicture}[baseline={([yshift=.4ex]current bounding box.center)}, scale=.74]
		\Tmmatrix{0}{0}{0}\umatrix{1}{0}{v_R^-}{1}\Tmmatrix{2}{0}{2}
		\Tmmatrix{3}{0}{4}\umatrix{4}{0}{u_R^-}{5}\Tmmatrix{5}{0}{6}\umatrix{6}{0}{v_L^-}{7}%
		\ytriangle{7}{0}{-}{i}
		\node at (-0.5,0) [left=-.3mm] {\small $a$};
	\end{tikzpicture}.~~~~
\end{eqnarray}
Paraparticle operators $\{\hat{\psi}^\pm_{i,a}\}$ constructed this way have several important properties. First, they commute with all individual terms in $\hat{H}_1$, therefore, their actions leave the zero-vortex sector $\Phi_0$ invariant. Second, as shown in the SI~\cite{Suppl}, although each paraparticle operator $\hat{\psi}^\pm_{i,a}$ is defined on a specific path, their actions in the zero-vortex sector $\Phi_0$ do not depend on the choice of the path, only on the endpoints. This is due to the special topological property of the zero vortex sector, and is reminiscent of the path-independence of the action of the string operators on the toric code ground states~\cite{kitaev2003fault}. Finally, in the zero vortex sector, the operators $\{\hat{\psi}^\pm_{i,a}\}$ satisfy the parastatistical CRs in Eq.~\ref{eq:fundamental_Rcommu}, justifying their names ``paraparticle operators''. These properties lead us to~\cite{Suppl} %
\begin{theorem}\label{thm:main_2DESM}
	In the zero-vortex sector, %
	$\hat{H}_2$ is mapped to the free paraparticle Hamiltonian 
	\begin{equation}\label{eq:freeparaH2}
		\hat{H}_2 =-\sum_{\langle ij\rangle, 1\leq a\leq m}  (J_{ij}\hat{\psi}^+_{j,a}\hat{\psi}^-_{i,a}+\mathrm{h.c.})-\sum_{l} \mu_l \hat{n}_l.
	\end{equation}
\end{theorem}
We expect that our 2D solvable spin models exhibit new chiral and gapless topological phases that are not exhibited by previous solvable models. 
To date, the only family of solvable models for chiral topological order in 2D is Kitaev's honeycomb model~\cite{kitaev2006anyons} and its generalizations~\cite{YaoKivelson,Chapman2020characterizationof,Wang2022IK}, whose gapped phases are classified by the 16-fold way~\cite{kitaev2006anyons}, depending on the Chern number $(\nu \mod 16)$ of the free fermion band. We expect that the gapped phases of our model are similarly classified by the Chern number of the free paraparticle band. When $\nu=0$, both Kitaev's honeycomb model and our models are in non-chiral quantum double phases, but the former only hosts $\mathbb{Z}_2$ Abelian anyons, while the latter host non-Abelian anyons already at $\nu=0$. We expect that our models host different chiral topological phases also at nonzero $\nu$, and different gapless topological phases when the free paraparticles have a gapless spectrum.

\paragraph{Particle exchange statistics in the 2D solvable model}
\begin{figure}
	\center{\includegraphics[width=0.65\linewidth]{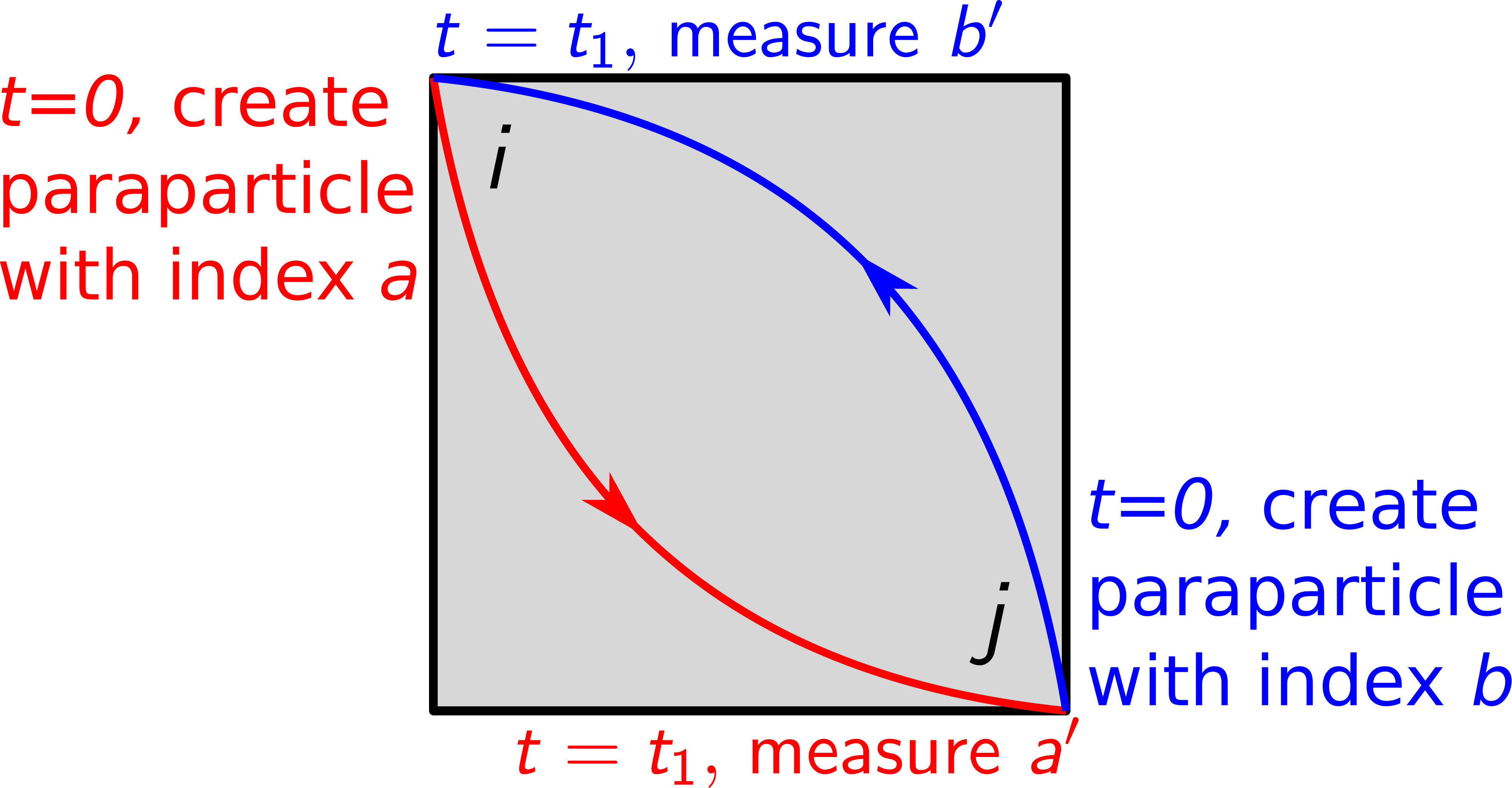} 
	}
	\caption{\label{fig:braiding2Dmodel} Illustration of paraparticle exchange in the 2D solvable spin model. The shaded square represents the 2D system with OBC as shown in Fig.~\ref{fig:mainlattice}. $i$ and $j$ label the black site in the upper left and lower right corners of the 2D lattice, respectively, where paraparticles can be locally created and measured.  The unitary exchange operator $\hat{E}_{ij}$ moves the paraparticles along the two colored paths, and the result of the exchange is given in Eq.~\eqref{eq:Uij_action}. }%
\end{figure}
We now illustrate the exchange statistics of the emergent paraparticles in the 2D solvable spin model, which reveals a striking physical difference between the  emergent paraparticles and ordinary fermions and bosons. 
 
Consider the paraparticle exchange process described in Fig.~\ref{fig:braiding2Dmodel}.
For simplicity, we consider the case when $-\mu_l$ is large, so that the ground state $\ket{G}$ of the 2D system has no paraparticles, i.e. $\hat{n}_l\ket{G}=0,~\forall l$.  
At $t=0$, we can apply local unitary operators %
on the ground state $\ket{G}$ to create a paraparticle at site $i$ and $j$, respectively, and obtain the state $|G;ia,jb\rangle\equiv \hat{\psi}^+_{i,a}\hat{\psi}^+_{j,b}\ket{G}$~\cite{Suppl}. %
Then we evolve the state $|G;ia,jb\rangle$ with $\hat{E}_{ij}$ that moves the paraparticles along the colored paths shown in Fig.~\ref{fig:braiding2Dmodel}~[$\hat{E}_{ij}$ can be constructed from a product of local unitaries of the form $e^{i\frac{\pi}{2} (\hat{e}_{kl}+\hat{e}_{lk})}$, where $\hat{e}_{kl}$ is mapped to a local 3-body interaction in the 2D model]. 
The result of this unitary exchange process is given by Eq.~\eqref{eq:braiding_derivation}, where $\ket{0}$ is understood as the ground state $\ket{G}$.  %
With the set-theoretical $R$-matrix in Eqs.~(\ref{eq:seth-R},\ref{eq:set-thR}), the final state is $-\ket{G;ib',ja'}$, where $(b',a')=r(a,b)$, and the labels $a',b'$ can be locally measured at the two corners~\cite{Suppl}. For example, if we start with $a=b=1$, we end up measuring $b'=4,a'=3$. That is, the auxiliary space of the paraparticles undergoes a non-trivial unitary rotation even though the two particles stay arbitrarily far apart with each other throughout the whole process. %
This is in stark contrast with fermions and bosons, in which case we would measure $a'=a$ and $b'=b$, i.e. the indices are simply carried with the particles without any change. 

In principle, the exchange process described above can also be done in the 1D spin model in Eq.~\eqref{eq:Hamil1Dspin}.  %
In this case, the paraparticles can also be created and measured at the two ends of the open chain, Eq.~\eqref{eq:braiding_derivation} still holds and the measurement result is the same. The major difference from the 2D case is that in 1D, the two paraparticles inevitably collide during the exchange, and the exchange statistics results from the interaction between the two paraparticles, which is sensitive to the microscopic details of the exchange operator $\hat{E}_{ij}$, and is not robust against local perturbations. By contrast, in 2D, the paraparticles can stay far away from each other throughout the exchange, and their exchange statistics has a topological nature independent of the detailed shape of the space-time trajectory of the particles, and is robust against all local perturbations when the particles are far away from the boundaries and from each other.

%	\title{Supplementary Information for\\ ``Particle exchange statistics beyond fermions and bosons''}
%\author{Zhiyuan Wang}
%\affiliation{Department of Physics and Astronomy, Rice University, Houston, Texas 77005,
%	USA}
%\affiliation{Rice Center for Quantum Materials, Rice University, Houston, Texas 77005, USA}
%\affiliation{Max-Planck-Institut f{\"{u}}r Quantenoptik, Hans-Kopfermann-Str. 1, 85748 Garching, Germany}
%\author{Kaden R.~A. Hazzard}
%\affiliation{Department of Physics and Astronomy, Rice University, Houston, Texas 77005,
%	USA}
%\affiliation{Rice Center for Quantum Materials, Rice University, Houston, Texas 77005, USA}
%\date{\today}
%
%\maketitle

\clearpage 

\begin{center}
	\textbf{\large Supplementary Information for ``Particle exchange statistics beyond fermions and bosons''}
\end{center}
\makeatletter
\def\renewtheorem#1{%
	\expandafter\let\csname#1\endcsname\relax
	\expandafter\let\csname c@#1\endcsname\relax
	\gdef\renewtheorem@envname{#1}
	\renewtheorem@secpar
}
\def\renewtheorem@secpar{\@ifnextchar[{\renewtheorem@numberedlike}{\renewtheorem@nonumberedlike}}
\def\renewtheorem@numberedlike[#1]#2{\newtheorem{\renewtheorem@envname}[#1]{#2}}
\def\renewtheorem@nonumberedlike#1{  
	\def\renewtheorem@caption{#1}
	\edef\renewtheorem@nowithin{\noexpand\newtheorem{\renewtheorem@envname}{\renewtheorem@caption}}
	\renewtheorem@thirdpar
}
\def\renewtheorem@thirdpar{\@ifnextchar[{\renewtheorem@within}{\renewtheorem@nowithin}}
\def\renewtheorem@within[#1]{\renewtheorem@nowithin[#1]}
\makeatother

\setcounter{equation}{0}
\setcounter{figure}{0}
\setcounter{table}{0}
\setcounter{page}{1}
\setcounter{section}{0}

\setcounter{secnumdepth}{1}
\setcounter{secnumdepth}{2}
\setcounter{secnumdepth}{3}
\makeatletter
\renewtheorem{theorem}{Theorem}[section]
\renewcommand{\theequation}{S\arabic{equation}}
\renewcommand{\thefigure}{S\arabic{figure}}

\renewcommand{\bibnumfmt}[1]{[S#1]}
\renewcommand{\thetable}{S\arabic{table}}
\renewcommand{\thesection}{S\arabic{section}}

This Supplementary Information fills in several technical details omitted in the main text. 
In Sec.~\ref{SI:relation_to_others} we explain the difference between parastatistics and other known types of particle statistics, and the difference between our theory of parastatistics and Green's theory~\cite{Green1952}.
In Sec.~\ref{SI:state_space} we present the detailed construction of the state space and derive the generalized exclusion statistics for all the $R$-matrices that appeared in the main text, and prove that our second quantization theory is well-defined for an arbitrary $R$-matrix, including non-unitary ones. %
In Sec.~\ref{SI:relation_first_quantization} we show the relation between the second quantization formulation of parastatistics and the  wavefunction formulation 
we discussed in the introduction part of the main text.
In Sec.~\ref{SI:1Dspin_model} and Sec.~\ref{SI:2DparaKDH} we present the detailed construction of the 1D solvable spin model defined in Eq.~(\ref{eq:Hamil1Dspin}) %
and the 2D solvable spin model defined in Eq.~\eqref{def:2DsolvableH}, respectively, including the definition of local operators $\{\hat{x}^\pm_{i,a},\hat{y}^\pm_{i,a}\}^m_{a=1}$ and the tensors $u_L,v_L,u_R,v_R$, and we prove that the MPO JWT defined in Eq.~\eqref{eq:JWT_string} and Eq.~\eqref{eq:JWT_string_2D} map the spin Hamiltonians into free paraparticle Hamiltonians.  
Finally in Sec.~\ref{SI:superselection} we discuss superselection rules and the observability of elementary paraparticles.
\section{Difference between parastatistics and other types of particle statistics}\label{SI:relation_to_others}
In the following, we briefly explain the difference between parastatistics and other known types of particle statistics, including non-Abelian anyons~\cite{Nayak2008NAAnyons}, parafermions~\cite{Fendley2014Freepara}, and other types of exclusion statistics. We also explain the difference between our theory of parastatistics and Green's theory~\cite{Green1952}.

For non-Abelian anyons in 2D, braiding two anyons also results in a matrix rotation on the internal space~(more precisely, the topologically protected degenerate fusion space) of a system with multiple anyonic excitations, similar to Eq.~\eqref{eq:wavefuntion_exchange_para}. However, for non-Abelian anyons, the matrices $\{R_{j}\}^{n-1}_{j=1}$ only satisfy the second equation in Eq.~\eqref{eq:SNbraidrelations}, but not the first one, and this is the reason why anyons cannot be consistently defined in 3+1D, where exchanging two identical particles twice should give back the original state.  
In 2D, parastatistics can be considered as a special case of non-Abelian statistics satisfying $R_{j}^2=\mathds{1}$; however, to the best of our knowledge, this special case was previously believed to contain only fermions and bosons. %
Another key difference between anyons and paraparticles is that there does not exist an exactly solvable free particle theory for genuine anyons~(meaning that $R_{j}^2\neq\mathds{1}$). By a ``solvable free particle theory of anyons'', we mean a many-body Hamiltonian describing a system of non-interacting anyons freely moving in space, similar to Eq.~\eqref{eq:u1_sym_H}, such that one can obtain the exact many body spectrum by solving the one particle spectrum. %

Parafermions~\cite{Fendley2014Freepara,Fendley2016review} are exotic emergent excitations that appear in 1D $\mathbb{Z}_n$-symmetric quantum spin chains and the 1D boundary of certain 2D topological phases, such as quantum Hall/superconductor hybrids~\cite{Fendley2016review}. They are defined by the following algebraic relations between parafermion operators that generalize the Clifford algebra of Majorana fermions:
\begin{eqnarray}\label{eq:def_parafermions}
	\hat{\psi}_{i}\hat{\psi}_{j}&=&\omega\hat{\psi}_{j}\hat{\psi}_{i},\text{ for } i<j,\nonumber\\
	\hat{\psi}_{i}^n&=&1, ~\forall i,
\end{eqnarray}
where $n\geq 2$ is an integer and $\omega$ is a primitive $n$-th root of unity. 
As we can see, despite the similarity in the names, parafermions are defined by a very different generalization of the second quantization of fermions compared to our generalization in Eq.~\eqref{eq:fundamental_Rcommu}, and the requirement $i<j$ in Eq.~\eqref{eq:def_parafermions} can only be consistently specified in 1D, so parafermions are intrinsically limited to 1D. Interestingly, for parafermions in 1D, there also exists an exactly solvable free particle theory ~\cite{Fendley2014Freepara}, where one can exactly obtain the full many-body spectrum by solving the single particle spectrum. However, the solvable Hamiltonians for free parafermions are always non-Hermitian and have complex energy spectrum. 

Green's theory of parastatistics~\cite{Green1952,Araki1961,Greenberg1965,LANDSHOFF196772,druhl1970parastatistics,Taylor1970b} is defined by a set of trilinear CRs between paraparticle creation and annihilation operators 
\begin{equation}\label{eq:CRGreen}
	\begin{aligned}
		& {\left[\left[\hat{\psi}^{\dagger}_k, \hat{\psi}_l\right]_{ \pm}, \hat{\psi}_m\right]_-=-2 \delta_{k m} \hat{\psi}_l,} \\
		& {\left[\left[\hat{\psi}_k, \hat{\psi}_l\right]_{ \pm}, \hat{\psi}_m\right]_-=0,}
	\end{aligned}
\end{equation}
where $[\hat{A},\hat{B}]_\pm=\hat{A}\hat{B}\pm \hat{B}\hat{A}$. This theory of parastatistics is also consistently defined in any spatial dimension, and
it was shown in Ref.~\cite{Taylor1970b} that the exchange statistics of paraparticles in this theory also realize higher dimensional representations of the symmetric group $S_N$, in the sense of Eqs.~(\ref{eq:wavefuntion_exchange_para},\ref{eq:SNbraidrelations}). %
Green's theory also includes exactly solvable theories of free paraparticles, but the exclusion statistics of Green's paraparticles is much harder to compute and is still not fully solved to date~\cite{Stoilova2020}. In particular, unlike in our formulation, in Green's theory, the grand partition function of the whole system does not factorize as a product of single mode partition functions~\cite{Stoilova2020}, making it challenging to compute its thermodynamic properties. Furthermore, it is not known whether the paraparticles defined by Green's theory can appear as emergent quasiparticles in condensed matter systems, in a way distinct from fermions and bosons. 

\begin{table*}[t]
	\centering
	\begin{tabular}{|c|c|c|c|c|}
		\hline
		Particle statistics & Non-Abelian anyons~\cite{Nayak2008NAAnyons}	&  Parafermions~\cite{Fendley2014Freepara,Fendley2016review} & Green's parastatistics~\cite{Green1952} & $R$-matrix parastatistics  \\
		\hline
		Definition & Braided fusion category & Eq.~\eqref{eq:def_parafermions} & Eq.~\eqref{eq:CRGreen} & Eq.~\eqref{eq:fundamental_Rcommu}\\
		\hline
		Braid group &$R_j^2\neq \mathds{1},~B_n$  & - & $R_j^2= \mathds{1},~S_n$ & $R_j^2= \mathds{1},~S_n$ \\ 
		\hline
		Spatial dimension & $d=2$ & $d=1$ & Any $d$ & Any $d$ \\
		\hline
		Free particle theory  & No  & Non-Hermitian & Hard to compute thermodynamics &  Yes\\ %
		\hline
		Emergent & Yes & Yes & Unknown & Yes \\
		\hline
	\end{tabular}
	\caption{\label{tab:comparison_statistics} Comparison between different types of particle statistics. The $R$-matrix parastatistics refers to the one we introduced in  the main text. The last row indicates whether a given type of particle statistics is known to emerge in locally interacting spin models. }
\end{table*}

In Tab.~\ref{tab:comparison_statistics} we summarize the comparison between the above three types of particle statistics and the $R$-matrix parastatistics we introduced in the main text. 
A caveat about terminology is that one should be careful about the use of ``paraboson'' and ``parafermion''  in the recent literature -- some of them~\cite{Aneva2005deformedGreen,Kanakoglou2007braidedGreen,alderete2017Greenexperiment,alderete2021experimental,Toppan2021Z2Z2,Toppan2021Inequivalent} refer to Green's parastatistics~\cite{Green1952} defined by Eq.~\eqref{eq:CRGreen}, some refer to parafermions~\cite{Fendley2014Freepara,Fendley2016review} defined by Eq.~\eqref{eq:def_parafermions}, and some refer to the $q$-deformed bosons and fermions~\cite{Biedenharn1989qboson}. The last one is limited to 1D, defined by $q$-deformed second quantization algebra of bosons and fermions,  and does not contain free particle theories. 

Furthermore, some generalizations of the Pauli exclusion principle have been proposed in the first quantization formulation. Ref.~\cite{Gentile1940} studied the thermodynamics of an ideal gas composed of particles with exclusion statistics $d_l=1$ for  $0\leq l\leq m$, and $d_l=0$ for $l>m$. Ref.~\cite{Haldane1991} proposed a definition of exclusion statistics that is well-defined in any spatial dimension and generalizes Pauli's principle, which applies to many interesting physical systems such as quasiparticles of the fractional quantum Hall effect~\cite{Laughlin1983}, spinons in gapless spin-$1/2$ antiferromagnetic chains~\cite{Haldane1991}, and quasiparticles in conformal field theory spectra~\cite{Schoutens1997CFT,Bouwknegt1999CFT}.  The generalization of Pauli's principle in Ref.~\cite{Haldane1991} is not compatible with second quantization beyond the simple case of fermions and bosons~\cite{Haldane1991}. In particular, the state counting formula in Eq.~(3) of Ref.~\cite{Haldane1991} does not apply to paraparticles introduced in this paper.

\section{The state space and exclusion statistics for a general $R$-matrix}\label{SI:state_space}
In the main text we derived the generalized exclusion statistics for the $R$-matrix in Ex.~\ref{ex:1m} in Tab.~\ref{tab:Hilbert_series}, and constructed an orthonormal basis for the many particle state space in Eq.~\eqref{eq:state_space_Ex3}. In this section we generalize these results to an arbitrary $R$-matrix. Specifically, in Sec.~\ref{SI:basis_state_space}
we construct a basis for the state space for an arbitrary $R$-matrix, which is orthonormal if $R$ is unitary, in Sec.~\ref{SI:action_parafield_basis} we define the action of the paraparticle creation and annihilation operator $\hat{\psi}^\pm_{i,a}$ on these basis states, and in Sec.~\ref{SI:exclusion_statistics_calc} we calculate exclusion statistics for other $R$-matrices we studied in the main text. In Sec.~\ref{SI:state_space_math} we construct the state space in a rigorous mathematical framework, and in particular, in Sec.~\ref{SI:proof_hermiticity} we show that even with a non-unitary $R$-matrix, we can still define a Hermitian inner product on the state space, with respect to which all physical observables are Hermitian.

\subsection{A basis for the state space}\label{SI:basis_state_space}
Before diving into details, let us first summarize the basic idea behind the construction of the state space in a nutshell~(this idea is formalized in a rigorous mathematical framework in Sec.~\ref{SI:state_space_math}). 
Analogous to the Fock space of fermions and bosons, there is a vacuum state $|0\rangle$ satisfying $\hat{\psi}^-_{i,a}\ket{0}=0$, and we always assume it is normalized as $\braket{0|0}=1$. The many-particle state space is spanned by all states of the form  $|\psi\rangle=\hat{\psi}^+_{i_1,a_1}\hat{\psi}^+_{i_2,a_2}\ldots \hat{\psi}^+_{i_n,a_n}|0\rangle$. 
(Note that these are not all linearly independent.) 
The second relation in Eq.~\eqref{eq:fundamental_Rcommu} imposes linear dependence relations on these states, which gives rise to the generalized exclusion statistics. 
The action of creation operators on this set of states is described straightforwardly by 
$\hat{\psi}^+_{i,a}|\psi\rangle=\hat{\psi}^+_{i,a}\hat{\psi}^+_{i_1,a_1}\ldots \hat{\psi}^+_{i_n,a_n}|0\rangle$. The action of annihilation operators $\hat{\psi}^-_{i,a}$ is uniquely determined by the first relation in Eq.~\eqref{eq:fundamental_Rcommu}, 
which allows us to move $\hat{\psi}^-_{i,a}$ all the way to the right until it hits $|0\rangle$~(which it annihilates). 

However, to actually compute the exclusion statistics and to work out an explicit matrix representation of $\hat{\psi}^\pm_{i,a}$, we need to first construct a basis for the state space, in a way that generalizes Eq.~\eqref{eq:state_space_Ex3} in the main text to an arbitrary $R$-matrix. 
This is what we do in the following.

Let $\{\Psi^\alpha_{a_1a_2\ldots a_n}\}_{\alpha=1}^{d_n}$ be a complete set of linearly independent solutions to the system of linear equations 
\begin{equation}\label{eq:V_n_basis}
	\sum_{a'_{j},a'_{j+1}}R^{a_ja_{j+1}}_{a'_ja'_{j+1}}\Psi_{a_1\ldots a'_ja'_{j+1}\ldots a_n}=\Psi_{a_1\ldots a_ja_{j+1}\ldots a_n}
\end{equation} 
for $j=1,2,\ldots,n-1$. Intuitively, Eq.~\eqref{eq:V_n_basis} requires that $\Psi_{a_1\ldots a_n}$ is an $R$-symmetric function, which in the case of fermions or bosons~($R=\pm 1$) reduces to totally symmetric or antisymmetric functions. 
A basis for the state space is constructed as the set of states of the form
\begin{equation}\label{eq:full_basis}
	|{}^{\alpha_1}_{n_1},{}_{n_2}^{\alpha_2},\ldots, {}_{n_N}^{\alpha_N}\rangle=\hat{\Psi}^{(1)+}_{n_1,\alpha_1}\hat{\Psi}^{(2)+}_{n_2,\alpha_2}\ldots\hat{\Psi}^{(N)+}_{n_N,\alpha_N}|0\rangle,%
\end{equation}
where the numbers $\{(n_i,\alpha_i)\}^N_{i=1}$ can be chosen independently for different modes~(with the only constraint being $1\leq \alpha_i\leq d_{n_i}$ for each $i$), and the  operator
\begin{equation}\label{eq:single_mode_creation}
	\hat{\Psi}^{(i)+}_{n,\alpha}\equiv \frac{1}{\sqrt{n!}}\sum_{a_1 a_2\ldots a_n}\Psi^\alpha_{a_1 a_2\ldots a_n}\hat{\psi}^+_{i,a_1}\hat{\psi}^+_{i,a_2}\ldots\hat{\psi}^+_{i,a_n}, %
\end{equation}
creates a multiparticle state in the mode $i$ with occupation number $n$~(eigenvalue of  $\hat{n}_i\equiv \hat{e}_{ii}$). 

When the $R$-matrix is unitary, we can consistently set $\hat{\psi}^+_{i,a}=(\hat{\psi}^-_{i,a})^\dagger$~\footnote{First, notice that $\hat{\psi}^+_{i,a}=(\hat{\psi}^-_{i,a})^\dagger$ is fully consistent with the fundamental CRs in Eq.~\eqref{eq:fundamental_Rcommu}, as taking the Hermitian conjugate on both sides leaves Eq.~\eqref{eq:fundamental_Rcommu} invariant~(maps the first line to itself and swaps the second and the third lines). Then it can be checked straightforwardly that the explicit matrix representation of $\hat{\psi}^\pm_{i,a}$ defined in Sec.~\ref{SI:action_parafield_basis}  indeed satisfies $\hat{\psi}^+_{i,a}=(\hat{\psi}^-_{i,a})^\dagger$. By contrast, if $R$ is not unitary, the relation $\hat{\psi}^+_{i,a}=(\hat{\psi}^-_{i,a})^\dagger$ is not consistent with  Eq.~\eqref{eq:fundamental_Rcommu} as taking the Hermitian conjugate on both sides leads to extra relations that result in an algebraic inconsistency, and the representation of $\hat{\psi}^\pm_{i,a}$ constructed in Sec.~\ref{SI:action_parafield_basis}  does not satisfy $\hat{\psi}^+_{i,a}=(\hat{\psi}^-_{i,a})^\dagger$. However, as we show in Sec.~\ref{SI:proof_hermiticity},  even for a non-unitary $R$, we can still define a Hermitian inner product on the Fock space such that $\hat{e}^\dagger_{ij}=\hat{e}_{ji}$ is  satisfied, and consequently all physical observables are Hermitian with respect to this inner product.
} and in this case, we require that the coefficients
$\{\Psi^\alpha_{a_1a_2\ldots a_n}\}_{\alpha=1}^{d_n}$ are normalized as 
\begin{equation}\label{eq:singlemodewf_normalization}
	\sum_{a_1,a_2,\ldots, a_n}\Psi^{\beta*}_{a_1a_2\ldots a_n}\Psi^\alpha_{a_1a_2\ldots a_n}=\delta_{\alpha\beta}.
\end{equation}
Then the basis in Eq.~\eqref{eq:full_basis} is orthonormal, i.e.  
\begin{equation}\label{eq:full_basis_orthonormal}
	\Braket{{}^{\beta_1}_{n^{\prime}_1},{}_{n^{\prime}_2}^{\beta_2},\ldots, {}_{n^{\prime}_N}^{\beta_N}|{}^{\alpha_1}_{n^{\vphantom{\prime}}_1},{}_{n^{\vphantom{\prime}}_2}^{\alpha_2},\ldots, {}_{n^{\vphantom{\prime}}_N}^{\alpha_N}}=\prod^N_{j=1}\delta_{n^{\vphantom{\prime}}_jn^{\prime}_j}\delta_{\alpha_j\beta_j},%
\end{equation}
see Sec.~\ref{SI:action_parafield_basis} for a proof. 

For non-unitary $R$-matrices, we show in Sec.~\ref{SI:state_space_math} that the basis states in Eq.~\eqref{eq:full_basis} are still linearly independent, and we define an alternative Hermitian inner product on the state space~[which is generally different from Eq.~\eqref{eq:full_basis_orthonormal}], with respect to which all physical observables are still Hermitian.

\subsection{Action of $\hat{\psi}^\pm_{i,a}$ on the basis states}\label{SI:action_parafield_basis}
The actions of $\{\hat{\psi}^\pm_{i,b}|1\leq b\leq m,1\leq i\leq N\}$ on the basis states in Eq.~\eqref{eq:full_basis} are uniquely determined by the fundamental CRs in Eq.~\eqref{eq:fundamental_Rcommu}:
the action of an annihilation operator $\hat{\psi}^-_{i,a}$ is determined by using the first relation in Eq.~\eqref{eq:fundamental_Rcommu} to move $\hat{\psi}^-_{i,a}$ all the way to the right until it hits $|0\rangle$, and the action of a creation operator $\hat{\psi}^+_{i,a}$ is determined by using the second relation in Eq.~\eqref{eq:fundamental_Rcommu} to move it to the right, passing all the $\hat{\psi}^+_{j,b}$ for $j<i$, and then combine it with $\hat{\Psi}^+_{n_i,\alpha}$. 

It turns out that the resulting matrix representation of $\{\hat{\psi}^\pm_{i,b}\}$ obtained this way is equivalent to using the MPO JWT in Eq.~\eqref{eq:JWT_string} to represent $\{\hat{\psi}^\pm_{i,b}\}$ as MPO string operators acting on a 1D spin chain, where the local operators $\hat{y}^\pm_{ja}$ and $\hat{T}^\pm_{j,ab}$ are explicitly defined in Sec.~\ref{SI:spin_model_def}. This is reminiscent of the familiar fact in the second quantization of fermions that the action of fermionic operators on the particle number basis involves a fermion minus sign, and implementing such a fermion minus sign computationally is equivalent to doing a JWT. For paraparticles, the fermion minus sign becomes the paraparticle $R$-matrix, and the JW string becomes an MPO string of $R$~[or more precisely the tensor $T^\pm$, which is constructed out of $R$ and $\{\Psi^\alpha_{a_1a_2\ldots a_n}\}_{\alpha=1}^{d_n}$ in Eq.~\eqref{eq:Tpm_matrix_element}].

The MPO representation of $\{\hat{\psi}^\pm_{i,b}\}$ allows us to prove the orthonormality condition in Eq.~\eqref{eq:full_basis_orthonormal}:  under the MPO JWT, the basis state $\ket{{}^{\alpha_1}_{n_1},{}_{n_2}^{\alpha_2},\ldots, {}_{n_N}^{\alpha_N}}$ of the Fock space defined in Eq.~\eqref{eq:full_basis} is mapped to the product state $\ket{n_1,\alpha_1}\otimes \ket{n_2,\alpha_2}\otimes\ldots\otimes\ket{n_N,\alpha_N}$ of the 1D spin chain, and the latter is orthonormal since the local basis $\ket{n,\alpha}$ of the on-site Hilbert space is shown to be orthonormal in Sec.~\ref{SI:spin_model_def}.

\subsection{Calculation of exclusion statistics and single mode partition functions}\label{SI:exclusion_statistics_calc}
\subsubsection{$R$-matrices in Tab.~\ref{tab:comparison_statistics}}
We here present the calculation of the numbers  $\{d_n\}_{n\geq 0}$ for the $R$-matrices in Exs.~\ref{ex:decoupled}-\ref{ex:1m1}. To this end, we need to solve Eq.~\eqref{eq:V_n_basis} for each $R$-matrix and for each particle number $n$. Note that Eq.~\eqref{eq:V_n_basis} does not put any restriction on $\Psi$ for $n=0$ and $n=1$, so we have $d_0=1, d_1=m$ for all the four families of $R$-matrices. The physical meaning of this is clear: we always have one vacuum state $|0\rangle$, and $m$ degenerate single particle states. For the $R$-matrix in Ex.~\ref{ex:decoupled}, Eq.~\eqref{eq:V_n_basis} sets the requirement that $\Psi_{a_1a_2\ldots a_n}$ is antisymmetric under the exchange of any two neighboring indices, e.g., $\Psi_{a_1a_2\ldots a_n}=-\Psi_{a_2a_1\ldots a_n}$. Therefore for each $n$, $\Psi$ has $\binom{m}{n}$ independent components, which can be chosen to be $\{\Psi_{a_1a_2\ldots a_n}~|~1\leq a_1<a_2<\ldots <a_n\leq m\}$, therefore, $d_n=\binom{m}{n}$ for $0\leq n\leq m$ and $d_n=0$ for $n>m$. For the $R$-matrix in Ex.~\ref{ex:Green}, Eq.~\eqref{eq:V_n_basis} still relates an arbitrary component $\Psi_{a_1a_2\ldots a_n}$ to an element in $\{\Psi_{a_1a_2\ldots a_n}~|~1\leq a_1<a_2<\ldots <a_n\leq m\}$, although potentially with a different sign factor, and we still have $\Psi_{a_1a_2\ldots a_n}=0$ if any two indices are equal. This leads to the same $d_n$ as in Ex.~\ref{ex:decoupled}. For the $R$-matrix in Ex.~\ref{ex:1m}, Eq.~\eqref{eq:V_n_basis} becomes $\Psi_{a_1a_2\ldots a_n}=-\Psi_{a_1a_2\ldots a_n}$, leading to $\Psi=0$ and therefore $d_n=0$ for any $n\geq 2$.  For the $R$-matrix in Ex.~\ref{ex:1m1}, Eq.~\eqref{eq:V_n_basis} with $n=2$ gives $\lambda_{ab}\sum_{c,d}\cc_{cd}\Psi_{cd}=2\Psi_{ab}$, and since $\mathrm{Tr}[\lambda \cc^T]=2$, this equation has a unique solution $\Psi_{ab}=\lambda_{ab}$~(up to a constant factor), therefore $d_2=1$. Moreover, Eq.~\eqref{eq:V_n_basis} with $n=3$ implies $\Psi_{abc}=\lambda_{ab}\phi_c=\phi'_a\lambda_{bc}$ for some vectors $\phi,\phi'$, which has no nonzero solution since $\lambda$ is invertible, leading to $d_n=0$ for $n\geq 3$~(the case for $n>3$ is proved by applying this argument to the first 3 indices of $\Psi_{a_1a_2\ldots a_n}$). 

The single mode partition function $z_R(x)$ can be calculated directly from the definition in Eq.~\eqref{eq:single_mode_Z}, the results are given in Tab.~\ref{tab:Hilbert_series}. 
In the mathematics literature $z_R(x)$~(where $x=e^{-\beta\epsilon}$) is called the \textit{Hilbert series} of the $R$-matrix~\cite{polishchuk2005quadratic}. There is a very useful identity relating the Hilbert series of the $R$-matrices $R$ and $-R$~(note that $-R$ also satisfies the YBE in Eq.~\eqref{eq:YBE} if $R$ does): $z_{R}(-x)z_{-R}(x)=1$, which allows us to compute the exclusion statistics $\{d_n\}_{n\geq 0}$ of $-R$ if the exclusion statistics of $R$ is known. For example, for the $R$-matrix in Ex.~\ref{ex:1m1}, we have $z_{-R}(x)=1/(1-mx+x^2)$, from which we obtain $d_0=1, d_1=m, d_2=m^2-1$, and $d_{n+1}=m d_n-d_{n-1}$ for $n\geq 1$. 

\subsubsection{The set-theoretical $R$-matrix in Eq.~\eqref{eq:seth-R}}\label{SI:set-thR}
The single mode partition function $z_R(x)=(1+x)^4$ of the set-theoretical $R$-matrix defined in Eqs.~(\ref{eq:seth-R},\ref{eq:set-thR}) can be proved either by directly solving Eq.~\eqref{eq:V_n_basis}, or by using the following fact  
\begin{fact}{(Proposition 1.7 in Ref.~\cite{etingof1999set})}
	Consider $R$ as a quantum gate acting on two neighboring qudits~($d=m=4$), and consider an arbitrary quantum circuit generated by the gates $R_{12}, R_{23},\ldots,R_{n-1,n}$ acting on a system of $n$ qudits. Then there exists a unitary transformation $\hat{U}$ that simultaneously transforms $R_{12}, R_{23},\ldots,R_{n-1,n}$ into the trivial swap gates, i.e. 
	\begin{equation}\label{eq:DTwistGlobalUnitary}
		U R_{j,j+1} U^\dagger=-X_{j,j+1},~j=1,2,\ldots,n-1,
	\end{equation}
	where $X$ is the two qudit swap gate. Such a unitary transformation can be constructed from Eq.~(1.16) in Ref.~\cite{etingof1999set}
	\begin{equation}
		U \ket{x_1, \ldots, x_n}=\ket{f_{x_n} f_{x_{n-1} \ldots f_{x_2}}\left(x_1\right), \ldots, f_{x_n}\left(x_{n-1}\right), x_n},
	\end{equation}
	where $x_1,x_2,\ldots,x_n\in \{1,2,3,4\}$ label basis states of the $n$ qudits, and $f_y(x)$ is the second component of $r(x,y)$. %
	Therefore we have $z_R(x)=z_{-X}(x)=(1+x)^4$.
\end{fact}

\subsection{Mathematical details on the structure of the state space}\label{SI:state_space_math}
The goal of this section is to provide a rigorous mathematical framework for our second quantization formulation of parastatistics, and in the process we prove some technical claims we made in the previous sections and in the main text. It can be skipped by most readers without impacting the understanding of physics.

In this section we construct the state space and define the action of $\hat{\psi}^\pm_{i,a}$ within an alternative, rigorous mathematical framework that analyzes the structure of the second quantization algebra~[defined by Eq.~\eqref{eq:fundamental_Rcommu} of the main text] and its representations. 
This framework proves several technical claims we made in the previous sections and the main text. Specifically, Secs.~\ref{SI:existence_vacuum}-\ref{SI:full_state_space} establishes that
(1) the action of $\hat{\psi}^\pm_{i,b}$ described in Sec.~\ref{SI:action_parafield_basis} is an irreducible representation of the second quantization algebra~\eqref{eq:fundamental_Rcommu}, and it is the unique irreducible representation subject to some physical constraints; %
(2) the basis states constructed in Eq.~\eqref{eq:full_basis} are linearly independent even for non-unitary $R$-matrices. Furthermore, in Sec.~\ref{SI:proof_hermiticity} we show that even with a non-unitary $R$-matrix, we can still define a Hermitian inner product on the state space, with respect to which all physical observables are Hermitian.

\subsubsection{Notations and definitions}
We begin by introducing some notations. Denote by $\mathcal{X}_{R,N}$ the unital associative algebra over $\mathbb{C}$ generated by  $\{\hat{\psi}^\pm_{i,b}|1\leq i \leq N, 1\leq b\leq m\}$ modulo all the relations in Eq.~\eqref{eq:fundamental_Rcommu}. Define $\mathcal{X}^+_{R,N}$ as the~(unital) subalgebra of $\mathcal{X}_{R,N}$ generated by all the creation operators $\{\hat{\psi}^+_{i,b}|1\leq i \leq N, 1\leq b\leq m\}$, and similarly $\mathcal{X}^-_{R,N}$ the~(unital) subalgebra of $\mathcal{X}_{R,N}$ generated by all the annihilation operators $\{\hat{\psi}^-_{i,b}|1\leq i \leq N, 1\leq b\leq m\}$. 

An important observation is that the algebra $\mathcal{X}_{R,N}$ can be obtained from $\mathcal{X}_{R,1}$ as
\begin{equation}\label{eq:XRNXR1}
	\mathcal{X}_{R,N}\cong \mathcal{X}_{\Pi\boxtimes R,1},
\end{equation} 
where  $\Pi\boxtimes R$ is the direct product $R$-matrix defined as
\begin{equation}\label{eq:product_R}
	(\Pi\boxtimes R)^{AB}_{CD}\equiv \Pi^{ij}_{kl} R^{ab}_{cd},
\end{equation}
where we group the spatial index $i=1,2,\ldots,N$ and the internal index $a$ in $\hat{\psi}^\pm_{i,a}$ into a single collective index: $A=(i,a), B=(j,b), C=(k,c)$ and $D=(l,d)$. $\Pi$ acts on the spatial part defined as $\Pi^{ij}_{kl}=\delta_{il}\delta_{jk}$, and $R$ acts on the internal part. It is straightforward to check that $\Pi\boxtimes R$ constructed this way also satisfies the YBE Eq.~\eqref{eq:YBE}, and Eq.~\eqref{eq:XRNXR1} can be checked by comparing the defining CRs of both sides. %
For this reason, in Secs.~\ref{SI:existence_vacuum} and \ref{SI:LWR} we focus on the algebra $\mathcal{X}_{R}\equiv \mathcal{X}_{R,1}$, but keep in mind that any claim we make on $\mathcal{X}_{R}$ applies equally well  to $\mathcal{X}_{R,N}$ by using the product $R$-matrix $\Pi\boxtimes R$. We will omit the mode labels $i,j$ and simply write $\hat{\psi}^\pm_a$ when there is no confusion. 

\subsubsection{Existence and uniqueness of vacuum state from physical requirements}\label{SI:existence_vacuum}
For the theory to make physical sense, the spectrum of the total particle number operator $\hat{n}$ should be bounded from below. This means that there exists at least one state $|n_{\min}\rangle$ with the smallest eigenvalue $n_{\min}$ of $\hat{n}$. Since $\hat{\psi}^-_{a}$ decreases the eigenvalue of $\hat{n}$ by $1$, the minimality of $n_{\min}$ requires that $\hat{\psi}^-_{a}|n_{\min}\rangle=0,\forall a$, since otherwise $\hat{\psi}^-_{a}|n_{\min}\rangle$ would be an eigenstate of $\hat{n}$ with eigenvalue $n_{\min}-1<n_{\min}$. Therefore $\hat{n}|n_{\min}\rangle=\sum_{a}\hat{\psi}^+_{a}\hat{\psi}^-_{a}|n_{\min}\rangle=0$, i.e., $n_{\min}=0$. We call this state the vacuum state, denoted by $|0\rangle$. %

It can be proven that in an irrep $V$ of $\mathcal{X}_{R}$, 
the vacuum state must be unique. Here is a sketch of the proof by contradiction: assume there exists two linearly independent  vacuum states, say  $|0\rangle,|0'\rangle\in V$. 
Then $V_0=\mathcal{X}^+_{R}|0\rangle$
would be invariant under the action of $\mathcal{X}_{R}$. To prove this, it is enough to show that $V_0$ is invariant under all the generators $\hat{\psi}^\pm_{a}$ of  $\mathcal{X}_{R}$: $\hat{\psi}^+_{a}$ leaves $V_0$ invariant since $\hat{\psi}^+_{a}\mathcal{X}^+_{R}\subseteq\mathcal{X}^+_{R}$, while $\hat{\psi}^-_{a}\mathcal{X}^+_{R}\subseteq \mathcal{X}^+_{R}\hat{\psi}^-_{a}+ \mathcal{X}^+_{R}$ according to the first relation in Eq.~\eqref{eq:fundamental_Rcommu}, so $\hat{\psi}^-_{a}V_0\subseteq V_0$. Therefore, $V_0$ is a subrepresentation of $V$. 
Furthermore, $|0'\rangle \notin V_0$ since the only state in $V_0$ annihilated by $\hat{n}$ is $|0\rangle$. Therefore, $V_0$ is a proper subrepresentation of $V$, contradicting the irreducibility of $V$.

\subsubsection{The state space generated by $|0\rangle$ and $\{\hat{\psi}^+_{a}\}$}\label{SI:LWR}
The algebra $\mathcal{X}_{R}$ is the special case of the quantum Weyl algebras~(QWAs)  $A_m(R)$ studied in Ref.~\cite{GIAQUINTO1995QWA} with $q=1$, by identifying $x_a$ with $\hat{\psi}^+_a$ and $\partial_a$ with $\hat{\psi}^-_a$,
and Thm.~1.5 in Ref.~\cite{GIAQUINTO1995QWA} provides the rigorous mathematical foundation for the construction of state space: %
\begin{theorem}{(Thm.~1.5 in Ref.~\cite{GIAQUINTO1995QWA})}\label{thm:PBW}
	There is a vector space isomorphism $\mathbb{C}_R\langle \hat{\psi}^+_a\rangle\otimes \mathbb{C}_R\langle \hat{\psi}^-_a\rangle \cong \mathcal{X}_{R}$, where $\mathbb{C}_{R}\langle \hat{\psi}^+_{a}\rangle$ is the unital associative algebra generated by  $\{\hat{\psi}^+_{b}| 1\leq b\leq m\}$, subject to the second relations in Eq.~\eqref{eq:fundamental_Rcommu}, and similarly $\mathbb{C}_{R}\langle \hat{\psi}^-_{a}\rangle$ is the unital associative algebra generated by  $\{\hat{\psi}^-_{b}|1\leq b\leq m\}$, subject to the third relations in Eq.~\eqref{eq:fundamental_Rcommu}.
\end{theorem}
This theorem extends the simpler fact that, as a vector space, $\mathcal{X}_{R}$ is spanned by $\mathcal{X}^+_{R}\otimes \mathcal{X}^-_{R}$, since for any monomial of $\hat{\psi}^+_1, \ldots, \hat{\psi}^+_m, \hat{\psi}^-_1, \ldots, \hat{\psi}^-_m$ in $\mathcal{X}_{R}$~(e.g. $\hat{\psi}^-_a \hat{\psi}^+_b \hat{\psi}^-_c \hat{\psi}^+_d$), one can always use the first relation in Eq.~\eqref{eq:fundamental_Rcommu} to ``normal order'' all $\hat{\psi}^+_a$s to the left and $\hat{\psi}^-_a$s to the right, leading to a sum of terms, each with at most $m$ creation and $m$ annihilation operators. The non-trivial aspect of this theorem is that $\mathcal{X}^+_{R}\cong \mathbb{C}_R\langle \hat{\psi}^+_a\rangle$, and $\mathcal{X}^-_{R}\cong \mathbb{C}_R\langle \hat{\psi}^-_a\rangle$, i.e. the relations in the first and third lines of Eq.~\eqref{eq:fundamental_Rcommu} do not imply any additional relations on the $\hat{\psi}^+_a$s other than the second line in Eq.~\eqref{eq:fundamental_Rcommu}.   See Ref.~\cite{GIAQUINTO1995QWA} for a detailed proof.

We now construct the state space as the representation space of $\mathcal{X}_{R}$, defined as the canonical left $\mathcal{X}_{R}$-module $\mathfrak{V}=\mathcal{X}_{R}/[\sum_a \mathcal{X}_{R}\hat{\psi}^-_a]$. More explicitly, $\mathfrak{V}$ is the left $\mathcal{X}_{R}$-module generated by a vacuum state $|0\rangle$ satisfying the relation $\hat{\psi}^-_a|0\rangle=0$ for all $a$, such that $\mathfrak{V}=\mathcal{X}_{R}|0\rangle$. %
Then Thm.~\ref{thm:PBW} immediately implies that~(see the comment at the end of Sec.~1 in Ref.~\cite{GIAQUINTO1995QWA}), as a vector space, $\mathfrak{V}=\mathcal{X}^+_{R}|0\rangle\cong\mathbb{C}_R\langle \hat{\psi}^+_a\rangle$. %
Furthermore, Thm.~3.2 of Ref.~\cite{GIAQUINTO1995QWA} proves that the representation $\mathfrak{V}$ of $\mathcal{X}_{R}$ is irreducible, and our discussion in Sec.~\ref{SI:existence_vacuum} implies that it is the only irrep of $\mathcal{X}_{R}$ with the spectrum of $n$ bounded from below. 

In the following we find a basis for the  state space $\mathfrak{V}\equiv \mathcal{X}^+_{R}|0\rangle$. 
We use the eigenvalues of the particle number operator $\hat{n}$ to decompose $\mathfrak{V}$  into a direct sum of eigenspaces of $\hat{n}$: $\mathfrak{V}=\bigoplus_{n\geq 0} \mathfrak{V}_n$.  Each subspace $\mathfrak{V}_n$ is spanned by states with fixed particle number 
\begin{equation}\label{def:V_n}
	\mathfrak{V}_n=\mathrm{span}\{ \hat{\psi}^+_{a_1}\hat{\psi}^+_{a_2}\ldots\hat{\psi}^+_{a_n}|0\rangle| 1\leq a_j\leq m,j=1,2,\ldots,n\}.
\end{equation}
Notice, however, due to the CR in Eq.~\eqref{eq:fundamental_Rcommu}, the states defined in the RHS of Eq.~\eqref{def:V_n} are linearly dependent. For example, the state $\hat{\psi}^+_{a}\hat{\psi}^+_{b}|0\rangle$ is the same as $\sum_{c,d}R^{cd}_{ab}\hat{\psi}^+_{c}\hat{\psi}^+_{d}|0\rangle$. 
A linearly independent basis for $\mathfrak{V}_n$ is established by the following theorem:%
\begin{theorem}\label{thm:state_space}
	The states $\{|n,\alpha\rangle\}_{\alpha=1}^{d_n}$  defined by Eqs.~(\ref{eq:V_n_basis}-\ref{eq:single_mode_creation})~(for the case $N=1$) form a complete, linearly independent basis for $\mathfrak{V}_n$.
\end{theorem}

We now sketch the proof of Thm.~\ref{thm:state_space}. Following our discussion in the previous paragraph, the $n$-particle space $\mathfrak{V}_n$ of a single mode~[defined in Eq.~\eqref{def:V_n}] can be identified with $\mathbb{C}^{(n)}_R\langle \hat{\psi}^+_a\rangle$, the subspace of $\mathbb{C}_R\langle \hat{\psi}^+_a\rangle$ spanned by all degree $n$  monomials in $\hat{\psi}^+_a$s. So it remains to be proven that $\mathbb{C}^{(n)}_R\langle \hat{\psi}^+_a\rangle$ is isomorphic~(as a vector space) to the space of solutions $\Psi_{a_1\ldots a_n}$ to Eq.~\eqref{eq:V_n_basis}. %
For convenience, we define a product vector space $\mathfrak{A}\equiv \mathfrak{a}^{\otimes n}$, %
where $\mathfrak{a}$ is an $m$-dimensional vector space with basis $\{v_1,v_2,\ldots, v_m\}$.  The tensor  $R^{ab}_{cd}$ defines a  linear map $R$ in the product space $\mathfrak{a}\otimes \mathfrak{a}$ as $R(v_c\otimes v_d)=\sum_{ab}R^{ab}_{cd}v_a\otimes v_b$, %
and this action is extended to $ \mathfrak{a}^{\otimes n}$ as %
\begin{equation}\label{def:Rjjp1}
	R_{j,j+1}=\underset{{(1)}}{\mathds{1}}\otimes\ldots\otimes \underset{(j-1)}{\mathds{1}}\otimes \underset{(j,j+1)}{R}\otimes\underset{(j+2)}{\mathds{1}}\otimes\ldots\otimes \underset{(n)}{\mathds{1}}.
\end{equation}  
Furthermore, we can associate a tensor $\Psi_{a_1\ldots a_n}$ to a vector in  $\mathfrak{a}^{\otimes n}$ through $\Psi=\sum_{a_1\ldots a_n}\Psi_{a_1\ldots a_n}v_{a_1}\otimes v_{a_2}\otimes\ldots \otimes v_{a_n}$. Then Eq.~\eqref{eq:V_n_basis} is equivalent to
\begin{equation}\label{eq:V_n_basis_SI}
	R_{j,j+1}\Psi=\Psi~(\text{in } \mathfrak{a}^{\otimes n}),~j=1,2,\ldots,n-1.
\end{equation}
In short, we need to prove that $\mathbb{C}^{(n)}_R\langle \hat{\psi}^+_a\rangle$ is isomorphic~(as a vector space) to the common eigenspace~(with eigenvalue $+1$) of all $R_{j,j+1}$. We have, as a vector space,
\begin{equation}\label{eq:def_Cn_as_quotient_space}
	\mathbb{C}^{(n)}_R\langle \hat{\psi}^+_a\rangle\cong \frac{\mathfrak{a}^{\otimes n}}{\sum^{n-1}_{j=1}[(\mathds{1}-R_{j,j+1})\mathfrak{a}^{\otimes n}]},
\end{equation}
since $\mathbb{C}_R\langle \hat{\psi}^+_a\rangle$ is, by definition, isomorphic to the quotient of the tensor algebra $T(\mathfrak{a})$ over the quadratic relations $R(\mathfrak{a}\otimes \mathfrak{a})=(\mathfrak{a}\otimes \mathfrak{a})$~\cite{Majid1990}, where $\mathfrak{a}$ is an $m$-dimensional vector space. %
Note that in Eq.~\eqref{eq:def_Cn_as_quotient_space}, $(\mathds{1}-R_{j,j+1})\mathfrak{a}^{\otimes n}$ for each $j$ is considered as a subspace of $\mathfrak{a}^{\otimes n}$, and $\sum^{n-1}_{j=1}$ means the sum of these subspaces. 
We now prove the following lemma: 
\begin{lemma}\label{lemma:quotient_space_vs_eigenspace}
	Let $H_1,H_2,\ldots,H_k$ be Hermitian matrices %
	acting on a Hilbert space $V$. Then  
	\begin{eqnarray}\label{def:quotient_space_vs_eigenspace}
		&&\mathrm{span}\{|\psi\rangle\in V~|~H_{j}|\psi\rangle=|\psi\rangle, 1\leq j\leq k\} \nonumber\\ 
		&\cong&	\frac{V}{\sum^{k}_{j=1}[(\mathds{1}-H_{j})V]}.
	\end{eqnarray}
\end{lemma}
\begin{proof}
	For a Hilbert space $V$, and a subspace $V_1\subseteq V$, the quotient space $V/V_1$ is isomorphic to the orthogonal complement $V_1^\perp$. Therefore we have
	\begin{eqnarray}\label{eq:quotient_space_vs_eigenspace}
		\frac{V}{\sum^{k}_{j=1}[(\mathds{1}-H_{j})V]}&\cong&\left\{\sum^{k}_{j=1}[(\mathds{1}-H_{j})V]\right\}^\perp\nonumber\\
		&=&\bigcap^{k}_{j=1}\left[(\mathds{1}-H_{j})V\right]^\perp,%
	\end{eqnarray}
	where in the second line we used $(V_1+ V_2)^\perp=V_1^\perp\cap V_2^\perp$. %
	But since $H_j$ is assumed to be Hermitian, $\left[(\mathds{1}-H_{j})V\right]^\perp$ is simply the eigenspace of $H_{j}$ with eigenvalue $+1$, since for any $\ket{u}\in \left[(\mathds{1}-H_{j})V\right]^\perp$, we have, by definition, $\braket{v|(1-H_j)|u}=0$, $\forall \ket{v}\in V$, implying $(1-H_j)\ket{u}$=0. 
	Therefore second line of Eq.~\eqref{eq:quotient_space_vs_eigenspace} is the same as the LHS of Eq.~\eqref{def:quotient_space_vs_eigenspace}.
\end{proof}
Although the $R$-matrices in our models are not always Hermitian, there always exists a Hermitian inner product on the space $\mathfrak{A}=\mathfrak{a}^{\otimes n}$ with respect to which the matrices $\{R_{j,j+1}\}^{n-1}_{j=1}$ are all Hermitian. This is because any finite-dimensional representation of a finite group is isomorphic to a unitary representation~(Theorem 4.6.2 in Ref.~\cite{etingof2011introduction}), and in our case the matrices $\{R_{j,j+1}\}^{n-1}_{j=1}$ generate the finite group $S_n$~(notice that if $R_{j,j+1}$ is unitary then it is Hermitian since $R_{j,j+1}^2=\mathds{1}$). Therefore, Lemma~\ref{lemma:quotient_space_vs_eigenspace} still applies, implying that the RHS of Eq.~\eqref{eq:def_Cn_as_quotient_space} is isomorphic to the common eigenspaces of $\{R_{j,j+1}\}^{n-1}_{j=1}$ defined by Eq.~\eqref{eq:V_n_basis}. This concludes the proof of Thm.~\ref{thm:state_space}. 

\subsubsection{Many particle state space}\label{SI:full_state_space}
We now prove that the states defined in  Eqs.~(\ref{eq:V_n_basis}-\ref{eq:single_mode_creation}) form a  linearly independent basis for the many particle state space $\mathcal{X}^+_{R,N}|0\rangle$ for any positive integer $N$.
We need the following lemma:
\begin{lemma}\label{lemma:structure_thm_X_RN}
	There is a vector space isomorphism $\mathcal{X}_{R,N}\cong \mathcal{X}_{R}^{\otimes N}$. In particular, $ \mathcal{X}_{R}$ is isomorphic to the subalgebra of $\mathcal{X}_{R,N}$ generated by $\{\hat{\psi}^\pm_{i,a}|1\leq a\leq m\}$, for any $i\in \{1,2,\ldots,N\}$. 
\end{lemma}
\begin{proof}
	By Thm.~\ref{thm:PBW}, we have~(as vector spaces) $\mathcal{X}_{R}\cong \mathbb{C}_R\langle \hat{\psi}^+_a\rangle\otimes \mathbb{C}_R\langle \hat{\psi}^-_a\rangle $, and $\mathcal{X}_{R,N}\cong \mathbb{C}_{\Pi\boxtimes R}\langle \hat{\psi}^+_a\rangle\otimes \mathbb{C}_{\Pi\boxtimes R}\langle \hat{\psi}^-_a\rangle $, so we only need to prove that~(as a vector space) $\mathbb{C}_{\Pi\boxtimes R}\langle \hat{\psi}^+_a\rangle\cong\mathbb{C}_{ R}\langle \hat{\psi}^+_a\rangle^{\otimes N}$. This can be proven by induction on $N$, where the induction step $N\to N+1$ can be proven in a similar way as Thm.~\ref{thm:PBW}. Alternatively, it is straightforward to show that $h_{\Pi\boxtimes R}(x)=h_R(x)^N$, and since for every $R$-matrix, $\dim \mathbb{C}_{ R}\langle \hat{\psi}^+_a\rangle=h_R(1)$, we have $\dim \mathbb{C}_{\Pi\boxtimes R}\langle \hat{\psi}^+_a\rangle=h_R(1)^N=\dim \mathbb{C}_{ R}\langle \hat{\psi}^+_a\rangle^{\otimes N}$, so as a vector space, $\mathbb{C}_{\Pi\boxtimes R}\langle \hat{\psi}^+_a\rangle\cong\mathbb{C}_{ R}\langle \hat{\psi}^+_a\rangle^{\otimes N}$.
\end{proof}
While Lemma~\ref{lemma:structure_thm_X_RN} seems natural, the non-trivial part is that the CRs~\eqref{eq:fundamental_Rcommu} involving  any other modes $\hat{\psi}^\pm_{j,a}$~(with $j\neq i$) do not give rise to any additional algebraic relations on $\{\hat{\psi}^\pm_{i,a}|1\leq a\leq m\}$. This is a rigorous justification that different modes are mutually independent. Lemma~\ref{lemma:structure_thm_X_RN}  along with Thm.~\ref{thm:state_space} immediately imply that the states defined in  Eqs.~(\ref{eq:V_n_basis}-\ref{eq:single_mode_creation}) form a  linearly independent basis for  $\mathcal{X}^+_{R,N}|0\rangle$.

\subsubsection{Proof of unitarity for theories based on non-unitary $R$-matrices}\label{SI:proof_hermiticity}
We now prove our claim in the main text %
that even with a non-unitary $R$-matrix, we can consistently define the Hermitian conjugate $\dagger$ on the states and operators such that $\hat{e}^\dagger_{ij}=\hat{e}_{ji}$, for $1\leq i,j\leq N$, which guarantees Hermiticity of Hamiltonians and unitarity of quantum time evolution. %
To define the Hermitian conjugate $\dagger$ of operators, we need to define a Hermitian inner product $\langle\ldots|\ldots\rangle$ on the state space, and then the Hermitian conjugate of an operator $\hat{O}$ is defined as  $\langle \Psi|\hat{O}^\dagger\Phi\rangle\equiv\langle \hat{O}\Psi|\Phi\rangle$, for any states $|\Psi\rangle,|\Phi\rangle$. In the following we first show that such an inner product can be consistently defined on the state space such that the induced Hermitian conjugate $\dagger$ satisfies  $\hat{e}^\dagger_{ij}=\hat{e}_{ji},~\forall i,j$, and then give a more explicit definition of this inner product.

To begin, we first notice that, with the CRs in Eq.~\eqref{eq:commu_Eab_Ecd}, the set of operators $\{\hat{e}_{ij}+\hat{e}_{ji},i(\hat{e}_{ij}-\hat{e}_{ji})|1\leq i,j\leq N\}$ spans a closed Lie algebra $\mathfrak{u}_N\cong \mathfrak{su}_N\oplus \mathfrak{u}_1$~(over the field of real numbers $\mathbb{R}$), where the $\mathfrak{u}_1$ part is $n\equiv \sum^N_{i=1} \hat{e}_{ii}$, and the $\mathfrak{su}_N$ part is spanned by  $\{\hat{e}_{ij}+\hat{e}_{ji},i(\hat{e}_{ij}-\hat{e}_{ji})|1\leq i<j\leq N\}$ along with $ \{\hat{e}_{ii}-\hat{e}_{i+1,i+1}|1\leq i\leq N-1\}$.
We now invoke the following theorem whose proof can be found in Ref.~\cite{bourbaki2005lie_chap9} 
\begin{theorem}\label{thm1}
	For each representation $\rho$ of a compact semisimple real Lie algebra $\mathfrak{g}$ on a finite dimensional $\mathbb{C}$-vector space $V$, there exists a Hermitian inner product on $V$ such that all $\rho(x) (x\in \mathfrak{g})$ are Hermitian. 
\end{theorem}
Since $\mathfrak{su}_N$ is a compact semisimple real Lie algebra, Thm.~\ref{thm1} guarantees the existence of a Hermitian inner product such that $\{\hat{e}_{ij}+\hat{e}_{ji},i(\hat{e}_{ij}-\hat{e}_{ji})|1\leq i<j\leq N\}$ and $\{\hat{e}_{ii}-\hat{e}_{i+1,i+1}|1\leq i\leq N-1\}$  are all Hermitian, as long as the state space is finite dimensional. But even if the state space is infinite dimensional, we will see later that the full state space can always be decomposed as a direct sum of finite dimensional irreducible representations~(irreps) of $\mathfrak{su}_N$, and Thm.~\ref{thm1} still applies to each irrep. As for the $\mathfrak{u}_1$ part, since $n$ is proportional to the identity operator in each irrep, with the proportionality constant being the total particle number, it follows that $n$ is also Hermitian. Since Thm.~\ref{thm1} implies %
that $\hat{e}_{ij}+\hat{e}_{ji} =\hat{e}_{ij}^\dagger+\hat{e}_{ji}^\dagger$ and $i(\hat{e}_{ij}-\hat{e}_{ji})=-i(\hat{e}_{ij}^\dagger-\hat{e}_{ji}^\dagger)$, it follows that $\hat{e}^\dagger_{ij}=\hat{e}_{ji}$, for all $1\leq i,j\leq N$.

The Hermitian inner product on the state space can be defined more explicitly as follows. The %
state space constructed in Eqs.~(\ref{eq:V_n_basis}-\ref{eq:single_mode_creation}) decomposes into a direct sum of different particle  number sectors, and $\hat{n}$ is proportional to identity in each sector. Each particle number sector further decomposes into a direct sum of irreps of  $\mathfrak{su}_N$. We set $\langle \Psi|\Phi\rangle=0$ if $|\Psi\rangle,|\Phi\rangle$ lie in different irreps~(i.e., inequivalent irreps or different copies of equivalent irreps) of $\mathfrak{su}_N$. In this way, $n$ is automatically Hermitian~(indeed, it is real and diagonal) and the problem reduces to defining $\langle \ldots|\ldots\rangle$ within each irrep of $\mathfrak{su}_N$.

We now show that within each irrep, the inner product between any two states is uniquely determined~(up to a multiplicative factor) by the requirement $\hat{e}^\dagger_{ij}=\hat{e}_{ji},~\forall i,j$. We show this in the framework of highest weight theory~\cite{humphreys_LA}. For every finite-dimensional irrep $V$ of a finite-dimensional semisimple Lie algebra $\mathfrak{g}$, there exists a unique~(up to a multiplicative constant) highest weight vector $|\Lambda\rangle$~(a $|\Lambda\rangle$ that is annihilated by all positive root operators $\hat{e}_{\alpha}|\Lambda\rangle=0$), and all other weight vectors $|\Lambda'\rangle$ in $V$ can be constructed by applying negative root operators on $|\Lambda\rangle$, i.e. $|\Lambda'\rangle=\prod_\alpha \hat{e}_{-\alpha}|\Lambda\rangle$, where the product is over some ordered set of positive roots. For the case of $\mathfrak{su}_N$, positive root operators are $\{\hat{e}_{ij}|1\leq i<j\leq N\}$, negative root operators are $\{\hat{e}_{ji}|1\leq i<j\leq N\}$, while $\{\hat{e}_{ii}-\hat{e}_{i+1,i+1}|1\leq i\leq N-1\}$ spans the Cartan subalgebra. Without loss of generality we can assume $\langle\Lambda|\Lambda\rangle=1$. %
Then for any two weight vectors $|\Lambda_1\rangle,|\Lambda_2\rangle\in V$, their inner product can be calculated as
\begin{eqnarray}\label{eq:Herm_inner_prod}
	\langle \Lambda_1 |\Lambda_2\rangle&=&\prod_{\alpha,\beta} \langle\Lambda | \hat{e}_{-\beta}^\dagger \hat{e}_{-\alpha}|\Lambda\rangle\nonumber\\
	&=& \langle\Lambda | \prod_{\alpha,\beta} \hat{e}_{\beta} \hat{e}_{-\alpha}|\Lambda\rangle,
\end{eqnarray} 
and the last line can be calculated by using the CRs between $\hat{e}_{\beta}$ and $ \hat{e}_{-\alpha}$~(to move all the positive root operators $\hat{e}_{\beta}$ to the right). Notice that there may be several different ways to represent $|\Lambda_{1,2}\rangle$ in the form $\prod_\alpha \hat{e}_{-\alpha}|\Lambda\rangle$, and consequently there are different ways to compute the same inner product $\langle \Lambda_1 |\Lambda_2\rangle$. Thm.~\ref{thm1} guarantees that all the different ways of computing $\langle \Lambda_1 |\Lambda_2\rangle$ give the same result.   %

\section{Relation between the second and the first quantization formulation}\label{SI:relation_first_quantization}
In this section we show the relation between the second quantized formulation of parastatistics and the first quantized wavefunction formulation %
we discussed in the introduction of the main text. %
To this end we first show the relation between the $R$-matrix $R^{ab}_{cd}$ and the coefficients $(R_{j})^I_J$ appearing in  Eq.~\eqref{eq:wavefuntion_exchange_para}.
Let the index $I$~(and similarly for $J$) be a collection of $n$ auxiliary indices $I=(a_1,a_2,\ldots,a_n)$ labeling the basis states of a product vector space $\mathfrak{A}\equiv \mathfrak{a}^{\otimes n}$~(the internal space of wavefunctions), where the basis of $\mathfrak{a}$ is $\{v_1,v_2,\ldots, v_m\}$.  Now let $(R_{j})^I_J$ be the matrix element of the linear mapping $R_{j,j+1}$ defined in Eq.~\eqref{def:Rjjp1}. 
With this choice of $R_{j}$, Eq.~\eqref{eq:wavefuntion_exchange_para} becomes~(take $n=3$ and $j=1$ for example) %
\begin{equation}\label{eq:wavefuntion_exchange_Rmat}
	\Psi^{a_1 a_2 a_3}(x_2,x_1,x_3)=\sum_{b_1,b_2} R^{a_1a_2}_{b_1 b_2 } \Psi^{b_1 b_2 a_3}(x_1,x_2,x_3).
\end{equation} 
Then all the relations in Eq.~\eqref{eq:SNbraidrelations} reduce to Eq.~\eqref{eq:YBE}. An isomorphism between the space of $n$-particle wavefunctions in the first quantization formulation and the subspace of $n$-particle states in the second quantization formulation is defined as follows: each $n$-particle wavefunction $\Psi^I(x_1,\ldots,x_n)$ satisfying Eq.~\eqref{eq:wavefuntion_exchange_para}~[with the $R$-matrix in Eq.~\eqref{def:Rjjp1}] corresponds to the $n$-particle state 
\begin{equation}\label{eq:Psi_state_full}
	|\Psi\rangle=\frac{1}{\sqrt{n!}}\sum_{\substack{I,x_1,\ldots,x_n}}\Psi^{I}(x_1,\ldots,x_n)\hat{\psi}^+_{x_1,a_1}\ldots\hat{\psi}^+_{x_n,a_n}|0\rangle,%
\end{equation}%
That Eq.~\eqref{eq:Psi_state_full} indeed defines an isomorphism between the two vector spaces can be seen as follows.
Note that Eq.~\eqref{eq:V_n_basis} and Eq.~\eqref{eq:full_basis}~(for the case $N=1$) with the $R$-matrix $\Pi\boxtimes R$ is the same as Eq.~\eqref{eq:wavefuntion_exchange_para} and Eq.~\eqref{eq:Psi_state_full} with the $R$-matrix $R$, respectively. Then Thm.~\ref{thm:state_space} applied to the %
algebra $\mathcal{X}_{\Pi\boxtimes R}\cong \mathcal{X}_{R,N}$
shows that a linearly independent basis for the space  of $n$-particle wavefunctions satisfying Eq.~\eqref{eq:wavefuntion_exchange_para} correspond to a linearly independent basis for the $n$-particle subspace $\mathfrak{V}_n$ of the Fock space $\mathcal{X}^+_{\Pi\boxtimes R}|0\rangle\cong \mathcal{X}^+_{R,N}|0\rangle$ via the relation Eq.~\eqref{eq:Psi_state_full}, thereby establishing the isomorphism.

The problem with the first quantization formulation is that it is very hard to guarantee locality in this formulation, and without locality such theories are hard to be realized as elementary particles or as emergent quasiparticle excitations in locally interacting systems~\footnote{Note that it is exactly for this reason that the Doplicher-Haag-Roberts~(DHR) no-go theorem~\cite{doplicher1971local,*doplicher1974local} does not apply to the first quantization formulation, since the former takes locality as a fundamental assumption, while the latter does not have locality built-in.}. 
Part of the difficulty comes from the fact that locality puts stringent restrictions on the representation of $S_n$ realized by $\{R_{j}\}^{n-1}_{j=1}$. For example, a necessary condition for locality is the cluster law introduced in some earlier works~\cite{hartle1969,Taylor1970a}~(their first quantization formulation of parastatistics is slightly different from ours presented in the main text but closely related, and the cluster law applies to our formulation as well). This implies that only some very special choice of the matrices $\{R_{j}\}^{n-1}_{j=1}$ in Eq.~\eqref{eq:wavefuntion_exchange_para} leads to a local quantum theory, for example, the one given by Eq.~\eqref{def:Rjjp1}.   %

\section{Details on the 1D spin model and the MPO JWT}\label{SI:1Dspin_model}
In this section we provide mathematical details on the 1D spin model defined in Eqs.~(\ref{eq:Hamil1Dspin}-\ref{eq:Hamil1Dspin_para}), including~(Sec.~\ref{SI:spin_model_def}) an explicit definition of the local spin operators $\{\hat{x}^\pm_{i,a},\hat{y}^\pm_{i,a}\}^m_{a=1}$ and (Sec.~\ref{SI:gJWT}) a tensor network proof of the key properties of the MPO JWT in Eq.~\eqref{eq:JWT_string}. %

\subsection{Model definition for an arbitrary $R$-matrix}\label{SI:spin_model_def}
\begin{figure}
	\center{\includegraphics[width=\linewidth]{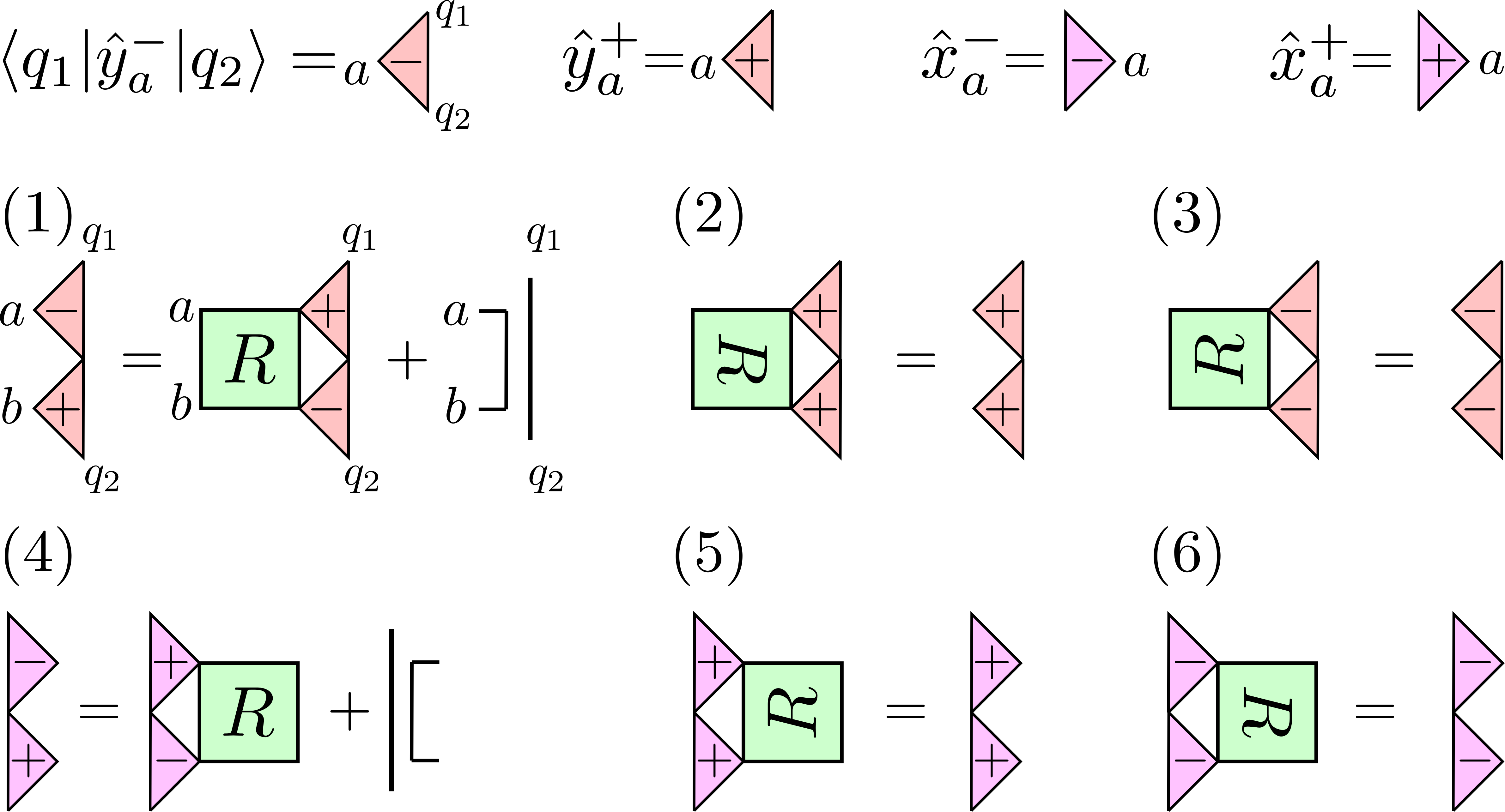}}
	\caption{\label{fig:XYalgebra} Graphical representation of the CRs between the local spin operators $\{\hat{x}^\pm_{a},\hat{y}^\pm_{a}\}^m_{a=1}$ in Eq.~\eqref{eq:XYQA}. The matrix elements of each operator $\{\hat{x}^\pm_{a},\hat{y}^\pm_{a}\}^m_{a=1}$ is a tensor~(represented by the triangles) with two quantum indices~(e.g. the indices $q_1$ and $q_2$ shown in figure) and one auxiliary index~(e.g. the index $a$), and the $R$-matrix~(represented by a square) is a tensor with four auxiliary indices.  
		Matrix multiplication goes from top to bottom in the quantum space and from left to right in the auxiliary space. }
\end{figure}

We first define the local spin operators $\{\hat{x}^\pm_{i,a},\hat{y}^\pm_{i,a}\}^m_{a=1}$ that appear in the Hamiltonian in Eq.~\eqref{eq:Hamil1Dspin}, for any given $R$-matrix. The Hilbert space of the whole system with $N$ sites in total is $\mathfrak{V}^{\otimes N}$, where $\mathfrak{V}$ is the Hilbert space for a single site. For any fixed $i=1,2,\ldots,N$, the operators $\{\hat{x}^\pm_{i,a},\hat{y}^\pm_{i,a}\}^m_{a=1}$ act locally on the $i$-th factor space, and 
they are constructed to satisfy the following algebraic relations~(from now on we omit the site label $i$)%
\begin{eqnarray}\label{eq:XYQA}
	\hat{y}^-_a \hat{y}^+_b&=&\sum_{c,d}R^{ac}_{bd} \hat{y}_c^+ \hat{y}_d^-+\delta_{ab},\nonumber\\%\label{eq:QAy1}\\
	\hat{y}^+_a \hat{y}^+_b&=&\sum_{c,d}R^{cd}_{ab} \hat{y}_c^+ \hat{y}_d^+,\nonumber\\%\label{eq:QAYp1}\\
	\hat{y}^-_a \hat{y}^-_b&=&\sum_{c,d}R^{ba}_{dc} \hat{y}_c^- \hat{y}_d^-,\nonumber\\%\label{eq:QAYp2}\\
	\hat{x}^-_a \hat{x}^+_b&=&\sum_{c,d}R^{ca}_{db} \hat{x}_c^+ \hat{x}_d^-+\delta_{ab},\nonumber\\%\label{eq:QAx1}\\
	\hat{x}^+_a \hat{x}^+_b&=&\sum_{c,d}R^{dc}_{ba} \hat{x}_c^+ \hat{x}_d^+,\nonumber\\%\label{eq:QAXp1}\\
	\hat{x}^-_a \hat{x}^-_b&=&\sum_{c,d}R^{ab}_{cd} \hat{x}_c^- \hat{x}_d^-,\nonumber\\%\label{eq:QAXp2}\\
	{}[\hat{x}^+_a,\hat{y}^+_b]&=&[\hat{x}^-_a,\hat{y}^-_b]=0,\nonumber\\%\label{eq:QAxy}
	\sum_{a} \hat{x}_a^+\hat{x}_a^-&=&\sum_{a}\hat{y}^+_a\hat{y}^-_a, %
\end{eqnarray}
While these CRs superficially resemble the CRs between paraparticle operators in Eq.~\eqref{eq:fundamental_Rcommu}, the difference is that the spin operators here are strictly local in that they commute on different sites, and therefore are in  principle realizable, while the paraparticle operators are generally non-local operators. The first 6 lines in Eq.~\eqref{eq:XYQA} are shown graphically in Fig.~\ref{fig:XYalgebra}.

We now define a local Hilbert space and a matrix representation of these local spin operators. 
The single site Hilbert space $\mathfrak{V}$ is  spanned by $\{\ket{n,\alpha}|1\leq \alpha\leq d_n,n\in\mathbb{Z}_{\geq 0}\}$, where $\{d_n\}_{n\geq 0}$ are the same numbers that define the generalized exclusion statistics introduced in the main text, and $\ket{n,\alpha}$ is defined as
\begin{equation}\label{eq:basisnormalization}
	\ket{n,\alpha}\equiv \frac{1}{\sqrt{n!}}\sum_{a_1 a_2\ldots a_n}\Psi^\alpha_{a_1 a_2\ldots a_n}\hat{y}^+_{a_1}\hat{y}^+_{a_2}\ldots\hat{y}^+_{a_n}\ket{0}, %
\end{equation}
where $\{\Psi^\alpha_{a_1a_2\ldots a_n}\}_{\alpha=1}^{d_n}$ is a complete set of linearly independent solutions to the system of linear equations~\eqref{eq:V_n_basis}. The matrix elements of $\hat{x}^\pm_{a},\hat{y}^\pm_{a}$ in this basis are defined as
\begin{eqnarray}\label{eq:XYmatrix_rep}
	\hat{y}^\pm_{a}\ket{n,\alpha}&=&\sum_{\beta=1}^{d_{n\pm 1}}Y^{\pm}_{a,\beta\alpha}\ket{n\pm 1,\beta},\nonumber\\
	\hat{x}^\pm_{a}\ket{n,\alpha}&=&\sum_{\beta=1}^{d_{n\pm 1}} X^{\pm}_{a,\beta\alpha}\ket{n\pm 1,\beta},
\end{eqnarray}
where the coefficients $ Y^{\pm}_{a,\beta\alpha},X^{\pm}_{a,\beta\alpha}$ are given by
\begin{eqnarray}\label{eq:XYExplicit}
	\sum_{\beta}Y^{+}_{a,\beta\alpha}\Psi^\beta_{a_0 a_1 \ldots a_n}&=&\!\frac{1}{\sqrt{n+1}}\bar{Y}^{a_0 a_1\ldots a_n}_{~a_{\hphantom{0}} b_1\ldots b_n} \Psi^\alpha_{b_1\ldots b_n},\nonumber\\
	\sum_{\beta}X^{+}_{a,\beta\alpha}\Psi^\beta_{a_n \ldots a_1 a_0}&=&\!\frac{1}{\sqrt{n+1}}\bar{X}^{a_n \ldots a_1 a_0}_{~\!b_n\ldots b_1a_{\hphantom{0}}} \Psi^\alpha_{b_n\ldots b_1},\nonumber\\
	\sum_{\beta}Y^-_{a,\beta\alpha}\Psi^\beta_{a_2\ldots a_n}&=&\sqrt{n}\Psi^\alpha_{a a_2\ldots a_n},\nonumber\\
	\sum_{\beta} X^-_{a,\beta\alpha}\Psi^\beta_{a_n\ldots a_2}&=&\sqrt{n}\Psi^\alpha_{a_n\ldots a_2 a},
\end{eqnarray}
where the tensors $\bar{Y},\bar{X}$ are defined as
\begin{eqnarray}\label{def:XYRtensor}
	\bar{Y}_{01\ldots n}&=&1+R_{01}+R_{12}R_{01}+\ldots+R_{n-1,n}\cdots R_{12}R_{01},\nonumber\\
	\bar{X}_{n\ldots 10}&=&1+R_{10}+R_{21}R_{10}+\ldots+R_{n,n-1}\cdots R_{21} R_{10},\nonumber\\
\end{eqnarray}
where we use the tensor notation introduced in Eqs.~(\ref{def:Rjjp1},\ref{eq:V_n_basis_SI}). %
It is straightforward to check that $\hat{x}^\pm_{a},\hat{y}^\pm_{a}$ defined this way satisfy all the CRs in Eq.~\eqref{eq:XYQA}, and Eq.~\eqref{eq:XYmatrix_rep} is consistent with Eq.~\eqref{eq:basisnormalization}.  In addition, we have the relation 
\begin{equation}
	{}[\hat{n},\hat{x}_a^\pm]=\pm \hat{x}^\pm_a,~~%
	{}[\hat{n},\hat{y}_a^\pm]=\pm \hat{y}^\pm_a,%
\end{equation}
where $\hat{n}\equiv \sum_{a} \hat{x}_a^+\hat{x}_a^-=\sum_{a}\hat{y}^+_a\hat{y}^-_a$.

We can now compute $\hat{T}^\pm_{ab}$ using their definition %
$\hat{T}^\pm_{ab}\equiv
\mp [\hat{y}^\pm_{a},\hat{x}^\mp_{b}]$,
and the  matrix elements of $\hat{x}^\pm_{a},\hat{y}^\pm_{a}$ in Eq.~\eqref{eq:XYExplicit}. The matrix elements of $\hat{T}^\pm_{ab}$ are defined similarly as
\begin{eqnarray}
	\hat{T}^\pm_{ab}\ket{n,\alpha}=\sum_{\beta=1}^{d_{n}}T^{\pm}_{ab,\beta\alpha}\ket{n,\beta},
\end{eqnarray}
where the coefficients $T^{\pm}_{ab,\beta\alpha}$ are 
\begin{eqnarray}\label{eq:Tpm_matrix_element}
	\sum_\beta T^+_{ab,\beta\alpha}\Psi^\beta_{a_1 a_2 \ldots a_n}&=&
	\sum_{a'_1,\ldots,a'_n}\Psi^\alpha_{a'_1,a'_2\ldots,a'_n}\nonumber\\
	&&\times \begin{tikzpicture}[baseline={([yshift=-.4ex]current bounding box.center)}, scale=.8]
		\TpRmatrix{0}{0}{a_1}{a}{a'_1}{}
		\TpRmatrix{1}{0}{a_2}{}{a'_2}{}
		\draw[dotted, thick] (3*\AL, 0) -- (5*\AL,0);
		\TpRmatrix{3}{0}{a_n}{}{a'_n}{b}
	\end{tikzpicture},\nonumber\\
	\sum_\beta T^-_{ab,\beta\alpha}\Psi^\beta_{a_1 a_2 \ldots a_n}&=&
	\sum_{a'_1,\ldots,a'_n}\Psi^\alpha_{a'_1 a'_2\ldots a'_n}\\
	&&\times\begin{tikzpicture}[baseline={([yshift=-.4ex]current bounding box.center)}, scale=.8]
		\TmRmatrix{0}{0}{a_1}{a}{a'_1}{}
		\TmRmatrix{1}{0}{a_2}{}{a'_2}{}
		\draw[dotted, thick] (3*\AL, 0) -- (5*\AL,0);
		\TmRmatrix{3}{0}{a_n}{}{a'_n}{b}
	\end{tikzpicture}.\nonumber
\end{eqnarray}
All the results above are valid for an arbitrary $R$-matrix satisfying Eq.~\eqref{eq:YBE} of the main text, including non-unitary ones. 
When the $R$-matrix is unitary, we require that $\{\Psi^\alpha_{a_1a_2\ldots a_n}\}_{\alpha=1}^{d_n}$ is normalized as in Eq.~\eqref{eq:singlemodewf_normalization}, %
and we define an inner product on the on-site Hilbert space $\bra{n',\beta}n,\alpha\rangle=\delta_{n'n}\delta_{\alpha\beta}$. Then one can check that the matrix representation of $\{\hat{x}^\pm_{i,a},\hat{y}^\pm_{i,a}\}^m_{a=1}$ defined in  Eq.~\eqref{eq:XYmatrix_rep} satisfies $\hat{x}_a^+=(\hat{x}_a^-)^\dagger$, $\hat{y}_a^+=(\hat{y}_a^-)^\dagger$.  Therefore the 1D spin model is Hermitian.
Furthermore, we have $\hat{T}_{ab}^+=(\hat{T}_{ab}^-)^\dagger$, leading to  $\hat{\psi}^+_{i,a}=(\hat{\psi}^-_{i,a})^\dagger$, where $\hat{\psi}^+_{i,a}$ are defined by the MPO JWT in Eq.~\eqref{eq:JWT_string}. In a similar way, one can show that the 2D model defined in Eq.~\eqref{def:2DsolvableH} is also Hermitian when $R$ is unitary. Unitary $R$-matrices have been classified up to a certain equivalence relation in Ref.~\cite{LECHNER2019106769}, their Hilbert series all have the form in Eq.~(5.7) of Ref.~\cite{LECHNER2019106769}.

\begin{figure}
	\center{\includegraphics[width=0.75\linewidth]{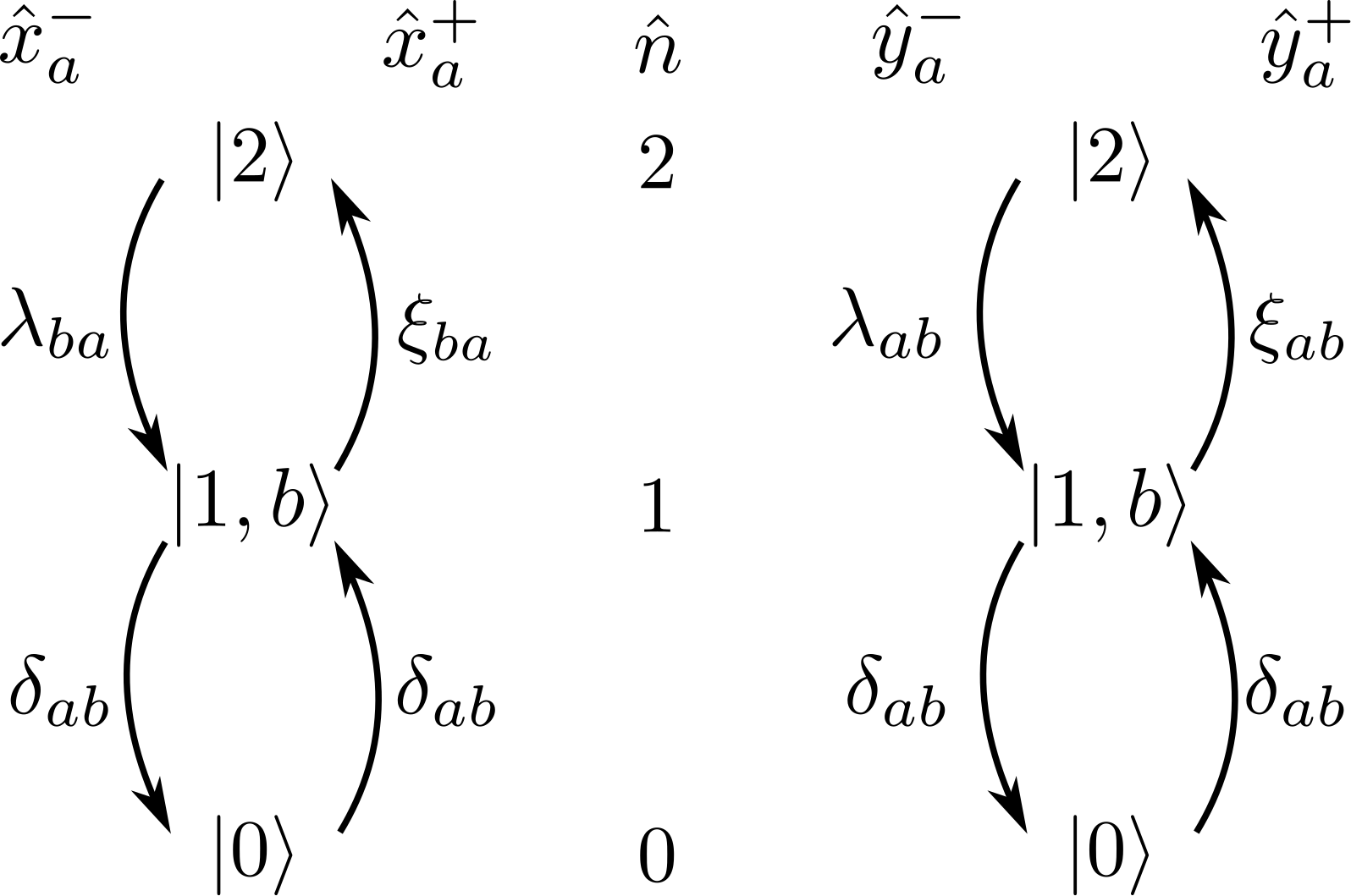}}
	\caption{\label{fig:xyaction} The action of the operators $\hat{x}^\pm_a,\hat{y}^\pm_a$ and $\hat{n}$ on the basis states $|0\rangle$, $\{|1,b\rangle\}^m_{b=1}$, $|2\rangle$, for the spin model corresponding to the $R$-matrix in Ex.~\ref{ex:1m1}. %
	}
\end{figure}

We can give the matrix elements of $\{\hat{x}^\pm_{a},\hat{y}^\pm_{a}\}^m_{a=1}$ more explicitly for the $R$-matrices given in Tab.~\ref{tab:Hilbert_series}. One finds that for the $R$-matrices in Ex.~\ref{ex:decoupled} and Ex.~\ref{ex:Green},  the corresponding spin models are free fermion solvable models that fall into the classification in Ref.~\cite{Chapman2020characterizationof}. In particular, Ex.~\ref{ex:Green} corresponds to $m$ decoupled chains of 1D $XY$ models, %
each of which maps to a 1D free fermion chain. %
The spin model for the $R$-matrix in Ex.~\ref{ex:1m} has been presented in the main text. 
For the $R$-matrix in Ex.~\ref{ex:1m1}, the local Hilbert space $\mathfrak{V}$ is $m+2$-dimensional, with basis states $|0\rangle, \{|1,b\rangle\}^m_{b=1},|2\rangle$; the non-zero matrix elements of $\hat{y}^\pm_a$ are $\hat{y}^+_a|0\rangle=|1,a\rangle$,  $\hat{y}^-_a|1,b\rangle=\delta_{ab}|0\rangle$,  $\hat{y}^+_a|1,b\rangle=\cc_{ab}|2\rangle$, and $\hat{y}^-_a|2\rangle=\sum_b \lambda_{ab}|1,b\rangle$; and the non-zero matrix elements of $\hat{x}^\pm_a$ are $\hat{x}^+_a|0\rangle=|1,a\rangle$,  $\hat{x}^-_a|1,b\rangle=\delta_{ab}|0\rangle$, $\hat{x}^+_a|1,b\rangle=\cc_{ba}|2\rangle$, and $\hat{x}^-_a|2\rangle=\sum_b \lambda_{ba}|1,b\rangle$.
The action of the operators $\hat{x}^\pm _i,\hat{y}^\pm_i$ and $n$ on the orthonormal basis states are shown in Fig.~\ref{fig:xyaction}. 

The spin model Hamiltonian in Eq.~\eqref{eq:Hamil1Dspin} is not Hermitian for the non-unitary $R$-matrix in Ex.~\ref{ex:1m1} for $m\geq 3$~\footnote{The model is still well-defined for $m=2$, where $\hat{H}$ is Hermitian. But that case is trivial: when $m=2$, $\hat{H}$ is equal to the sum of two decoupled chains of XY models.}, since $\hat{x}^+_a\ne (\hat{x}^-_a)^\dagger,\hat{y}^+_a\ne (\hat{y}^-_a)^\dagger$. However, $\hat{H}$ is parity-time symmetric~\cite{Bender_2007,el2018non}, and  all its eigenvalues are real. %
To be precise, let $P$ be the parity operator that generates the chain reflection symmetry, and let $T$ be the time-reversal symmetry, which, in our spin model, is simply complex conjugation. Using the explicit representation of the matrices $\lambda_{ab},\cc_{ab}$ in Tab.~\ref{tab:Hilbert_series}, %
we see that $\lambda_{ab}^*=\lambda_{ba},\cc^*_{ab}=\cc_{ba}$, and therefore, the time-reversal operation $T$ simply swaps the operators $\hat{x}^\pm_a\leftrightarrow \hat{y}^\pm_a$ in the Hamiltonian $\hat{H}$, which can subsequently be undone by $P$. Thus, $\hat{H}$ is invariant under the combined operation $PT$. As we already see from the exact solution of the spectrum, all eigenvalues of $\hat{H}$ are real.
This kind of PT-symmetric Hamiltonian can still define unitary quantum dynamics~\cite{Bender_2007}, so it is interesting to conceive an experimental platform that realizes the non-Hermitian Hamiltonian in Eq.~\eqref{eq:Hamil1Dspin} with this kind of emergent paraparticle. %
It is also interesting to investigate if the parastatistics in Ex.~\ref{ex:1m1} can alternatively be realized in a Hermitian spin model.  %
\subsection{Generalized Jordan-Wigner transformations}\label{SI:gJWT}
\begin{figure}
	\center{\includegraphics[width=\linewidth]{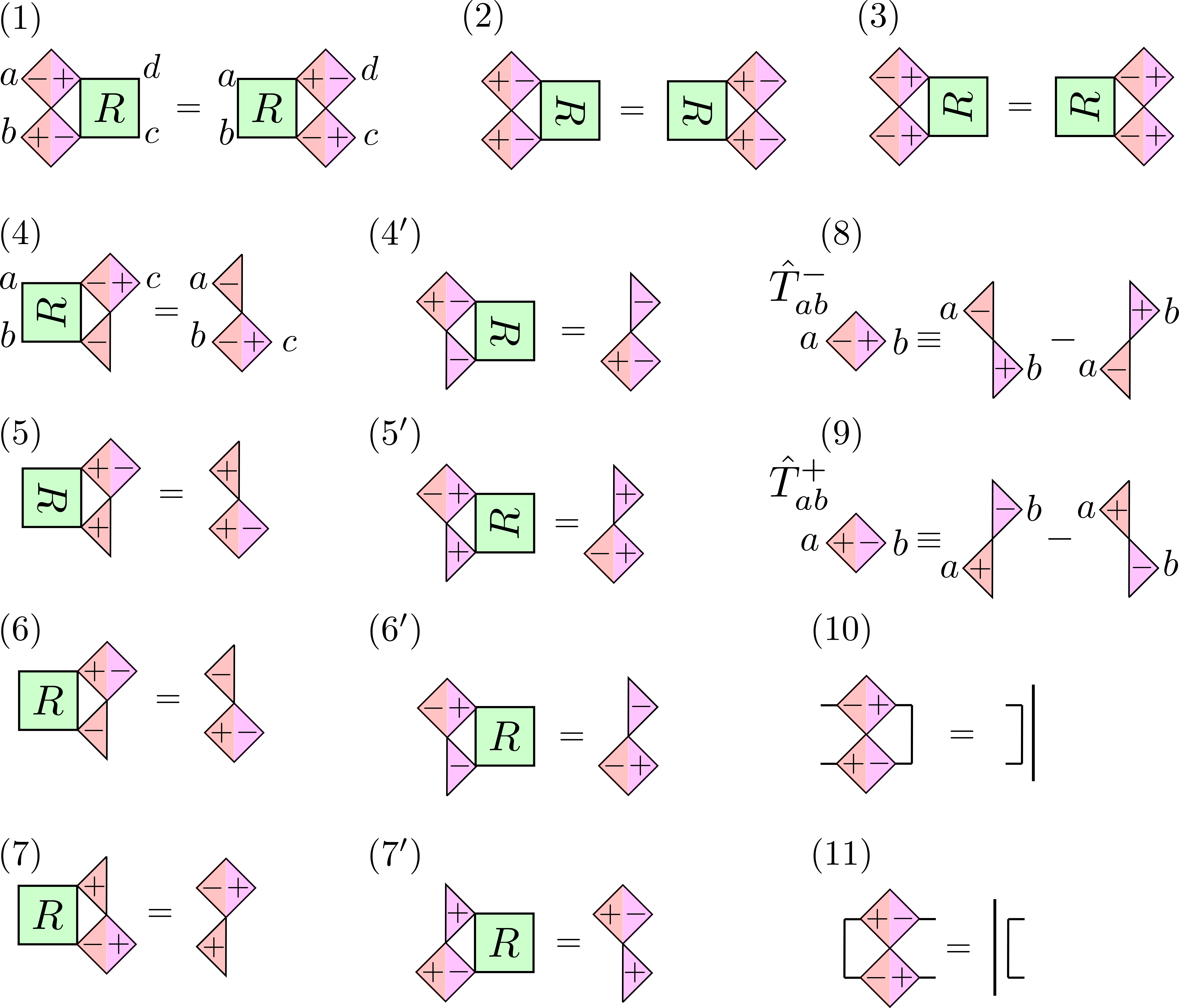}}
	\caption{\label{fig:RSS}Graphical representation of the CRs between the local spin operators $\hat{T}^\pm_{ab}$ and $\{\hat{x}^\pm_{a},\hat{y}^\pm_{a}\}^m_{a=1}$.}
\end{figure}
We now prove that the emergent paraparticles' creation and annihilation operators defined by the MPO JWT in Eq.~\eqref{eq:JWT_string} do satisfy the parastatistical CRs in Eq.~\eqref{eq:fundamental_Rcommu}, and the spin Hamiltonian in Eq.~\eqref{eq:Hamil1Dspin} is mapped to the free paraparticle Hamiltonian in Eq.~\eqref{eq:Hamil1Dspin_para}. An important first step is to prove the algebraic relations between the local spin operators $\hat{T}^\pm_{ab}$ and $\{\hat{x}^\pm_{a},\hat{y}^\pm_{a}\}^m_{a=1}$ as shown graphically in Fig.~\ref{fig:RSS}. 

\subsubsection{Proof of CRs in Fig.~\ref{fig:RSS}}
Indeed, all the CRs in Fig.~\ref{fig:RSS} can be proved by straightforward computations using the explicit definition of $\hat{x}^\pm_{a},\hat{y}^\pm_{a}$ and $\hat{T}^\pm_{ab}$ given above. A smarter proof strategy is given below.

First, we use Fig.~S3.8 and Fig.~S3.9 as the definitions of the tensors
$T^-=\begin{tikzpicture}[baseline={([yshift=-.6ex]current bounding box.center)}, scale=.8]
	\Tmmatr{0}{0}
\end{tikzpicture}$ and $T^+=\begin{tikzpicture}[baseline={([yshift=-.6ex]current bounding box.center)}, scale=.8]
	\Tpmatr{0}{0}
\end{tikzpicture}$, respectively. 
Then Fig.~S3.7 can be proved easily using Fig.~S2.2 and $[\hat{x}^+_a,\hat{y}^+_b]=0$, and Fig.~S3.7' is proved similarly. Then Fig.~S3.5 is proved as follow. Denote this equation as $\hat{l}_{A}=\hat{r}_{A}$, where $A$ is a collective label for all the open indices. Notice that\\
(1). This equation holds when acting on $\ket{0}$: $\hat{l}_{A}\ket{0}=\hat{r}_{A}\ket{0}$;\\
(2). Both sides of this equation transform in the same way when we commute them with $\hat{x}^+_a$, i.e.
\begin{eqnarray}
	\hat{l}_{A}\hat{x}^+_a=\sum_{B,b} w^{Bb}_{Aa}\hat{x}^+_{b}\hat{l}_{B},\nonumber\\
	\hat{r}_{A}\hat{x}^+_a=\sum_{B,b} w^{Bb}_{Aa}\hat{x}^+_{b}\hat{r}_{B},
\end{eqnarray}
where $w^{Bb}_{Aa}$ are some constant coefficients.
\\
We can therefore conclude that $\hat{l}_A=\hat{r}_A$ in the whole space since the whole space is spanned by states of the form $\hat{x}^+_{a_1}\ldots \hat{x}^+_{a_n}\ket{0}$.
Fig.~S3.5' is proved in the same way, the only difference is that in step (2) above we use the fact that both sides  transform in the same way when we commute them with $\hat{y}^+_a$, since Fig.~S3.7 tells us how $\hat{T}^-$ commutes with $\hat{y}^+_a$.

Now that S3.5, S3.5', S3.7, and S3.7' are proved, we know how $T^{\pm}$ commutes with $\hat{x}^+_a,\hat{y}^+_a$. All the remaining equations in Fig.~\ref{fig:RSS} can be proved in the same way using steps (1) and (2) above, i.e. by showing that the equation is true when acting on $\ket{0}$, and that both sides transform in the same way when we commute them with either $\hat{x}^+_a$ or $\hat{y}^+_a$, implying that the equation holds on the whole space.

\subsubsection{The spin-paraparticle mapping via the MPO JWT  %
}
\begin{figure}
	\center{\includegraphics[width=0.5\linewidth]{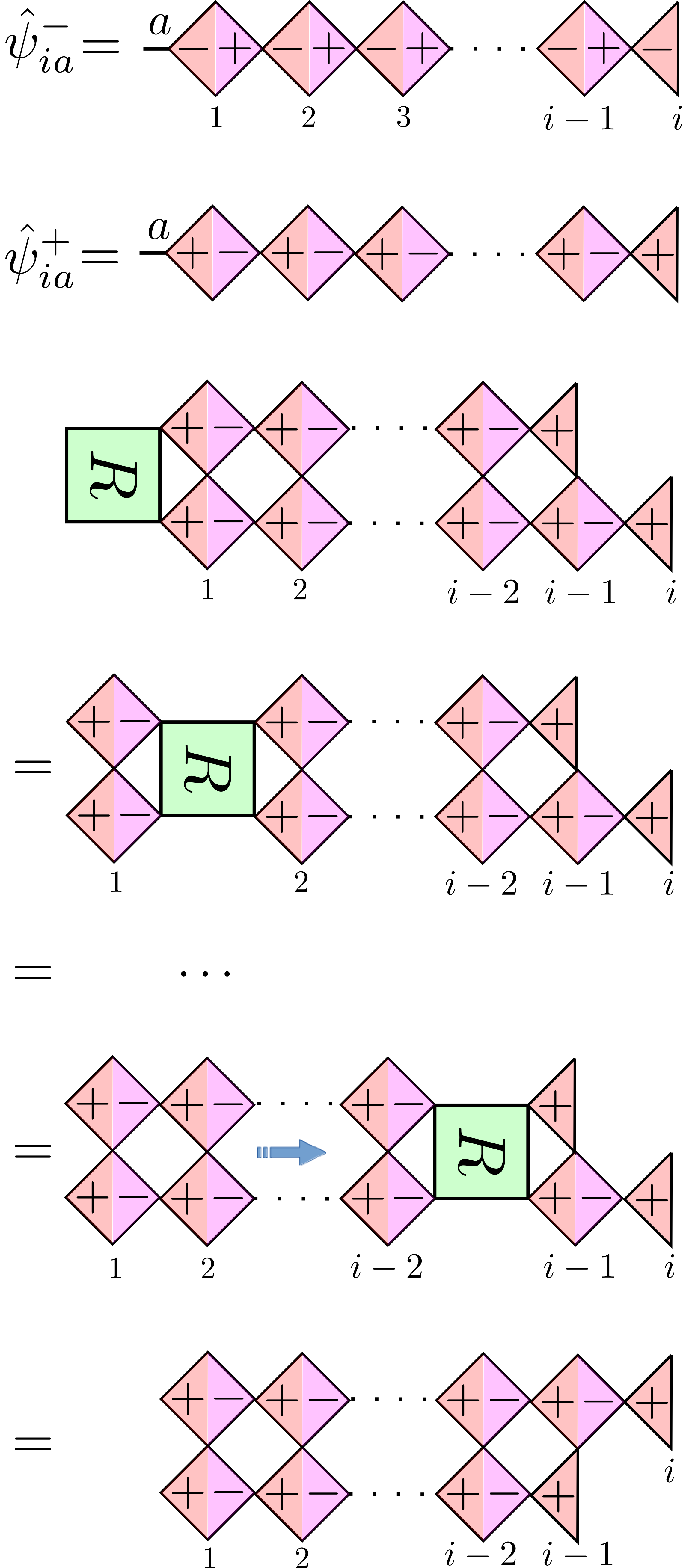}}
	\caption{\label{fig:JW} The graphical proof of the first relation in Eq.~\eqref{eq:fundamental_Rcommu}, using the definition of $\hat{\psi}^\pm_{i,a}$ in Eq.~\eqref{eq:JWT_string} and the algebraic relations in Fig.~\ref{fig:RSS}.}
\end{figure}

With all those relations shown in Fig.~\ref{fig:RSS}, Eq.~\eqref{eq:fundamental_Rcommu} can be proven. For example, in Fig.~\ref{fig:JW} we show the proof of the first parastatistical CR in Eq.~\eqref{eq:fundamental_Rcommu} for the $\hat{\psi}^\pm_{i,a}$ defined in terms of the spin operators in Eq.~\eqref{eq:JWT_string} and the algebraic relations in Fig.~\ref{fig:RSS}. Other relations in Eq.~\eqref{eq:fundamental_Rcommu} are proven in a similar way. Furthermore, one can insert Eq.~\eqref{eq:JWT_string} into Eq.~\eqref{eq:Hamil1Dspin_para} to reproduce Eq.~\eqref{eq:Hamil1Dspin}, using the last two relations in Fig.~\ref{fig:RSS} and a similar graphical manipulation as in Fig.~\ref{fig:JW}. This proves the exact mapping from the 1D spin model to free paraparticles.

\section{The 2D solvable spin model with emergent paraparticles}\label{SI:2DparaKDH}
In this section we provide technical details for the 2D solvable spin models with emergent free paraparticles, introduced in the Methods of the main text. 
Specifically, in Sec.~\ref{SI:modeldef} we define the tensors $u^\pm_L,u^\pm_R,v^\pm_L,v^\pm_R$ that appear in the three body interaction terms, in Sec.~\ref{SI:facts_2DHamiltonian} we prove two important properties of the solvable spin Hamiltonian, and then in Sec.~\ref{SI:2DMPOJWT} we prove important properties of the 2D MPO JWT which eventually lead to free paraparticle representation in Thm.~\ref{thm:main_2DESM}.  %
Finally, in Sec.~\ref{SI:paracreation_measure} we show that paraparticles in the 2D solvable spin model can be locally created and measured at special points on the boundary. 
\subsection{Definition of the tensors $u_L,u_R,v_L,v_R$}\label{SI:modeldef}

Let \begin{tikzpicture}[baseline={([yshift=-.6ex]current bounding box.center)}, scale=.8]
	\umatr{0}{1}{u^+}
\end{tikzpicture} and \begin{tikzpicture}[baseline={([yshift=-.6ex]current bounding box.center)}, scale=.8]
	\umatr{0}{1}{u^-}
\end{tikzpicture}
be tensors with 4 indices, where the horizontal ones are parastatistical indices~(taking values $1,\ldots, m$) and the vertical~(taking values $1,\ldots, Q$) are quantum, and suppose that they satisfy the tensor equations 
\begin{eqnarray}\label{eq:uvInverse}
	\begin{tikzpicture}[baseline={([yshift=.4ex]current bounding box.center)}, scale=.8]
		\umatrix{0}{1}{u^+}{}
		\umatrix{0}{0}{u^-}{}
		\indexflowin{0}{1}{1}{}
		\indexflowout{0}{0}{1}{}
		\deltatensorR{-0.5}{1}
	\end{tikzpicture}~=~ 
	\begin{tikzpicture}[baseline={([yshift=.4ex]current bounding box.center)}, scale=.8]
		\deltavert{0}{0}
		\deltatensorR{0.5}{0.5}
	\end{tikzpicture}~~,\quad
	\begin{tikzpicture}[baseline={([yshift=.4ex]current bounding box.center)}, scale=.8]
		\umatrix{0}{1}{u^-}{}
		\umatrix{0}{0}{u^+}{}
		\indexflowin{0}{1}{-1}{}
		\indexflowout{0}{0}{-1}{}
		\deltatensor{0.5}{1}
	\end{tikzpicture}~=~ 
	\begin{tikzpicture}[baseline={([yshift=.4ex]current bounding box.center)}, scale=.8]
		\deltavert{0}{0}
		\deltatensor{-0.5}{0.5}
	\end{tikzpicture}~,
\end{eqnarray}
where arrows are used to distinguish between input and output indices when we group the parastatistical and quantum indices to view $u^+$ and $u^-$ as $mQ\times mQ$ matrices, and we use the convention that vertical indices always go from top to bottom by default. More precisely, with matrix notation, we have
\begin{equation}\label{eq:matrixvstensorindexing}
	[u^+]^{b\beta}_{a\alpha}=
	\begin{tikzpicture}[baseline={([yshift=-.6ex]current bounding box.center)}, scale=.8]
		\umatr{0}{0}{u^+}
		\node  at (0-1*\AL,0) [left] {\footnotesize $a$};
		\node  at (0+1*\AL,0) [right] {\footnotesize $b$};
		\quantumindices{0}{0}{\beta}{\alpha}
	\end{tikzpicture},\quad [u^-]^{a\alpha}_{b\beta}=
	\begin{tikzpicture}[baseline={([yshift=-.6ex]current bounding box.center)}, scale=.8]
		\umatr{0}{0}{u^-}
		\node  at (0-1*\AL,0) [left] {\footnotesize $a$};
		\node  at (0+1*\AL,0) [right] {\footnotesize $b$};
		\quantumindices{0}{0}{\alpha}{\beta}
	\end{tikzpicture}.
\end{equation} 
When we collect indices in this way, Eq.~\eqref{eq:uvInverse} simply means that $u^+$ and $u^-$ are matrix inverse of each other. We use $u$ to collectively denote the tensors $u^+,u^-$. We will also use the notation $\hat{u}^+_{ab}$ to denote the quantum operator $\begin{tikzpicture}[baseline={([yshift=-.6ex]current bounding box.center)}, scale=.8]
	\umatr{0}{1}{u^+}
	\node  at (0-1*\AL,1) [left] {\footnotesize $a$};
	\node  at (0+1*\AL,1) [right] {\footnotesize $b$};
\end{tikzpicture}$ and similarly $\hat{u}^-_{ab}$ to denote the operator $\begin{tikzpicture}[baseline={([yshift=-.6ex]current bounding box.center)}, scale=.8]
	\umatr{0}{1}{u^-}
	\node  at (0-1*\AL,1) [left] {\footnotesize $a$};
	\node  at (0+1*\AL,1) [right] {\footnotesize $b$};
\end{tikzpicture}$.

\begin{definition} 
	Let $R$ and $R'$ be two involutive solutions to the YBE with the same $m$~(i.e., same dimension). 
	We say that the tensors $u,v$ satisfy the $(R,R')$ commutation relation, denoted as $u\xleftrightarrow{RR'}v$, 
	if the following systems of relations hold: 
	\begin{eqnarray}\label{eq:uuRSuugraphical}
		\begin{tikzpicture}[baseline={([yshift=.4ex]current bounding box.center)}, scale=.8]
			\umatrix{0}{1}{u^+}{}
			\umatrix{0}{0}{v^+}{q}
			\indexflowin{0}{1}{1}{1}
			\indexflowin{0}{0}{1}{2}
			\RmatrixRo{-1}{0.5}{R}{270}
			\indexflowout{-1}{1}{-1}{1}
			\indexflowout{-1}{0}{-1}{2}
		\end{tikzpicture}
		&=& 
		\begin{tikzpicture}[baseline={([yshift=.4ex]current bounding box.center)}, scale=.8]
			\umatrix{0}{1}{v^+}{}
			\umatrix{0}{0}{u^+}{q}
			\indexflowout{0}{1}{-1}{1}
			\indexflowout{0}{0}{-1}{2}
			\RmatrixRo{1}{0.5}{R'}{270}
			\indexflowin{1}{1}{1}{1}
			\indexflowin{1}{0}{1}{2}
		\end{tikzpicture}~~,\nonumber\\
		\begin{tikzpicture}[baseline={([yshift=.4ex]current bounding box.center)}, scale=.8]
			\umatrix{0}{1}{u^-}{}
			\umatrix{0}{0}{v^-}{q}
			\indexflowout{0}{1}{1}{2}
			\indexflowout{0}{0}{1}{1}
			\RmatrixRo{-1}{0.5}{R}{90}
			\indexflowin{-1}{1}{-1}{2}
			\indexflowin{-1}{0}{-1}{1}
		\end{tikzpicture}&=& 
		\begin{tikzpicture}[baseline={([yshift=.4ex]current bounding box.center)}, scale=.8]
			\umatrix{0}{1}{v^-}{}
			\umatrix{0}{0}{u^-}{q}
			\RmatrixRo{1}{0.5}{R'}{90}
			\indexflowout{1}{1}{1}{2}
			\indexflowout{1}{0}{1}{1}
			\indexflowin{0}{1}{-1}{2}
			\indexflowin{0}{0}{-1}{1}
		\end{tikzpicture}~~,\nonumber\\
		\begin{tikzpicture}[baseline={([yshift=.4ex]current bounding box.center)}, scale=.8]
			\umatrix{0}{1}{v^+}{}
			\umatrix{0}{0}{u^-}{q}
			\RmatrixRo{-1}{0.5}{R}{0}
			\indexflowin{-1}{1}{-1}{1}
			\indexflowout{-1}{0}{-1}{1}
			\indexflowin{0}{1}{1}{2}
			\indexflowout{0}{0}{1}{2}
		\end{tikzpicture}
		&=& 
		\begin{tikzpicture}[baseline={([yshift=.4ex]current bounding box.center)}, scale=.8]
			\umatrix{0}{1}{u^-}{}
			\umatrix{0}{0}{v^+}{q}
			\RmatrixRo{1}{0.5}{R'}{0}
			\indexflowin{0}{1}{-1}{1}
			\indexflowout{0}{0}{-1}{1}
			\indexflowin{1}{1}{1}{2}
			\indexflowout{1}{0}{1}{2}
		\end{tikzpicture}~~,\nonumber\\
		\begin{tikzpicture}[baseline={([yshift=.4ex]current bounding box.center)}, scale=.8]
			\umatrix{0}{1}{u^+}{}
			\umatrix{0}{0}{v^-}{q}
			\RmatrixRo{-1}{0.5}{R}{0}
			\indexflowin{-1}{1}{-1}{1}
			\indexflowout{-1}{0}{-1}{1}
			\indexflowin{0}{1}{1}{2}
			\indexflowout{0}{0}{1}{2}
		\end{tikzpicture}&=& 
		\begin{tikzpicture}[baseline={([yshift=.4ex]current bounding box.center)}, scale=.8]
			\umatrix{0}{1}{v^-}{}
			\umatrix{0}{0}{u^+}{q}
			\RmatrixRo{1}{0.5}{R'}{0}
			\indexflowin{0}{1}{-1}{1}
			\indexflowout{0}{0}{-1}{1}
			\indexflowin{1}{1}{1}{2}
			\indexflowout{1}{0}{1}{2}
		\end{tikzpicture}~.
	\end{eqnarray}
\end{definition}
\begin{remark}\label{rmk:Ruvimplication}
	If we view $u^\pm$ and $v^\pm$ as a $mQ\times mQ$ matrices~(with index convention explained above), Eq.~\eqref{eq:uuRSuugraphical} can be rewritten as the following matrix equations
	\begin{eqnarray}\label{eq:uuRSuumatrix}
		u^+_{1q}v^+_{2q}R_{12}&=&R'_{12}v^+_{1q}u^+_{2q},\nonumber\\
		R_{12} u^-_{2q}v^-_{1q}&=&v^-_{2q}u^-_{1q}R'_{12},\nonumber\\
		v^+_{2q}R_{12}u^-_{2q}&=&u^-_{1q}R'_{12}v^+_{1q},\nonumber\\
		u^+_{2q}R_{12}v^-_{2q}&=&v^-_{1q}R'_{12}u^+_{1q},
	\end{eqnarray}
	where both sides of the equation act on a $m\times m\times Q$ dimensional product vector space $V_1\otimes V_2\otimes V_q$, and the subscripts indicate which factor space a tensor acts on. It is straightforward to see that the four equations in Eq.~\eqref{eq:uuRSuumatrix} are equivalent to each other under the invertibility condition~\eqref{eq:uvInverse}. 
	We present all four for readers' convenience, since they all are important for the 2D MPO JWT. %
\end{remark}
\begin{remark}\label{rmk:TRRT}
	Recall that in defining the 1D MPO JWT  we introduced  the tensors $T^+=\begin{tikzpicture}[baseline={([yshift=-.6ex]current bounding box.center)}, scale=.8]
		\Tpmatr{0}{0}
	\end{tikzpicture}$ and  $T^-=\begin{tikzpicture}[baseline={([yshift=-.6ex]current bounding box.center)}, scale=.8]
		\Tmmatr{0}{0}
	\end{tikzpicture}$. If we use the same tensor indexing convention as $u^\pm$ explained above, then their CRs  shown in Fig.~\ref{fig:RSS} can be succinctly rewritten as $T\xleftrightarrow{RR}T$.
\end{remark}
Now let $u_L,v_L,u_R,v_R$ be invertible~[i.e. Eq.~\eqref{eq:uvInverse} holds for $u=u_L,v_L,u_R,v_R$] tensors satisfying the diagram of CRs shown in Fig.~\ref{fig:CRuLvLuRvR}. Such tensors can always be constructed~\footnote{Note that when $R'=X$, Eq.~\eqref{eq:uuRSuugraphical} is very similar to Eq.~(1) of Ref.~\cite{wang2024hopf}, where $R$ corresponds to the solvable gate $U$, and the tensors $u,v$ correspond to the tensors $\rho,v$. Indeed, one can use Eq.~(2) of Ref.~\cite{wang2024hopf} to construct the tensors $u,v$ here if we know the underlying Hopf algebra for the $R$-matrix. } for any $R$ matrix associated with a finite dimensional $\mathbb{C}^*$-triangular Hopf algebra~\cite{Radford1993MQHA,Majid1995BookFoundationQG,etingof1998THAconstruction,etingof2000semisimpleTHAclassification}, and in this construction $Q$ is the dimension of the Hopf algebra. In particular, the set-theoretical $R$ matrix defined in Eq.~\eqref{eq:seth-R} is associated with a 64-dimensional Hopf algebra $\mathcal{H}_{64}$~\footnote{The minimal $\mathbb{C}^*$ triangular Hopf algebra $\mathcal{H}_{64}$ is constructed using the method introduced in Ref.~\cite{etingof1998THAconstruction}, and it plays a key role in the construction of the tensors $u_L,v_L,u_R,v_R$. Our construction of $u_L,v_L,u_R,v_R$ from a minimal triangular Hopf algebra is partially motivated by the relation between MPO and Hopf algebra introduced in Ref.~\cite{molnar2022matrix}.  We here mention that it may also be interesting to realize emergent paraparticles in higher dimensional fermionic systems, where triangular Hopf superalgebras~\cite{andruskiewitsch2001THAChevalley,etingof2001classificationTHAchevalley} may be useful tools in constructing exactly solvable models~(this is partially motivated by the connection between Lie superalgebras~\cite{Kac1977LieSuper,Rittenberg1978,Rittenberg1978a} and Green's parastatistics described in some recent works~\cite{Toppan2021Z2Z2,Toppan2021Inequivalent}). }, so in this case we have $m=4$ and $Q=64$. The detailed mathematical construction is quite technical, which we omit here, but in the accompanying Mathematica code~\cite{MmaCode}, we give numerical representations of these tensors~[in the  code, these tensors are stored as $mQ\times mQ$ matrices, using the indexing convention in Eq.~\eqref{eq:matrixvstensorindexing}%
], and verify that the satisfy all the CRs in Fig.~\ref{fig:CRuLvLuRvR}. 
\begin{figure}
	\centering
	\begin{tikzpicture}[baseline={([yshift=.ex]current bounding box.center)}, scale=.5]
		\node (uL) [] {\large $u_L$};
		\node (vL) [right= 3cm of uL] {\large $v_L$};
		\node (vR) [below= 2cm of uL] {\large $v_R$};
		\node (uR) [below= 2cm of vL] {\large $u_R$};
		\draw [very thick, <->, >=latex'] (uL) --  node[above=0.15cm] {$XX$} (uR);
		\draw [very thick, <->, >=latex'] (vL) --  node[below=0.15cm] {$XX$} (vR);
		\draw [very thick, <->, >=latex'] (uL) --  node[above] {$RX$}(vL);
		\draw [very thick, <->, >=latex'] (uR) --  node[below] {$RX$} (vR);   
		\draw [very thick, <->, >=latex'] (uR) --  node[right] {$XR$} (vL);
		\draw [very thick, <->, >=latex'] (uL) --  node[left] {$XR$} (vR);
		\draw [very thick, <->, >=latex'] (uL) to [out=90,in=180,looseness=13] node [below right] {$RR$} (uL); 
		\draw [very thick, <->, >=latex'] (vL) to [out=0,in=90,looseness=13] node [below left] {$RR$} (vL);
		\draw [very thick, <->, >=latex'] (vR) to [out=180,in=270,looseness=13] node [above right] {$RR$} (vR);
		\draw [very thick, <->, >=latex'] (uR) to [out=270,in=0,looseness=13] node [above left] {$RR$} (uR);
	\end{tikzpicture}
	\caption{\label{fig:CRuLvLuRvR} The diagram of CRs satisfied by the tensors $u_L,v_L,u_R,v_R$. The two sided arrow  $u\xleftrightarrow{RR'}v$ means that $u,v$ satisfy the $(R,R')$ CR defined in Eq.~\eqref{eq:uuRSuugraphical}. 
		A self loop on $u$ with relation $RR$ means  $u\xleftrightarrow{RR}u$. $X$ is the two qudit swap gate~($X$ has the same dimension as $R$).}
\end{figure}
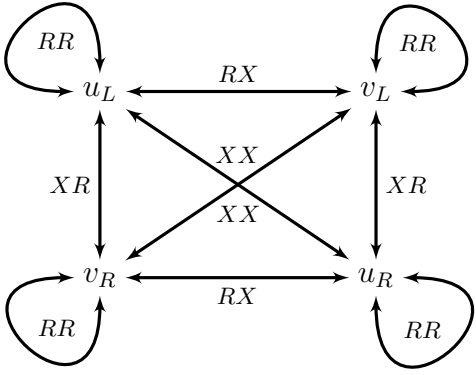
\begin{remark}\label{rmk:signconvention}
	Notice that Eq.~\eqref{eq:uuRSuugraphical} is invariant under a sign flip of either the $u$ tensor or the $v$ tensor, therefore we have a freedom to choose the signs of the tensors $u_L,v_L,u_R,v_R$. Indeed, we can choose their signs independently at each  white site of the 2D lattice. The mapping to free paraparticles is valid regardless of this sign choice.  However, this sign choice does affect the spectrum of the resulting free paraparticle Hamiltonian $\hat{H}_2$ in Eq.~\eqref{eq:freeparaH2}, since the signs of the tunneling constants of the free paraparticles depend on this sign choice. In our specific model, it turns out that the  tensors  $u_L,v_L,u_R,v_R$  we construct from the Hopf algebra $\mathcal{H}_{64}$ 
	satisfy
	\begin{equation}\label{eq:uvtensorsignconvention}
		v^+_Lu_L^{-}v^+_R u_R^{-}=-1.
	\end{equation}
	which is verified in the accompanying Mathematica code~\cite{MmaCode}. If we directly use these tensors to construct the 2D solvable spin model in Eq.~\eqref{def:2DsolvableH}, the resulting free paraparticle Hamiltonian will have a $\pi$-flux on each square plaquette of the lattice~(including all the white, gray, and colored plaquettes in Fig.~\ref{fig:mainlattice}), which brings some unnecessary inconvenience for our discussions later. We therefore  use our freedom of choosing the signs of $u_L,v_L,u_R,v_R$  and flip the signs of $v_L$ and $u_L$ on each horizontal triangle in Fig.~\ref{fig:mainlattice}~(a triangle is horizontal if its longest edge is horizontal; for example the $v_R$-triangle with vertices $0,1,2$ in Fig.~\ref{fig:mainlattice} is horizontal). We use this sign convention throughout this paper: whenever we mention a $u_L$ or $v_L$ tensor on a horizontal triangle, we mean the sign-flipped one. Notice that this sign-flip also changes the signs of the conserved loop terms $\hat{A}_\nu$ and $\hat{B_p}$, since each of them involves exactly one of $u_L,v_L$. This is why in Eq.~\eqref{def:2DsolvableH}, we define $\hat{H}_1=\sum_{\nu}\hat{A}_\nu+\sum_p \hat{B}_p$, so that with the additional sign-flip, our definition actually agrees with the definition $\hat{H}=-\sum_{\nu}\hat{A}_\nu-\sum_p \hat{B}_p$ in the quantum double model literature~(see also the note after Fact~\ref{fact:paravacuum}).
\end{remark}

\subsection{Important properties of the solvable Hamiltonian}\label{SI:facts_2DHamiltonian}
We begin by proving a few simple facts about the model defined in Eq.~\eqref{def:2DsolvableH} of the main text.
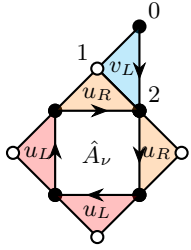
\begin{figure}[h]
	\centering
	\begin{tikzpicture}[baseline={([yshift=.4ex]current bounding box.center)}, scale=.8]
		\hoptriangleE{0}{2*\hL}{cyan}{v_L}{-1}
		\elemDiam{0}{0}
		\StringLabels{{0/2/1/above left,1/1/2/above right,1/3/0/above right}}{black}
		\node at (0,0) {$\hat{A}_\nu$};
	\end{tikzpicture}
	\caption{\label{fig:proof_loop_conservation} Example of two operators that commute: an 8-body term $\hat{A}_\nu$ and a 3-body term $v_L$. (See text for proof.)
	}
\end{figure}
\begin{fact}\label{fact:loopconservation}
	All loop terms in $\hat{H}_1$ mutually commute, and commute with each individual 3-body term in $\hat{H}_2$. Therefore, the loop terms are conserved quantities and eigenstates of $\hat{H}$ can be labeled by their common eigenvalues.  
\end{fact}
\begin{proof}
	This is proved by elementary tensor network manipulations. As an example, we prove the commutativity between the red loop term $\hat{A}_\nu$ and the cyan $v_L$ triangle term $\hat{h}_{02}$ on its top, as shown in Fig.~\ref{fig:proof_loop_conservation}.  Denote by $\hat{A}_\nu=\hat{A}_\nu^++\hat{A}_\nu^-$ and $\hat{h}_{02}=\hat{h}_{02}^++\hat{h}_{02}^-$, where $\hat{A}_\nu^+$ and  $\hat{h}_{02}^+$ are the terms involving $u^+_R$ and $v^+_L$, respectively, and $\hat{A}_\nu^-$ and  $\hat{h}_{02}^-$ are their Hermitian conjugate. We first prove that $[\hat{A}_\nu^+,\hat{h}_{02}^+]=0$. To this end, it is enough to prove that their  actions on the overlapping sites~(labeled $1,2$ in Fig.~\ref{fig:proof_loop_conservation}) commute. We have
	\begin{eqnarray}
		\begin{tikzpicture}[baseline={([yshift=.4ex]current bounding box.center)}, scale=.8]
			\umatrix{0}{1}{v_L^+}{}
			\umatrix{0}{0}{u_R^+}{1}
			\Tpmatrix{1}{0}{2}
			\ytriangle{1}{1}{+}{}
			\node  at (-0.5,1) [left] {\footnotesize $a$};
			\node  at (-0.5,0) [left] {\footnotesize $b$};
		\end{tikzpicture}=
		\begin{tikzpicture}[baseline={([yshift=.4ex]current bounding box.center)}, scale=.8]
			\umatrix{0}{1}{v_L^+}{}
			\umatrix{0}{0}{u_R^+}{1}
			\Tpmatrix{2}{1}{}
			\ytriangle{2}{0}{+}{2}
			\RmatrixRo{1}{0.5}{R}{270}
			\node  at (-0.5,1) [left] {\footnotesize $a$};
			\node  at (-0.5,0) [left] {\footnotesize $b$};
		\end{tikzpicture}
		=
		\begin{tikzpicture}[baseline={([yshift=.4ex]current bounding box.center)}, scale=.8]
			\pimatrix{-1}{0.5}
			\umatrix{0}{1}{u_R^+}{}
			\umatrix{0}{0}{v_L^+}{1}
			\Tpmatrix{1}{1}{}
			\ytriangle{1}{0}{+}{2}
			\node  at (-1.5,1) [left] {\footnotesize $a$};
			\node  at (-1.5,0) [left] {\footnotesize $b$};
		\end{tikzpicture},\nonumber%
	\end{eqnarray}	
	where we used the CR between $u_R$ and $v_L$ defined in Eq.~\eqref{eq:uuRSuugraphical} and Fig.~\ref{fig:CRuLvLuRvR}, and the CR between $\hat{y}^+$ and $\hat{T}^+$ in Fig.~\ref{fig:RSS}. Similarly, we have 
	\begin{eqnarray}
		\begin{tikzpicture}[baseline={([yshift=.4ex]current bounding box.center)}, scale=.8]
			\umatrix{0}{1}{v_L^-}{}
			\umatrix{0}{0}{u_R^+}{1}
			\Tpmatrix{1}{0}{2}
			\ytriangle{1}{1}{-}{}
			\node  at (-0.5,1) [left] {\footnotesize $a$};
			\node  at (-0.5,0) [left] {\footnotesize $b$};
		\end{tikzpicture}=
		\begin{tikzpicture}[baseline={([yshift=.4ex]current bounding box.center)}, scale=.8]
			\umatrix{0}{1}{v_L^-}{}
			\umatrix{0}{0}{u_R^+}{1}
			\Tpmatrix{2}{1}{}
			\ytriangle{2}{0}{-}{2}
			\RmatrixRo{1}{0.5}{R}{0}
			\node  at (-0.5,1) [left] {\footnotesize $a$};
			\node  at (-0.5,0) [left] {\footnotesize $b$};
		\end{tikzpicture}
		=
		\begin{tikzpicture}[baseline={([yshift=.4ex]current bounding box.center)}, scale=.8]
			\pimatrix{-1}{0.5}
			\umatrix{0}{1}{u_R^+}{}
			\umatrix{0}{0}{v_L^-}{1}
			\Tpmatrix{1}{1}{}
			\ytriangle{1}{0}{-}{2}
			\node  at (-1.5,1) [left] {\footnotesize $a$};
			\node  at (-1.5,0) [left] {\footnotesize $b$};
		\end{tikzpicture}, \nonumber
	\end{eqnarray}
	which proves that  $[\hat{A}_\nu^+,\hat{h}_{02}^-]=0$. By taking Hermitian conjugates, we obtain $[\hat{A}_\nu^-,\hat{h}_{02}^+]=[\hat{A}_\nu^-,\hat{h}_{02}^-]=0$, therefore, $[\hat{A}_\nu,\hat{h}_{02}]=0$.
	The commutativity between the loop terms and other triangle terms and between different loop terms is proved similarly. 
\end{proof}
\begin{fact}\label{fact:paravacuum}
	The 2D spin model can be viewed as a significant generalization of Kitaev's quantum double model as follows:
	the space of states in which all qudits on the black dots are in the state $|0\rangle$ is invariant under all individual terms in $\hat{H}_1$ and $\hat{H}_2$. In this sector, $\hat{H}_2$ vanishes and $\hat{H}_1$ reduces to Kitaev's quantum double model~\cite{kitaev2003fault,Buerschaper2009QDMSN,Beigi2011QDMBoundary,Buerschaper2013HATC,Meusburger2017QDMHopfGauge,Jia2023WHAQDM,Cowtan2023QDMBoundary} constructed from a minimal $\mathbb{C}^*$-triangular Hopf algebra $\mathcal{H}_{64}$. In particular, it has a unique ground state with the open boundary condition shown in Fig.~\ref{fig:mainlattice}, where the system size $L$ is odd~($L$ is the total number of black sites in one direction, e.g. $L=7$ in Fig.~\ref{fig:mainlattice}).   %
\end{fact}
\begin{proof}
	We have $\hat{x}^-_a|0\rangle=\hat{y}^-_a|0\rangle=0$ and $\hat{T}^\pm_{ab}|0\rangle=\delta_{ab}\ket{0}$, therefore in this sector all the 3 body interactions vanish, the black dots completely decouple, and the 8 body interactions reduce to 4 body interaction between the white dots. In the accompanying code~\cite{MmaCode} we check in a small lattice with $L=3$ that they are the vertex and plaquette terms in Kitaev's quantum double model based on the Hopf algebra $\mathcal{H}_{64}$, and that $\hat{H}_1$ has a unique ground state~(these claims can all be proved mathematically, which we present in a future work). 
	
	[Note: %
	The precise relation to Kitaev's quantum double model based on $\mathcal{H}_{64}$ is explained as follow. 
	In our model the loop terms $\hat{A}_\nu, \hat{B}_p$ both have eigenvalues $\{-4,-2,0,+2,+4\}$, while in the literature~\cite{kitaev2003fault,Buerschaper2013HATC}  Kitaev's quantum double model is often written as sum of local projectors. However, if we replace $\hat{H}_1$ in Eq.~\eqref{def:2DsolvableH} by  $\hat{H}'_1=-\sum_\nu f(\hat{A}_\nu/2)-\sum_p f(\hat{B}_p/2)$, where    %
	$f(x)=(x+1)x(x-1)(x-2)/4!$ 
	such that $f(\hat{A}_\nu/2)$ and $ f(\hat{B}_p/2)$ are projectors to the lowest eigenstate of  $\hat{A}_\nu$ and $\hat{B}_p$, respectively, then $\hat{H}'_1$ exactly reproduces Kitaev's quantum double model Hamiltonian in the sector where all the black sites are in the state $\ket{0}$. Since $\hat{A}_\nu$ and $\hat{B}_p$ are both conserved, and the ground state subspaces of $\hat{H}_1$ and $\hat{H}'_1$ are exactly the same, using either  $\hat{H}_1$ or $\hat{H}'_1$ in Eq.~\eqref{def:2DsolvableH} lead to the same conclusions. Therefore in this paper, we take the relatively simple choice $\hat{H}_1$, and all the existing knowledge about the ground state of Kitaev's quantum double model still applies to our model.] 
\end{proof}

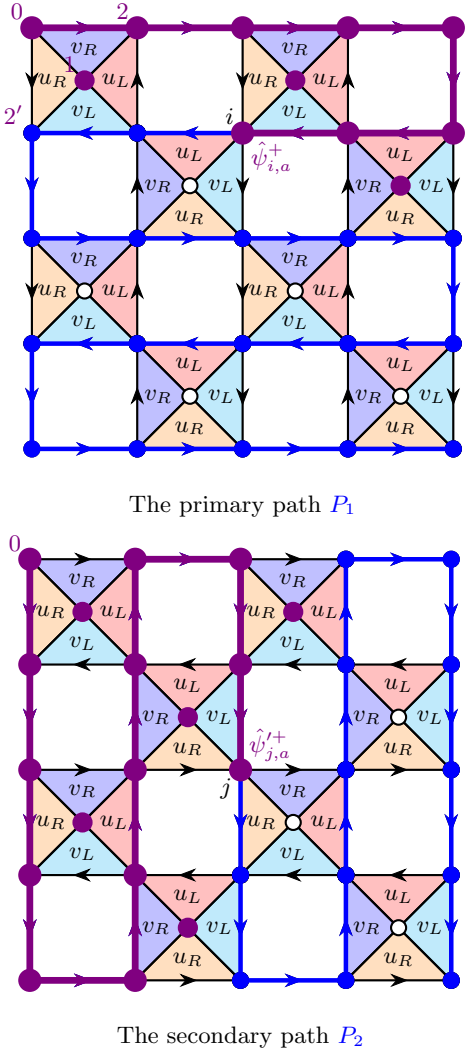
\begin{figure}[h]
	\centering
	\begin{tikzpicture}[baseline={([yshift=.4ex]current bounding box.center)}, scale=1.]
		\OBClattice{0}{0}{1}
		\FullPrimaryPath
		\ParticlePath{{0/0,2/0,4/0,6/0,8/0,8/-2,6/-2,4/-2}}{parapurple}
		\IsoDots{{1/-1,5/-1,7/-3}}{parapurple}%
		\node  at (4*\hL,-2*\hL)[above left] {$i$};
		\node  at (4*\hL,-2*\hL)[below right] {${\color{parapurple}\hat{\psi}^+_{i,a}}$};
		\node  at (4*\hL,-8*\hL)[below=0.5cm] {The primary path ${\color{blue}P_1}$};
		\StringLabels{{0/0/0/above left,2/0/2/above left,1/-1/1/above left,0/-2/2'/above left}}{parapurple}
	\end{tikzpicture}
	\quad
	\begin{tikzpicture}[baseline={([yshift=.4ex]current bounding box.center)}, scale=1.]
		\OBClattice{0}{0}{1}
		\FullSecondaryPath
		\ParticlePath{{0/0,0/-2,0/-4,0/-6,0/-8,2/-8,2/-6,2/-4,2/-2,2/0,4/0,4/-2,4/-4}}{parapurple}
		\IsoDots{{1/-1,1/-5,3/-3,3/-7,5/-1}}{parapurple}
		\node  at (4*\hL,-4*\hL)[below left] {$j$};
		\node  at (4*\hL,-4*\hL)[above right] {${\color{parapurple}\hat{\psi}'^+_{j,a}}$};
		\node  at (4*\hL,-8*\hL)[below=0.5cm] {The secondary path ${\color{blue}P_2}$};
		\StringLabels{{0/0/0/above left}}{parapurple}
	\end{tikzpicture}
	\caption{\label{fig:primarypath} Example of a paraparticle operator defined on the primary path $P_1$~(top) and the secondary path $P_2$~(bottom). The paraparticle operator $\hat{\psi}^\pm_{i,a}\equiv\hat{\psi}^\pm_{i,a}(P_1)$ %
		is a matrix product string operator acting on all the sites on $P_1$ starting at the lattice origin $0$ all the way up to the site $i$, as well as all the auxiliary sites adjacent to the path $P_1$~(all sites on which $\hat{\psi}^\pm_{i,a}$ acts nontrivially are colored purple). The paraparticle operator $\hat{\psi}'^\pm_{j,a}\equiv\hat{\psi}^\pm_{j,a}(P_2)$ on the secondary path $P_2$ is defined similarly. 
	}
\end{figure}

\subsection{The 2D MPO JWT and the proof of Thm.~\ref{thm:main_2DESM}}\label{SI:2DMPOJWT}
In the Methods we defined the paraparticle creation and annihilation operators $\hat{\psi}^\pm_{i,a}$ via a 2D MPO JWT of the form Eq.~\eqref{eq:JWT_string_2D}, on an arbitrary directed path in the lattice connecting the lattice origin to the site $i$. In this section we prove several important properties of these operators which eventually lead to Thm.~\ref{thm:main_2DESM} in the Methods, including their commutativity with the local terms in $\hat{H}_1$, and that their actions in the zero-vortex sector are path-independent and satisfy the fundamental CRs in Eq.~\eqref{eq:fundamental_Rcommu}.

Consider the primary path $P_1$ shown in the left of Fig.~\ref{fig:primarypath}, which travels through all the black sites of the entire lattice. We denote by $\hat{\psi}^\pm_{i,a}\equiv\hat{\psi}^\pm_{i,a}(P_1)$ the paraparticle operators acting on the substring of $P_1$ starting from the lattice origin $0$ and ending at site $i$, as shown in Fig.~\ref{fig:primarypath}, and
the paraparticle operators $\hat{\psi}'^\pm_{j,a}\equiv\hat{\psi}^\pm_{j,a}(P_2)$ on the secondary path $P_2$ are defined similarly. The following fact can be proved in an identical way as in the 1D case
\begin{fact}\label{fact:primary_para_mapping}
	The set of all paraparticle operators $\{\hat{\psi}^\pm_{i,a}\}$ on $P_1$ satisfy the CRs in Eq.~\ref{eq:fundamental_Rcommu} of the main text. For any edge $\langle ij\rangle\in P_1$, we have
	\begin{eqnarray}\label{eq:paratunneling}
		\scalebox{0.8}{\begin{tikzpicture}[baseline={([yshift=.4ex]current bounding box.center)}, scale=.8]
				\xtriangle{0}{0}{-}{i}\umatrix{1}{0}{w^+}{}\ytriangle{2}{0}{+}{j}
				\node at (1,-0.5) [below] {\small $k$};
		\end{tikzpicture}}&=&\sum^m_{a=1} \hat{\psi}^+_{j,a}\hat{\psi}^-_{i,a},\nonumber\\
		\scalebox{0.8}{\begin{tikzpicture}[baseline={([yshift=.4ex]current bounding box.center)}, scale=.8]
				\xtriangle{0}{0}{+}{i}\umatrix{1}{0}{w^-}{}\ytriangle{2}{0}{-}{j}
				\node at (1,-0.5) [below] {\small $k$};
		\end{tikzpicture}}&=&\sum^m_{a=1} \hat{\psi}^+_{i,a}\hat{\psi}^-_{j,a},
	\end{eqnarray}
	for $w=u_L,u_R,v_L$, or $v_R$. %
	A similar result holds for the secondary path $P_2$. Moreover, on every site $i$, we have
	\begin{equation}
		\hat{n}_{i}\equiv \sum^m_{a=1} \hat{\psi}^+_{i,a}\hat{\psi}^-_{i,a}= \sum^m_{a=1} \hat{y}^+_{i,a}\hat{y}^-_{i,a}= \sum^m_{a=1} \hat{\psi}'^+_{i,a}\hat{\psi}'^-_{i,a}.
	\end{equation}
\end{fact}
The space of states mentioned in Fact~\ref{fact:paravacuum} correspond to paraparticle vacuum, since we have $\hat{\psi}^-_{i,a}\ket{0}=\hat{\psi}'^-_{i,a}\ket{0}=0$, therefore, $\hat{n}_i\ket{0}=0$. The following fact describes a construction of all other states in the Hilbert space: %
\begin{fact}\label{fact:state_space_construction}
	The paraparticle operators $\{\hat{\psi}^\pm_{i,a}\}$ commute with all individual terms in $\hat{H}_1$. Moreover, the Hilbert space is spanned by states of the form
	\begin{equation}\label{eq:generic_state_form}
		\hat{\psi}^+_{i_1,a_1}\hat{\psi}^+_{i_2,a_2}\ldots \hat{\psi}^+_{i_n,a_n}\ket{\Omega},
	\end{equation}
	where $n$ is a non-negative integer, $i_1,\ldots,i_n$ are labels of the black sites~(not necessarily distinct), $a_1,\ldots,a_n\in\{1,2,\ldots,m\}$, and  %
	$\ket{\Omega}$ has no paraparticles (is $\ket{0}$ on all black dots) and is a common eigenstate of all the loop operators. 
\end{fact}
\begin{proof}
	That $\{\hat{\psi}^\pm_{i,a}\}$ commute with all loop terms is proved in a similar way as in Fact~\ref{fact:loopconservation}. Since $\{\hat{A}_\nu,\hat{B}_p,\hat{n}_i\}$ mutually commute, the full Hilbert space is spanned by their common eigenstates. For any common eigenstate $\ket{\Psi}$,  %
	let $n_\Psi$ be the total number of paraparticles in $\ket{\Psi}$, i.e. $\hat{n}\ket{\Psi}=n_\Psi\ket{\Psi}$, where $\hat{n}=\sum_i\hat{n}_i$ is the total particle number operator. In the following we use induction on $n_\Psi$ to prove that  $\ket{\Psi}$ is a linear combination of states of the form~\eqref{eq:generic_state_form}.  %
	
	The induction hypothesis is trivially true for $n_\Psi=0$. Now assume that the hypothesis is true for any common eigenstate $\ket{\Psi}$ with $n_\Psi=k$. %
	For a common eigenstate $\ket{\Psi}$ with $n_\Psi=k+1$, let $j$ be a black site such that $\hat{n}_j\ket{\Psi}=n_j\ket{\Psi}>0$. Then we have
	\begin{eqnarray}\label{eq:Psi_induction_step}
		\ket{\Psi}&=&\frac{1}{n_j}\hat{n}_j\ket{\Psi}\nonumber\\
		&=&\frac{1}{n_j}\sum^m_{a=1} \hat{\psi}^+_{j,a}\hat{\psi}^-_{j,a}\ket{\Psi}.%
	\end{eqnarray} 
	It is clear that each state $\hat{\psi}^-_{j,a}\ket{\Psi}$ is either zero, or a common eigenstate of $\{\hat{A}_\nu,\hat{B}_p,\hat{n}_i\}$ with total paraparticle number $n_\Psi-1=k$. By the induction hypothesis, $\hat{\psi}^-_{j,a}\ket{\Psi}$ is a linear combination of states of the form~\eqref{eq:generic_state_form}. Then the second line of Eq.~\eqref{eq:Psi_induction_step} implies that $\ket{\Psi}$ is also a linear combination of states of the form~\eqref{eq:generic_state_form}, which completes the induction step. Therefore, the induction hypothesis is true for any $n_\Psi\in \mathbb{Z}_{\geq 0}$. 
\end{proof}

The next lemma shows that $\{\hat{\psi}^+_{i,a}\}$  and $\{\hat{\psi}'^+_{i,a}\}$ defined on different paths also satisfy the canonical CR in Eq.~\eqref{eq:fundamental_Rcommu} of the main text:
\begin{lemma}\label{lemma:CRcreationP1P2}
	The paraparticle creation operators defined on different paths $P_1$ and $P_2$ satisfy the CRs
	\begin{equation}\label{eq:2DparaCR}
		\hat{\psi}^+_{i,a}(P_1)\hat{\psi}^+_{j,b}(P_2)=\sum_{cd}R^{cd}_{ab} \hat{\psi}_{j,c}^+(P_2) \hat{\psi}_{i,d}^+(P_1).
	\end{equation}
\end{lemma}
[Indeed, Lemma~\ref{lemma:CRcreationP1P2} and the subsequent Lemma~\ref{lemma:path-indep} hold for any two paths $P_1$ and $P_2$ in the lattice that start from the lattice origin $0$, but for the purpose of proving Thm.~\ref{thm:main_2DESM} in the main text, it is enough to only consider the case when $P_1$ and $P_2$ are the primary and second paths shown in Fig.~\ref{fig:primarypath}.]

It is not hard to be convinced about the correctness of Lemma~\ref{lemma:CRcreationP1P2} by considering some simple cases, as the treatment is very similar to the tensor network manipulations in the proof of Fact~\ref{fact:loopconservation}. Below we show the proof of a simple case
where both $i$ and $j$ are adjacent to the lattice origin $0$. The labeling of sites is shown in  Fig.~\ref{fig:primarypath}, so $i=2$ and $j=2'$ in this case. We have a tensor graphical derivation of Eq.~\eqref{eq:2DparaCR}:
\begin{eqnarray}\label{eq:2DparaCRproof}
	\begin{tikzpicture}[baseline={([yshift=.4ex]current bounding box.center)}, scale=.8]
		\RmatrixRo{-2}{0.5}{R}{270}
		\Tpmatrix{-1}{1}{}
		\Tpmatrix{-1}{0}{0}
		\umatrix{0}{1}{v_R^+}{}
		\umatrix{0}{0}{u_R^+}{1}
		\ytriangle{1}{1}{+}{}
		\draw[very thick]  (0.5,0) to (1.5,0);
		\node at (1,-0.5) [below] {\footnotesize $2$};
		\ytriangle{2}{0}{+}{2'}
	\end{tikzpicture}&=&
	\begin{tikzpicture}[baseline={([yshift=.4ex]current bounding box.center)}, scale=.8]
		\RmatrixRo{-1}{0.5}{R}{270}
		\Tpmatrix{-2}{1}{}
		\Tpmatrix{-2}{0}{0}
		\umatrix{0}{1}{v_R^+}{}
		\umatrix{0}{0}{u_R^+}{1}
		\ytriangle{1}{1}{+}{}
		\draw[very thick]  (0.5,0) to (1.5,0);
		\node at (1,-0.5) [below] {\footnotesize $2$};
		\ytriangle{2}{0}{+}{2'}
	\end{tikzpicture}\nonumber\\
	&=&
	\begin{tikzpicture}[baseline={([yshift=.4ex]current bounding box.center)}, scale=.8]
		\Tpmatrix{-1}{1}{}
		\Tpmatrix{-1}{0}{0}
		\umatrix{0}{0}{v_R^+}{1}
		\umatrix{0}{1}{u_R^+}{}
		\ytriangle{2}{1}{+}{}
		\draw[very thick]  (0.5,1) to (1.5,1);
		\node at (2,-0.5) [below] {\footnotesize $2'$};
		\ytriangle{1}{0}{+}{2}
	\end{tikzpicture},\nonumber
\end{eqnarray}
where in the first step we used $T\xleftrightarrow{RR}T$, and in the second step we used $u_R\xleftrightarrow{RX}v_R$.  The full proof of Lemma~\ref{lemma:CRcreationP1P2} that takes into account all possible cases is quite tedious: one begins by proving certain CRs satisfied by the MPO JW strings of $\hat{\psi}^+_{i,a}$ using induction on the string length, where the base case is given by the CRs of $u_L,u_R,v_L,v_R$~(which are essentially length-1 strings) in Fig.~\ref{fig:CRuLvLuRvR}, then one uses the CRs of the JW strings along with the CRs in Fig.~\ref{fig:RSS} to prove Eq.~\eqref{eq:2DparaCR}, in a way similar to the graphical proof of the 1D case shown in Fig~\ref{fig:JW} and the proof of the simple case shown above. We omit the full proof in this paper.

The following lemma proves that the actions of $\hat{\psi}^\pm_{i,a}$ and $\hat{\psi}'^\pm_{i,a}$  coincide in the zero-vortex sector $\Phi_0$: %
\begin{lemma}\label{lemma:path-indep}
	The action of paraparticle creation and annihilation operators $\hat{\psi}^\pm_{i,a}(P)$ in the zero-vortex sector does not depend on the path $P$, i.e. $\hat{\psi}^\pm_{i,a}(P_1)\ket{\Psi}=\hat{\psi}^\pm_{i,a}(P_2)\ket{\Psi},\forall \ket{\Psi}\in \Phi_0$.
\end{lemma}
\begin{proof}
	Let $\ket{G}\in\Phi_0$ be the ground state of $\hat{H}_1$ with $n_i=0$ on every site. It can be shown that the action of $\hat{\psi}^+_{i,a}$ on $\ket{G}$ reduces to a ribbon operator in Kitaev's quantum double model~\cite{kitaev2003fault,Yan2022Ribbon}, and is known to be path independent~\cite{kitaev2003fault}~(we also verify this computationally on a small lattice in the accompanying code~\cite{MmaCode}), i.e. $\hat{\psi}^+_{i,a}\ket{G}=\hat{\psi}'^+_{i,a}\ket{G}$. Then $\hat{\psi}^+_{i,a}-\hat{\psi}'^+_{i,a}$ annihilates all other states in the zero vortex sector as well, since, due to Fact~\ref{fact:state_space_construction}, all other states in the zero-vortex sector can be created by applying products of $\hat{\psi}^+_{i,a}$ on $\ket{G}$, and due to the CR in Lemma~\ref{lemma:CRcreationP1P2}, $\hat{\psi}^+_{i,a}-\hat{\psi}'^+_{i,a}$ can be moved all the way to the right to annihilate $\prod_{i,a}\hat{\psi}^+_{i,a}\ket{G}$. Therefore, $\hat{\psi}^+_{i,a}=\hat{\psi}'^+_{i,a}$ in the zero vortex  sector, and by taking Hermitian conjugate, $\hat{\psi}^-_{i,a}=\hat{\psi}'^-_{i,a}$ as well. 
\end{proof}
From the definition of the two paths $P_1$ and $P_2$ shown in Fig.~\ref{fig:primarypath}, we see that every edge of the lattice  %
either lies on $P_1$ or $P_2$~(or both). 
From Fact~\ref{fact:primary_para_mapping}, if an edge $\braket{ij}\in P_1$, we have
$\hat{h}_{ij}=\sum_{a} J_{ij}\hat{\psi}^+_{j,a}\hat{\psi}^-_{i,a}+\mathrm{h.c.}$ Otherwise $\braket{ij}\in P_2$, and we have $\hat{h}_{ij}=\sum_{a} J_{ij}\hat{\psi}'^+_{j,a}\hat{\psi}'^-_{i,a}+\mathrm{h.c.}$ But Lemma~\ref{lemma:path-indep} claims that the actions of $\{\hat{\psi}^+_{i,a}\}$  and $\{\hat{\psi}'^-_{i,a}\}$ on the zero-vortex sector are exactly the same, therefore, we conclude that in the zero-vortex sector, every 3-body term $\hat{h}_{ij}$  is mapped to $\sum_{a} J_{ij}\hat{\psi}^+_{j,a}\hat{\psi}^-_{i,a}+\mathrm{h.c.}$, completing the proof of Thm.~\ref{thm:main_2DESM}. 

\subsection{Creation and measurement of the paraparticles}\label{SI:paracreation_measure}
When we describe the paraparticle exchange process in Methods, we claim that the paraparticles in the 2D solvable spin model can be locally created and measured at the upper left and lower right corners of the 2D lattice with OBC, as shown in Fig.~\ref{fig:mainlattice}. We prove this claim in the following. We first prove that paraparticles can be locally created at the two corners~(sites $i$ and $j$ in Fig.~\ref{fig:mainlattice}) by applying local operators $y^+_{i,a},y^+_{j,b}$ to the ground state $\ket{G}$. More precisely, we prove the identity
\begin{equation}\label{eq:yypsipsi}
	\hat{y}^+_{i,a}\hat{y}^+_{j,b}\ket{G}=\hat{\psi}^+_{i,a}\hat{\psi}^+_{j,b}\ket{G}\equiv|G;ia,jb\rangle.
\end{equation}
From Eq.~\eqref{eq:JWT_string_2D}, we see that $\hat{\psi}^+_{i,a}=\hat{y}^+_{i,a}$, and
$\hat{\psi}^+_{j,b}=\sum_c \hat{W}_{bc}\hat{y}^+_{j,c}$, where $\hat{W}_{bc}$ is the MPO Jordan-Wigner string connecting sites $i$ and $j$, so we only need to prove that $\hat{W}_{bc}\ket{G}=\delta_{bc}\ket{G}$~(note that $[\hat{W}_{bc},\hat{y}^+_{j,c}]=0$). %
From Fact~\ref{fact:state_space_construction}, we know that $\hat{W}_{bc}$ commutes with all the loop terms $\hat{A}_\nu,\hat{B}_p$ in $\hat{H}_1$, therefore $ \hat{W}_{bc}\ket{G}$ is still a ground state of $\hat{H}_1$. Furthermore, it is straightforward to check that $[\hat{n}, \hat{W}_{bc}]=0$, leading to $\hat{n}\hat{W}_{bc}\ket{G}=0$, i.e., $\hat{W}_{bc}\ket{G}$ has no paraparticles. It follows that $\hat{W}_{bc}\ket{G}$ is also a ground state  of the system. Since the ground state is unique in OBC shown in Fig.~\ref{fig:mainlattice}~(Fact.~\ref{fact:paravacuum}), we have $\hat{W}_{bc}\ket{G}=W_{bc}\ket{G}$, where $W_{bc}$ are some constant numbers. It can be proved that $W_{bc}=\delta_{bc}$~\footnote{This is very natural from a physical viewpoint, because $W_{bc}=\delta_{bc}$ means that when there is only one paraparticle excitation, the index $b$ of the paraparticle does not change when we transport it from the upper left to the lower right corner using $\hat{E}_{ij}$: $\hat{y}^+_{i,b}\ket{G}=\hat{\psi}^+_{i,b}\ket{G}\xrightarrow{\hat{E}_{ij}}\hat{\psi}^+_{j,b}\ket{G}=\hat{y}^+_{i,b}\ket{G}$.  The proof involves technical results about the quantum double ground state, which we do not prove in this paper, but we verify it in a small system in the accompanying code~\cite{MmaCode}}. Therefore Eq.~\eqref{eq:yypsipsi} holds.
Since we assumed in the Methods that the ground state $\ket{G}$ has no paraparticles, i.e., $\hat{n}_l\ket{G}=0$ at any black site $l$, %
the qudit at $l$ is disentangled from the rest of the system and is in the state $\ket{0}$. At any black site $l$, we can find local unitary operators $\hat{U}_a$ that ``implement'' the action of $\hat{y}_a^+$, i.e.,  $\hat{U}_a\ket{0}=\hat{y}_a^+\ket{0}$, leading to  $\hat{U}_{l,a}\ket{G}=\hat{y}_{l,a}^+\ket{G}$. Therefore, we have $|G;ia,jb\rangle=\hat{y}^+_{i,a}\hat{y}^+_{j,b}\ket{G}=\hat{U}_{i,a}\hat{U}_{j,b}\ket{G}$, i.e., the two paraparticle state $|G;ia,jb\rangle$ can be created by applying local unitary operators at the two corners. 

We now prove that the paraparticle indices $a',b'$ can be locally measured at the two corner sites $i$ and $j$ in the final state $\ket{G;ib',ja'}=\hat{\psi}^+_{i,b'}\hat{\psi}^+_{j,a'}\ket{G}=\hat{y}^+_{i,b'}\hat{y}^+_{j,a'}\ket{G}$. 
At any black site $l$, we define the local operator 
\begin{equation}\label{eq:def_color_operator}
	\hat{c}_l=\sum^4_{c=1} c~\hat{y}^+_{l,c}\hat{y}^-_{l,c}. 
\end{equation}
The result of measuring $\hat{c}_i$ at site $i$ is computed as
\begin{eqnarray}
	\hat{c}_i\ket{G;ib',ja'}&=&\sum_{c}c~ \hat{y}^+_{i,c}\hat{y}^-_{i,c}\hat{y}^+_{i,b'}\hat{y}^+_{j,a'}\ket{G}\nonumber\\
	&=&b'~ \hat{y}^+_{i,b'}\hat{y}^+_{j,a'}\ket{G}\nonumber\\
	&=&b'\ket{G;ib',ja'},
\end{eqnarray}
where we used the CR in Eq.~\eqref{eq:XYQA} and the commutativity between $\hat{y}^-_{i,c}$ and $\hat{y}^+_{j,a'}$, and that $\hat{y}^-_{i,c}\ket{G}=0,~\forall c$. Therefore, measuring $\hat{c}_i$ at site $i$ gives a definite result $b'$. Similarly, measuring $\hat{c}_j$ at site $j$ gives $a'$. This completes the proof.

\section{Superselection rules and the observability of parastatistics}\label{SI:superselection}
As we mentioned at the end of the main text, it is straightforward to incorporate relativity into our second quantized theory to get a relativistic quantum field theory of elementary paraparticles. 
Nevertheless, in such a theory there are superselection rules that fundamentally constrain the observability of the exclusion and exchange statistics of paraparticles. In this section we explain this issue in detail, and then
discuss how our proposed realization of emergent paraparticles in condensed matter systems  breaks these superselection rules, which motivates routes to construct theories of elementary paraparticles that are observably distinct from fermions and bosons, in a way evading the no-go theorems~\cite{doplicher1971local,*doplicher1974local}. 

Superselection rules arise because the full state space of our second quantized theory  is a direct sum of exponentially many subspaces, such that
any physical observable has zero matrix element between states of different subspaces. Each subspace is called a superselection sector. 
An immediate consequence of the  superselection rules is that they forbid quantum transitions~(by any local unitary evolution) and thermalization between different superselection sectors. If the system is initialized in one sector, it will stay in that sector forever. This prevents the distinct thermodynamics of free paraparticles~(and hence their exclusion statistics) to be physically observed, since the correct thermodynamic description of the system in equilibrium is through the partition function $Z_\pi=\mathrm{Tr}_\pi[e^{-\beta \hat{H}}]$, where $\mathrm{Tr}_\pi$ means summing over all states in a specific sector $\pi$, and the result is generally different from that in Fig.~\ref{fig:Tempdep} obtained by averaging over the whole space~(all sectors). 
In our second quantized theory of paraparticles, it can be shown that the thermal expectation values of all physical observables in a specific sector are the same as some system of ordinary fermions and bosons, meaning that paraparticles in our second quantized theory cannot be distinguished from ordinary particles by local measurements.
This is reminiscent of the famous Doplicher-Haag-Roberts~(DHR) no-go theorem~\cite{doplicher1971local,*doplicher1974local}, which states, roughly, that any given superselection sector of a paraparticle system is equivalent to a given fixed particle number sector of a system of fermions and bosons. %
This problem does not arise for emergent paraparticles in our quantum spin models defined in Eqs.~\eqref{eq:Hamil1Dspin} and \eqref{def:2DsolvableH}, which have no such superselection rules, since any two states of the full Hilbert space can be connected by some local spin operators $\{\hat{x}^\pm_{i,a},\hat{y}_{i,a}^\pm\}$. %
The non-trivial exclusion and exchange statistics of paraparticles can therefore be physically observed in our solvable spin models. For example, adding an infinitesimal perturbation by such local  operators $\{\hat{x}^\pm_{i,a},\hat{y}_{i,a}^\pm\}$ will induce thermalization between sectors without perturbing the thermodynamic behavior, allowing the distinct thermodynamics of free paraparticles shown in Fig.~\ref{fig:Tempdep} to be physically observed. This is similar to how interactions are necessary to thermalize an ideal gas. Similarly, as we demonstrated in Methods and  
Sec.~\ref{SI:paracreation_measure}, the nontrivial exchange statistics of paraparticles can be physically observed in our 2D spin model, where we exploited the important fact that the local operators $\{\hat{x}^\pm_{i,a},\hat{y}_{i,a}^\pm\}$ allow local creation and measurement of paraparticles at the boundary of the system. 

The way emergent paraparticles in the quantum spin models in Eqs.~\eqref{eq:Hamil1Dspin} and \eqref{def:2DsolvableH} evade the conclusions of the DHR no-go theorem gives us an important hint on the relevance of parastatistics to elementary particles. Despite being a rigorous result, the DHR no-go theorem makes several technical assumptions on the physical systems being considered, one of which is the DHR condition~\cite{doplicher1971local,doplicher1974local,haag2012book}, which essentially %
assumes that all excitations are created by local operators. Yet as we see in Eqs.~\eqref{eq:JWT_string} and \eqref{eq:JWT_string_2D}, the quasiparticles in our spin models are created by 
non-local string operators, %
thereby rendering inapplicable the  DHR theorem in a similar way as the anyonic excitations in Kitaev's toric code model~\cite{kitaev2003fault,naaijkens2017quantum}, whose creation operators are attached by $Z_2$ gauge strings. Although our spin models are non-relativistic, %
one can potentially introduce other local observables %
that are compatible with causality and relativistic covariance but which break the superselection rules, like $\{\hat{x}^\pm_{i,a},\hat{y}_{i,a}^\pm\}$ in the spin models. A promising direction is to consider paraparticles coupled to gauge fields, as the DHR theorem does not apply to quantum gauge theories~\cite{haag2012book}, such as Kitaev's toric code model~($Z_n$ gauge theory) and 
Chern-Simons theories~\cite{Witten1989}, where anyons %
emerge. %
\end{document}